\documentclass[Times,twocolumn,tighten,deluxetable]{aastex63}
\usepackage{graphicx}
\usepackage{multirow}
\usepackage{txfonts}
\usepackage{natbib}
\usepackage{url}
\usepackage{hyperref}
\usepackage{longtable}
\usepackage{tabularx}
\usepackage{lipsum}
\usepackage{array}
\usepackage{rotating}
\usepackage{changepage}

\accepted{for publication in ApJ}

\shorttitle{X-ray Properties of Swift J151857.0-572147}
\shortauthors{K. Chatterjee et al.}

\begin{document}

\title{Interpreting the Spectro-Temporal Properties of the Black Hole Candidate Swift J151857.0-572147 during its First Outburst in 2024}

\correspondingauthor{Kaushik Chatterjee}
\email{mails.kc.physics@gmail.com, kaushik@ynu.edu.cn}

\author[0000-0002-6252-3750]{Kaushik Chatterjee}
\affiliation{South-Western Institue For Astronomy Research, Yunnan University, University Town, Chenggong, Kunming 650500, China}

\author[0009-0002-3317-5213]{S. Pujitha Suribhatla}
\affiliation{University of Rome Tor Vergata, Via Cracovia, 90, 00133 Roma RM, Italy}
\affiliation{Indian Institute of Astrophysics, II Block, Koramangala, Bengaluru 560034, Karnataka, India}

\author[0000-0003-0793-6066]{Santanu Mondal}
\affiliation{Indian Institute of Astrophysics, II Block, Koramangala, Bengaluru 560034, Karnataka, India}

\author[0000-0002-7782-5719]{Chandra B. Singh}
\affiliation{South-Western Institue For Astronomy Research, Yunnan University, University Town, Chenggong, Kunming 650500, China}

%%%%%%%%%%%%%%%%%%%%%%%%%%%%%%%%%%%%%%%%%%%%%%%%%%%%%%%%%%%%%%%%%%%%%%%%%%%%%%%%%%%%%%%%%%%%%%%%%%%%%%%%%%%%%%%%%%%%%%%%%%%%%%%%%%%%%%%%%%%%%%%%%%%%%%%%%%%%%%%%

\begin{abstract}
 
The transient Galactic black hole candidate Swift J151857.0-572147 went through an outburst in March 2024 for the first time. Using publicly archived {\it Insight}-HXMT data, we 
have analyzed the timing and spectral properties of the source. We have extracted the properties of the quasi-periodic oscillations (QPOs) by fitting the power density spectrum, 
which inferred that the QPOs are of type C. We have detected QPOs up to $\sim48$ keV using an energy dependence study of the QPOs. High-frequency QPO is not observed during this 
period. We also conclude that the oscillations of the shock in transonic advective accretion flows may be the possible reason for the origin of the QPOs. In the broad energy band 
of $2-100$ keV, simultaneous data from the three onboard instruments of \textit{Insight}-HXMT were used to perform spectral analysis. Different combinations of models, including 
broken power-law, multi-color disk blackbody, interstellar absorption, non-relativistic reflection in both neutral and ionized medium, and relativistic reflection were used to 
understand the spectral properties during the outburst. We discovered that at the beginning of the analysis period, the source was in an intermediate state and later transitioning 
toward the soft state based on the spectral parameters. It has a high hydrogen column density, which could be due to some local absorption to the source.

\end{abstract}


%%%%%%%%%%%%%%%%%%%%%%%%%%%%%%%%%%%%%%%%%%%%%%%%%%%%%%%%%%%%%%%%%%%%%%%%%%%%%%%%%%%%%%%%%%%%%%%%%%%%%%%%%%%%%%%%%%%%%%%%%%%%%%%%%%%%%%%%%%%%%%%%%%%%%%%%%%%%%%%%

\section{Introduction}

Black hole X-ray binaries (BHXRBs) are quite common and important astronomical binary systems. Since accretion serves as the power source in these systems, it is crucial to understand 
them (Frank, King \& Raine 2002). In BHXRBs, there is a black hole (BH) at the center and a companion star orbiting around. They are classified into two main types based on the companion's 
mass: low-mass BHXRBs (LMBHXRBs) and high-mass BHXRBs (HMBHXRBs) (Remillard \& McClintock 2006). In LMBHXRBs, the compact object is more massive and known as the primary whereas, the 
companion is the secondary object. Transient and persistent sources are the other categories into which BHXRBs are divided, based on the type of variability in their outbursts. While 
BH transient sources occasionally exceed detection levels and primarily remain in quiescence, or the dormant state ($L < 10^{32}$ ~erg/s; Hannikainen et al. 2005 and references therein), 
the flux or count rate of persistent sources remain higher than the detection level most of the time ($L > 10^{36}$ erg/s; Chen et al. 1997 and references therein). Transient sources 
experience outbursts that can endure for several weeks or even months (Tetarenko et al. 2016). Though the population of transient HMXBs is increasing, the majority of reported transients 
are LMXBs (McClintock et al. 2013; Remillard \& McClintock 2006 for a review as well). Debnath et al. (2010) classified the BH outbursts into two main categories based on their nature: 
slow rise slow decay (SRSD) and fast rise slow decay (FRSD). Zhang et al. (2019) divided outbursts into several types, such as glitter, reflare, multipeak, mini-outburst, or new-outburst, 
based on their rebrightening characteristics.

The multicolor thermal black body and the hard non-thermal power-law components combine to form the spectrum of a BH. The origin of the hard component can be explained by the Comptonizing 
region, also known as the `Compton Cloud', which is the repository of hot electrons (Thorne \& Price 1975; Sunyaev \& Titarchuk 1980, 1985). Over the years, many models have been put 
forward to explain the composite spectrum of BHXRBs. For example, the standard disk model (Shakura \& Sunyaev 1973, or SS73), the thick disk model (Paczynski \& Witta 1980), the advection 
dominated accretion flow or ADAF (Ichimaru 1977, Narayan \& Yi 1994), two-component advective flow or TCAF model (Chakrabarti \& Titarchuk 1995), etc. to explain variabilities in BHXRBs. 
The soft component is modeled as the radiation coming from the disk, which was explained well by the SS73 disk. However, the model could not describe the harder part of the spectrum. The 
thick disk model could explain the hard power-law part of the spectrum. This model is relevant for high luminosity states, where radiation pressure dominates. However, advection was not 
taken into consideration in this model. In the ADAF model, the energy generated due to viscous dissipation is advected to the BH and is radiatively inefficient. This model produces a power-law 
spectrum, however, the direct spectral fitting and explaining the spectral and timing properties are still lacking. Alternatively, the TCAF model solves flow equations and couples them with 
the radiative transfer processes to explain both spectral and timing properties simultaneously (Chakrabarti \& Titarchuk 1995, Molteni et al. 1996, Mondal \& Chakrabarti 2013, Chakrabarti 
et al. 2015).

During the rising phase of an outburst, BH binaries show four distinct BH spectral states starting from hard state (HS) to hard intermediate state (HIMS), soft intermediate state (SIMS), 
and soft state (SS) (see Remillard \& McClintock 2006, for a review). An opposite sequence follows in the declining phase. Briefly, a BH's spectral state transition forms a hysteresis 
loop from HS to SS (in the rising phase) and SS to HS (in the declining phase). There is no significant difference in the energy spectrum between HIMS and SIMS. SIMS is classified mostly 
based on the presence of type-B QPOs. A source that experiences all four of the aforementioned spectral states during an outburst, is known to have a complete outburst. If the source does 
not go to a soft state, it is known to have a failed outburst (Tetarenko et al. 2016).

Understanding temporal aspects is crucial to understanding the dynamics of the accreting flow around the BHs. The light curves exhibit variabilities in short timescales, in the 
order of milliseconds to seconds, during an outburst, particularly in the high-energy bands. Variabilities like broadband noise and narrow characteristics in the PDS, can be observed by 
the Fourier transformation of the light curve (van der Klis 1989; Belloni et al. 2002). A power-law function is used to describe the broadband noise, which is dispersed over a wide frequency 
range. The broadband noise can also be fitted with a set of zero-centered \textit{Lorentzians} (Belloni et al. 2002). \textit{Lorentzian} profiles can be used to describe the QPOs, 
which is a power peak in the restricted frequency range (Psaltis et al. 1999; Nowak 2000; Belloni et al. 2002). Because of their geometrical origin, low-frequency QPOs (LFQPOs) are frequently 
detected in BHXRBs. Types A, B, and C are the three categories into which LFQPOs are divided, based on some characteristics such as frequency ($\nu$), $Q$-value ($=\nu/\delta\nu$, where 
$\delta\nu$ is the full width at half maximum or FWHM), (\%) RMS, etc. (Casella et al. 2005). High-frequency QPOs can also be seen in BHXRBs, although it is quite rare. Several models have 
been put forward to understand the origin of QPOs, such as the shock oscillation model (Molteni et al. 1996; Chakrabarti et al. 2005, 2008, 2015), magneto-acoustic waves (Titarchuk et al. 
1998), accretion-ejection instability (Tagger \& Pellat 1999), Lense-Thirring precession (Stella et al. 1999; Ingram et al. 2009), precessing inner flow model (Ingram et al. 2009), 
corrugation modes (Kato \& Fukue 1980; Wagoner 1999; Kato 2001; Tsang \& Butsky 2013), pressure or accretion rate modes (Cabanac et al. 2010), variable Comptonization or vKompth 
model (Karpouzas et al. 2021; Bellavita et al. 2022; Garcia et al. 2022; and references therein), outflow model (Reig et al. 2003; Giannios et al. 2004; Kylafis et al. 2020; and references 
therein), JED-SAD model (Ferreira et al. 1997, 2022; Petruci et al. 2008; Marcel et al. 2019 and references therein), reltrans model (Ingram et al. 2019; Mastroserio et al. 2021; and 
references therein). However, to date, the origin of QPOs is still a topic of debate.

One of the characteristics of the TCAF model (Chakrabarti \& Titarchuk 1995) is the oscillation of the shock. According to the TCAF model, matter supplied by the companion star can have a 
Keplerian and a sub-Keplerian distribution of angular momentum. The Keplerian component creates a geometrically thin and optically thick accretion disk on the equatorial plane and 
flows in on a viscous timescale because of its high viscosity. As the critical viscosity of this matter increases, the accretion disk moves inward. The sub-Keplerian one falls 
radially in on a free-fall timeline and has a less viscous accretion flow. This matter resides both above and below the Keplerian disk. The optically thin sub-Keplerian component 
produces a shock front at the location where both the centrifugal and gravitational forces balance each other. The shock front is the boundary layer, also known as the CENtrifugal 
pressure-supported BOundary Layer (CENBOL), which behaves as a so-called Compton cloud. The Keplerian component can explain the soft multicolor blackbody component. The CENBOL region 
up-scatters the soft photons from the disk and makes them as hard power-law photons. The shock forms farther away from the BH at the beginning of the outburst and the Keplerian disk starts 
forming, therefore HS is observed. The blackbody photons start increasing as the Keplerian disk moves inward over time. Thereby, cooling increases as a great number of soft photons 
intercept the CENBOL. Throughout the process, more photons are released, the flux rises, and the source becomes softer. The CENBOL is completely quenched in a high soft state, and 
the disk approaches the inner most stable orbit. The spectrum becomes soft as a result of only the disk photons now contributing to the radiation. Such profiles of spectral state 
evolution have been observed in several sources in the TCAF scenario (Mondal et al. 2014, Debnath et al. 2015, Jana et al. 2016, Chatterjee et al. 2020, 2021, 2023).

Furthermore, this model explains the QPO properties in addition to the spectral features and their evolutions. When the infall and cooling timescales become comparable, a bigger CENBOL 
($\sim$ a few 100 $r_S$, where $r_S$ is the Schwartzschild radius) in the HS may result in LFQPOs after satisfying the resonance condition of the oscillation of the shock. The CENBOL 
slowly shrinks in size as the outburst progresses, and the QPO's frequency increases (Molteni et al. 1996, Chakrabarti et al. 2015, Garain et al. 2014). Cooling takes over the heating 
timescale as the SS is approached, the CENBOL gets smaller and eventually quenched, and there is no oscillation. As a result, there is no QPO in the SS (see Mondal et al. 2014, 
Debnath et al. 2015, Chakrabarti et al. 2015). During the whole evolution path, for different sizes of the CENBOL, different types of QPOs are observed.

While examining the correspondence between the spectral and timing properties solely in terms of the features of the light curve, such as the hardness ratio (HR), and the hardness 
intensity diagram (HID), a strong association is seen (Homan et al. 2001; Fender et al. 2004; Motta et al. 2011). Additionally, accretion rate ratio intensity diagrams (ARRIDs), can also 
be used to understand the complete cycle of an outburst from the fundamental accretion flow parameters (Mondal et al. 2014, Jana et al. 2016, Chatterjee et al. 2020). The interlinks between 
spectral and temporal features from a purely observational ground can also be addressed using the RMS-intensity diagram, or RID (Munoz-Darias et al., 2011) and HRD (Belloni et al., 2005).

First identified by {\it Swift/XRT} as a GRB (GRB 20240303A; Kennea et al. 2024), the new Galactic transient Swift J151857.0-572147 was found in Swift Trigger 1218452 
\href{https://gcn.nasa.gov/circulars/35849}{(GCN 35849)}. But thereafter, it was determined to be a Galactic transient due to its constant brightness and location in the Galactic plane. 
The RA and Dec of the source were determined to be RA(J2000) = $15h 18m 57.00s$ and Dec(J2000) = $-57d 21^{'} 47.9^{''}$ based on the optimal source localization utilizing XRT instantaneous 
on-board localization (Kennea et al. 2024). On March 4, 2024, during 15 minutes, from 02:13:13.3 to 02:28:08.9 (MJD 60373.1), follow-up radio observations were conducted using the MeerKAT 
telescope at $1.28 GHz$ (L-band) with a bandwidth of $856$ MHz at a flux density of $10$ mJy (Carotenuto et al., 2024; Cowie et al., 2024). The source's nature was identified as consistent 
with an X-ray binary in the hard state by using the inverted radio spectrum ($f(\nu) \propto \nu^\alpha$, where $\nu \sim +0.5$) in conjunction with the photon index. This suggested that 
the source might be a black hole or a neutron star. On March 9, 2024, from UT 10:35:10 to UT 11:06:20 (MJD 60378.45), the Australia Telescope Compact Array (ATCA) simultaneously recorded 
radio observations at frequencies of $5.5$ and $9$ GHz (Saikia et al., 2024). Additionally, their investigation confirmed the source to be a Galactic black hole. Target of Opportunity 
(ToO)  was carried out on this source with an exposure of 1000s by Swift/XRT following the ATCA, as reported by \href{https://www.astronomerstelegram.org/?read=16519}{Del Santo et al. (2024)}. 
According to Del Santo et al. (2024), it was discovered that the combination of the phenomenological disk black body (diskbb) and power-law  (po) models describes the spectrum 
quite well. These discoveries also confirmed that the source is a black hole. The source was detected by INTEGRAL serendipitously on March 8, 9, 10, and 11 of 2024 (Sguera 2024). The 60cm 
Robotic Eye Mount (REM) telescope observed the source in both optical and near-infrared wavelengths as part of the monitoring program of GRBs (Baglio et al. 2024). Optical measurements of 
the source were also carried out by the Las Cumbers Observatory (LCO) network (Saikia et al. 2024). 

From the {\it Swift/XRT} spectral modeling, Kennea et al. (2024) found a column density of $N_H = 5.6 \pm 0.06 \times 10^{22}$ ~cm$^{-2}$. Additionally, they observed a power-law photon 
index of $\Gamma = 1.78 \pm 0.02$. While Burridge et al. (2024) reported that the source's distance was $4.48^{+0.67}_{-0.47}$ kpc, with an HI absorption towards it, the absence of positive 
velocity absorption lines towards other sources in the field of the HI absorption for this source puts an upper limit on the distance as $15.64^{+0.77}_{-0.60}$ kpc. Mondal et al. (2024) 
reported the mass of the source to be $\sim9.2\pm1.6-10.5\pm1.8 M_\odot$, estimated using the JeTCAF (Jet in TCAF) model (Mondal \& Chakrabarti 2021), where the mass of the BH is a parameter 
and the distance to the source is a scale factor taken care of by the model normalization. Authors also estimated the possible disk inclination of $\sim35^\circ \pm 7^\circ-46^\circ\pm15^\circ$ 
with an average spin parameter of $0.65$ estimated using the KERRBB model (Li et al. 2005) for a given distance of 10 kpc. Since there is no confirmed distance estimation for this source, 
the authors have adopted an average value from the broad range of proposed distances in the literature. Therefore, the estimation of the spin may vary for different source distance values, 
requiring further modeling with a confirmed source distance.

\section{Observation and Data Reduction and Analysis}

This source has recently been observed by Swift satellite and confirmed to be a black hole candidate on March 10, 2024 (Del Santo et al. 2024). After its discovery and confirmation, 
it was monitored by various other X-ray satellites, for example, NICER, NuSTAR, IXPE, etc. We use X-ray data from China's first X-ray satellite mission {\it Insight}-HXMT (Zhang et al. 2020). 
After the onset of the outburst, 7 observation IDs were available publicly when we started our analysis. We list the data in Table 1 below.

Each of these observation IDs has multiple exposures (up to 14 for some). While listing our analysis results, we will list all those exposure IDs with MJD. Using raw data from all these obs 
IDs, we first produced science-analyzable, cleaned data and then performed our analysis. We discuss data reduction and analysis in the following subsections.

\subsection{Data Reduction}

Following the on-demand retrieval of level-1 data from the repository, we generated cleaned level-2 data for scientific study. The raw data cleaning procedure was carried out as follows. 
With the \href{http://hxmten.ihep.ac.cn/software.jhtml}{HXMTDAS}\footnote{http://hxmt.org/index.php/usersp/dataan} (version 2.05) software, we execute the {\fontfamily{pcr}\selectfont 
hpipeline} command using appropriate input and output directories. For each of the three instruments, this pipeline executes a series of automatic commands. However, there are a few 
prerequisites that must be met. Specific parameters were established to achieve good time interval (GTI), such as elevation angle $> 10^\circ$, geomagnetic cutoff rigidity $> 8 ~GV$, 
pointing offset angle $< 0.04^\circ$, and distance from the South Atlantic Anomaly (SAA) $> 600$~s. To facilitate background analyses, each telescope carries large and small field-of-view 
(FOV) detectors. The small FOV detectors are more suitable for pointing observation as they have a lower probability of source contamination\footnote{http://hxmten.ihep.ac.cn/AboutHxmt.jhtml}. 
Together, these commands extract, clean, and produce science products that are ready for analysis. The \href{http://hxmten.ihep.ac.cn/SoftDoc/501.jhtml}{HXMT 
Manual}\footnote{http://hxmten.ihep.ac.cn/SoftDoc/501.jhtml} contains a detailed discussion on this. The spectra for the HE, ME, and LE instruments are generated using the particular 
commands {\fontfamily{pcr}\selectfont hespecgen}, {\fontfamily{pcr}\selectfont mespecgen}, and {\fontfamily{pcr}\selectfont lespecgen}. On the other hand, the light curve files for the 
three instruments are created using the commands {\fontfamily{pcr}\selectfont helcgen}, {\fontfamily{pcr}\selectfont melcgen}, and {\fontfamily{pcr}\selectfont lelcgen}. Appropriate 
response files are generated by {\fontfamily{pcr}\selectfont herspgen}, {\fontfamily{pcr}\selectfont merspgen}, and {\fontfamily{pcr}\selectfont lerspgen}. The commands 
{\fontfamily{pcr}\selectfont hebkgmap}, {\fontfamily{pcr}\selectfont mebkgmap}, and {\fontfamily{pcr}\selectfont lebkgmap} for instruments HE, ME, and LE, respectively, were used to do 
the background subtraction for both the timing and spectral data. We group the spectrum using the {\fontfamily{pcr}\selectfont grppha} task of FTOOLS to a minimum of 30 counts per bin 
for $\chi^2$ fit-statistics in {\fontfamily{pcr}\selectfont XSPEC}. Additionally, to generate appropriate light curves for PDS generation and QPO search, we adjusted the time bin size 
to $0.01$~s. To search for high-frequency QPOs (HFQPOs), we also produced $1$~ms time-binned light curves for all the available exposures. The HE light curve covers a broad range of 
$27-250$~keV. To check the energy dependence of QPOs, we produced $0.01$~s time-binned HE light curves in seven different energy bands ($27-35$, $35-48$, $48-67$, $67-100$, $100-150$, 
$150-200$, $200-250$~keV). Along with this, we also cut light curves in the $48-250$ keV energy band for all the HE exposures. The reason for this will be discussed in later sections.

Detailed analysis using these cleaned light curves and spectra files is discussed in the next subsection.

\begin{table}
\scriptsize
% \addtolength{\tabcolsep}{0.5pt}
 \centering
 \caption{List of Data used. Column 1 lists each observation ID. In columns 2 and 3, we give the start and end date and time of each observation ID. Column 4 gives the exposure time of 
 each observation ID.}
 \label{tab:table1}
 \begin{tabular}{cccccc}
 \hline
 Obs. Id.$^{[1]}$  &    Start UT$^{[2]}$    &     End UT$^{[2]}$    &   Exp. (s)$^{[3]}$ \\
        (1)        &          (2)           &         (3)           &         (4)        \\
\hline
   P0614374001     &  2024-03-04 20:08:55   &  2024-03-06 02:13:33  &        108278      \\
   P0614374002     &  2024-03-06 02:13:31   &  2024-03-08 01:43:18  &        170987      \\
   P0614374003     &  2024-03-08 01:43:22   &  2024-03-10 02:48:37  &        176715      \\
   P0614374004     &  2024-03-10 02:48:37   &  2024-03-12 00:47:47  &        165550      \\
   P0614374005     &  2024-03-12 10:14:15   &  2024-03-12 19:53:12  &         34737      \\
   P0614374006     &  2024-03-13 09:59:33   &  2024-03-15 09:32:39  &        171186      \\
   P0614374008     &  2024-03-17 12:09:52   &  2024-03-17 21:39:51  &         34199      \\
\hline 
 \end{tabular}
\end{table}

\subsection{Data Analysis}

We conducted spectral and temporal research on the black hole candidate (BHC) Swift J151857.0-572147's very first outburst in 2024. First, we created $0.01 ~s$ time-binned light curves using 
data from the LE, ME, and HE modules of the HXMT. The fast Fourier transformation (FFT) in the {\fontfamily{pcr}\selectfont powspec} task of the {\fontfamily{pcr}\selectfont XRONOS} package 
in the {\fontfamily{pcr}\selectfont HEASoft} software was used to construct the power density spectrum (PDS) based on those light curves. The data from each observation was split up into many 
intervals, with 8192 new bins in each interval. To create the final PDS, the PDS for each interval must first be generated and then averaged. The PDS is normalized using the Leahy normalization 
(Leahy et al. 1983). A geometrical rebinning of -1.02 is applied. We used these procedures to look for LFQPOs. Initially, the analysis was done without subtracting the white noise. 
With the concern that it may affect the QPO properties, we have rechecked the analysis by considering the white noise subtraction by running the {\fontfamily{pcr}\selectfont powspec} task to 
produce the PDS using $norm=-1$. In both the cases, using the combination of either \textit{power-law}, \textit{constant}, and \textit{Lorentzian} or multiple \textit{Lorentzian} models, we 
fit all the features of the full PDS continuum from 0.01 to 50 Hz, including the fundamental QPO and harmonic (if present) features. We also estimated several QPO properties such as frequency 
($\nu_{qpo}$), full-width at half maximum (FWHM), and normalization. We found that the normalization of the fundamental QPO barely changes while we use white noise subtraction to that when we 
don't consider white noise subtraction. The normalization value stays well within the error range. The effect of white noise is negligible due to the fact that the signal-to-noise ratio of the 
narrow fundamental QPO feature is high enough to make the contribution from the white noise negligible. We report the results in the next section. We found a sharp peak at the position of $2 
\times \nu_{qpo}$~Hz for some observations, which happen to be the harmonic of the fundamental QPO. We used an additional \textit{Lorentzian} model to fit the harmonic peak. We have extracted 
their properties from the \textit{Lorentzian} model fitting. We fitted the PDS of all the exposures for three energy bands LE ($2-10$~keV), ME ($10-35$~keV), and HE ($27-250$~keV) of the 
listed 7 observations (Table 1). We discuss them later in the result section.

We also studied the energy dependence of the PDS using only the HE light curves. As mentioned above, we extracted $0.01 s$ time-binned HE light curves into 7 different energy bands (mentioned 
in the data reduction section) as HE covers a large energy range. We searched for only those exposures in which LFQPO was present at the full energy band. Using those 7 light curves separately, 
we produced PDS in the same way as mentioned above. Using the same model approach, we extracted QPO information like $\nu_{qpo}$, FWHM, and normalization. We also did the same for the $48-250$
keV HE light curve.

Using these QPO properties ($\nu_{qpo}$, FWHM, norm), we also estimated some properties of the QPOs, like Q-value, RMS (\%), that help designate their nature. We have also estimated the 
QPO significance in the result section following Sreehari et al. (2019).

We also used all three modules (LE, ME, and HE) for spectral analysis, fitting the broadband data in the $2-100$ keV energy range. First, we tried to do the spectral analysis using a combination 
of simple {\fontfamily{pcr}\selectfont disk blackbody} and {\fontfamily{pcr}\selectfont power-law} models. However, we did not find an acceptable fit. This is discussed in the Appendix. 
The combinations of {\fontfamily{pcr}\selectfont disk blackbody}, {\fontfamily{pcr}\selectfont broken power-law} models yielded the best fit for the data, according to our search. We have employed 
the {\fontfamily{pcr}\selectfont tbabs} model for interstellar absorption. Since we are simultaneously fitting all three modules, we have included a {\fontfamily{pcr}\selectfont constant} to 
normalize the three resultant fittings. For LE instrument, we have frozen the constant value to 1 and let the other two constants vary for ME and HE instruments. These are given in each table of 
the spectral fitting results in the Appendix section, where {\fontfamily{pcr}\selectfont constant1} and {\fontfamily{pcr}\selectfont constant2} are for the ME and HE instruments,
respectively. The following is our best model fit combination: {\fontfamily{pcr}\selectfont constant*tbabs*(diskbb + broken power-law)}. We take this as our Model-1. 
We also tried to analyze spectral data using reflection model {\fontfamily{pcr}\selectfont pexrav}. For that, our model combination reads as: {\fontfamily{pcr}\selectfont constant*tbabs*(diskbb 
+ pexrav)}. We take this as our Model-2. We also tried to use the reflection model {\fontfamily{pcr}\selectfont pexriv} which accounts for ionized medium. Thus, {\fontfamily{pcr}\selectfont
constant*tbabs*(diskbb + pexriv)} reads as our Model-3. We also performed the spectral fitting using the relativistic {\fontfamily{pcr}\selectfont relxill} model (Dauser et al. 2016), 
using the model combination as {\fontfamily{pcr}\selectfont constant*tbabs(diskbb + relxill)}. We take this as Model-4. Systematic errors were added to perform spectral analysis, as suggested 
in the HXMT manual. Not every exposure ID of the specified observation IDs was subjected to spectral analysis. We did not spectrally analyze every exposure, compared to the time analysis. Table
2 indicates the spectrally analyzed exposures with a `*'. This is because: in the case of timing analysis, we observed variations in timing properties in a single day, but in the case of spectrum 
analysis, the properties do not significantly change over a short period. We include them in the section on results.

\section{Results}

We discuss our results from the timing and spectral analysis in the following subsections. However, before going into the analysis results, we discuss the variation of the flux of the source 
during the outburst first below.

\subsection{Timing Properties}

First, we will discuss the outburst evolution from the light curve profiles and hardness ratio, and then we will discuss our analysis of QPOs.

\subsubsection{Outburst Profile, and Hardness Ratio}

Although the BHC Swift J151857.0-572147 was observed by the MAXI/GSC instrument, it was not recognized as a new source by them. The source is located at $\sim 0.2^\circ$ from the 
source Cir X-1. Although the facility could identify the brightening of the source, the two sources could not be resolved seperately. In Fig. 1, we show the location of the two sources in 
the upper panel. It can be noticed that the two sources are located very close to each other. The lower panel of the figure shows increased activity due to the outburst of Swift J151857.0-572147. 
According to Peng et al. (2024), Cir X-1 was active in the soft state during this time. Thus, the contribution from Cir X-1 could have contribution to the low energy part (i.e., LE) 
of the light curve. 

\begin{figure}[h]
%\vspace{0.8cm}
\centering
\includegraphics[width=8.5truecm, angle=0]{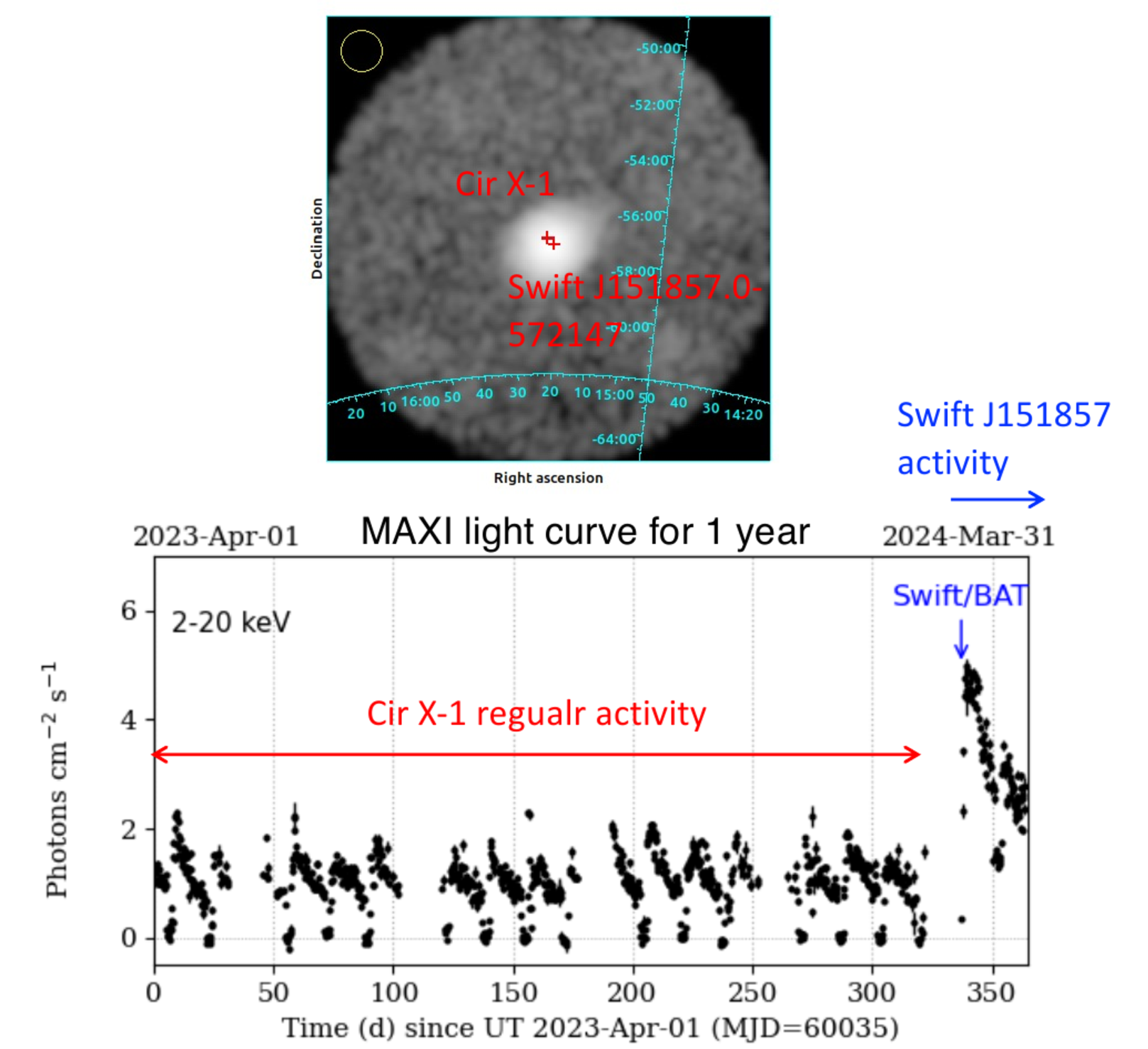}
\caption{MAXI/GSC field of view and source activity for the sources Cir X-1 and Swift J151857.0-572147 (Credits: MAXI Team).}
\end{figure}

\begin{figure}[!h]
\centering
\vbox{
\includegraphics[width=7.5truecm,angle=0]{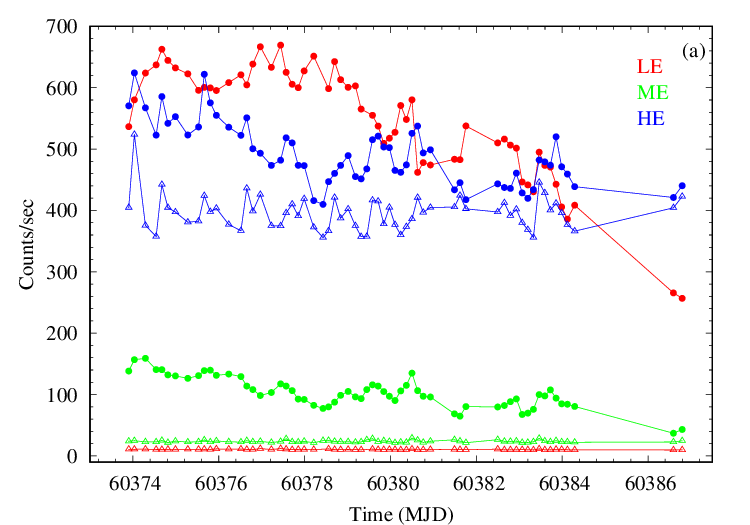}
\hskip 0.5cm
\includegraphics[width=7.5truecm,angle=0]{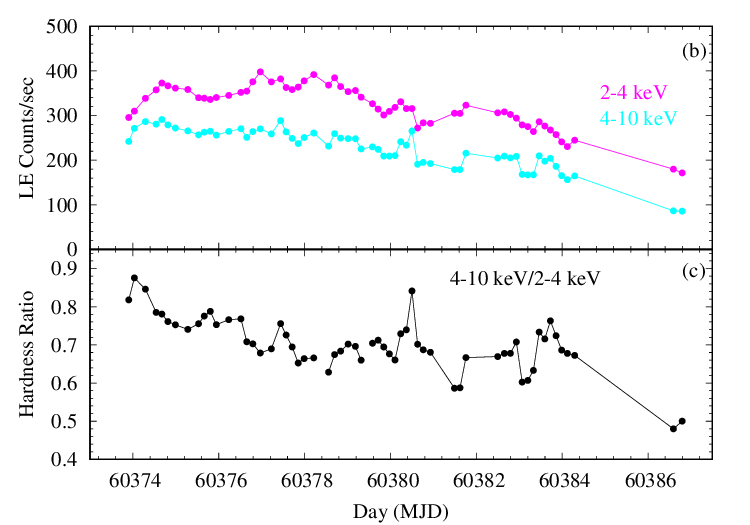}}
\caption{Variation of {\it Insight}-HXMT (a) source and background count rates in LE, ME, and HE bands, (b) $2-4$~keV$ and $4-10 ~keV LE count rates, (c) hardness ratio (HR; $4-10 ~keV/2-4$~keV) 
         with time.}
\end{figure}

In Fig. 2, we show the variation of the count rates for around 15 days. The count rates are extracted using LE, ME, and HE light curves of HXMT in the $2-10$, $10-35$, and $27-250$~keV energy 
bands. In panel (a), we show the variation of those source and background count rates for the three bands (in respective colors). Red is for LE, while green and blue colors are used to represent 
ME and HE bands. The filled circle (of each color) lines represent the source counts, whereas the triangle-shaped lines represent the background count rates. As can be noticed, the HE background 
count rate was quite high and was almost comparable to the source count rate. This could be due to the combination of two reasons e.g., the large effective area of HE and the close 
proximity of the source to Cir X-1. However, as mentioned earlier, the contribution from Cir X-1 could be in low energy as it was in the soft state. Thus, it is hard to comprehend the actual 
reason behind it. The other two bands showed a significant difference in count levels between source and background. In Table 2, we list the start, end, and average MJDs of all our analyzed 
exposures. We also list the source and background count rates for LE, ME, and HE in Table 2. In panel (b) of Fig. 2, we show the count rates in $2-4$, $4-10$~keV energy bands, which are extracted 
using LE light curves. In panel (c), the HR is plotted using the ratio of the LE count rates of $4-10$ to $2-4$~keV. 

From the light curves, we see that the source has high count rates in all three bands. For Insight-HXMT, a count rate of 800, 500, and 800 counts/sec from LE, ME, and HE corresponds 
to the flux equivalent to 1Crab. These flux values give the idea about the brightness of the source. From the variation, HR gives a rough idea that the source had already moved past its hard 
state as {\it Insight}-HXMT started monitoring the source. As time progressed, spectral nature progressed from intermediate to softer states. However, we need timing and spectral analysis 
results to designate this firmly. We discuss them in the next two subsections.

\subsubsection{Low-Frequency Quasi Periodic Oscillations}

\begin{figure}[!h]
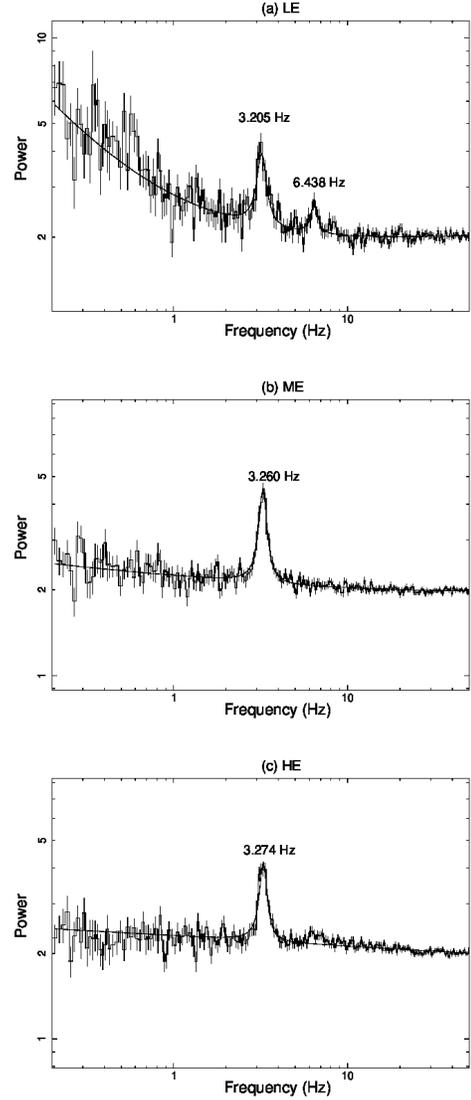

\centering
\vbox{
\hspace{0.6cm}
\includegraphics[width=5.0truecm,angle=270]{0p01s-pds_le_P061437400101_powspec_no-white-noise-sub.eps}
\hspace{-1.6cm}
\includegraphics[width=4.7truecm,angle=270]{0p01s-pds_le_P061437400101_xspec_no-white-noise-sub.eps}}
\caption{Model fitted power density spectrum (PDS) continuum in $0.01-50 ~Hz$ for the LE band for fitting in (a) {\fontfamily{pcr}\selectfont powspec} and (b) {\fontfamily{pcr}\selectfont
	 xspec}. The best fit is achieved using a combination of a set of models: \textit{power-law}, \textit{constant}, and \textit{Lorentzian}.}
\end{figure}

We have created the PDS to analyze QPOs using the $0.01$ sec time-binned light curves from all three bands (LE, ME, and HE), as stated in \S2. In Fig. 3, we show the best model-fitted 
PDS continuum for the LE band, for both fittings in (a) {\fontfamily{pcr}\selectfont powspec}, and (b) {\fontfamily{pcr}\selectfont xspec} for the observation ID P0614374001 (exposure ID: 
P061437400101-20240304-01-01). First, we used the combination of \textit{power-law}, \textit{constant}, and \textit{Lorentzian} models in {\fontfamily{pcr}\selectfont powspec}. After the best
model fitting, the same is followed in {\fontfamily{pcr}\selectfont xspec}. From the {\fontfamily{pcr}\selectfont xspec} fitting panel, one can also observe the goodness of fitting, as well as
the contributions from various fearues of the PDS, including the fundamental QPO and harmonic. While, both the QPO and harmonic were present in the LE band, the harmonic nature was absent in 
ME and was not very prominent in HE, as observed from the fittings in all the bands. The QPO and harmonic have a 1:2 ratio in frequency with the $\nu_{qpo} \sim 3.19 \pm 0.02$ and $\nu_{harmonic} 
\sim 6.43 \pm 0.04$~Hz, respectively. The fundamental and harmonic QPO in this exposure has an FWHM of $0.35 \pm 0.05$, $0.39 \pm 0.13$ and normalization of $1.84 \pm 0.23$, $0.57 \pm 0.14$, 
respectively. We noticed the presence of fundamental QPOs in the PDS of each of the three energy bands. We first checked all the exposures for the observation ID {\fontfamily{pcr}\selectfont 
P0614374001}. From our fitting, we first extracted the basic QPO information, which is QPO frequency ($\nu_{qpo})$, full-width at half-maximum (FWHM or LW), and QPO normalization (LN). 

We found that QPO was present in most of the exposures of this observation ID. It was present for all exposures in the ME band and was absent in the last LE band and second and fifth HE band. 
Also, the QPO frequency evolved within a short period of $\sim 1.5$ days of the duration of this observation ID P0614374001. Thus, we checked for QPOs for all the exposures. At the 
start of our analysis period, fundamental QPO was present in almost all of the exposures. The $\nu_{qpo}$ was $\sim$ 3.2 for all three bands on MJD 60373.9, and it increased as the outburst 
progressed. Then after some days, it decreased, and then again showed an increasing trend. Then, it again decreased and increased and decreased and continued this way. The highest frequency 
in the LE band was 8.1 Hz on MJD 60376.9, on which both the light curves in the ME and HE bands were not created by the {\fontfamily{pcr}\selectfont hpipeline} command. The highest frequency 
in the ME band was 8.97 Hz on MJD 60377.9, on which the LE and ME light curves were not produced. In the HE band, $\nu_{qpo}$ was the highest on MJD 60379.3 with a value of 6.82 Hz. We show 
the variations of the QPO frequency during our full analysis period in Fig. 4(a-c) for (a) LE, (b) ME, and (c) HE. In Table 3, we listed the values of $\nu_{qpo}$ in columns 2, 3, \& 4 for LE, 
ME, and HE. Although for the exposure P061437400103-20240305-02-01, there was a presence of a harmonic nature in the HE band, we did not fit it as the noise was high and the harmonic was like 
a broad \textit{Lorentzian} feature. We did not find harmonic for any other exposures of any other observation ID in any of the three bands.

\begin{figure}[!h]
%\vspace{0.8cm}
  \centering
    \includegraphics[width=7.0cm]{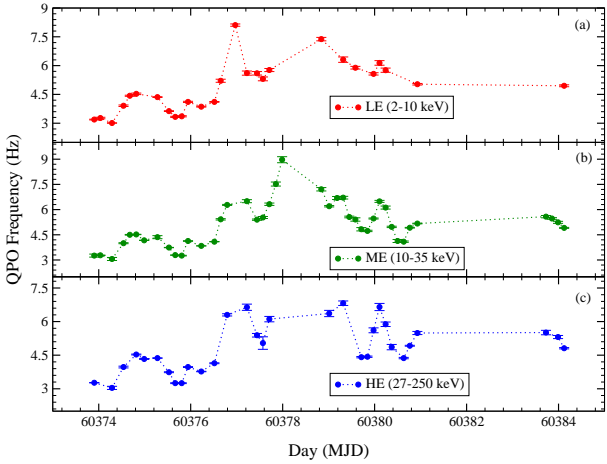}
    \caption{Evolution of QPO frequency with time during the whole period of analysis for (a) LE, (b) ME, and (c) HE.}
\end{figure}

\begin{figure}[!h]
%\vspace{0.8cm}
  \centering
    \includegraphics[width=8.0cm]{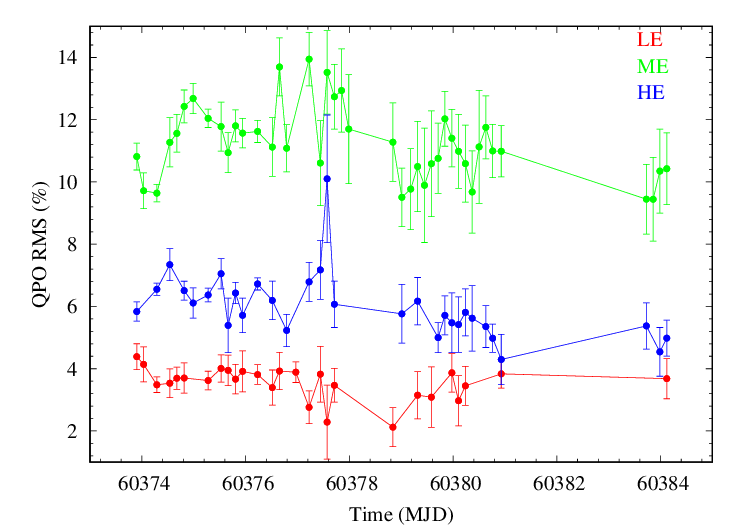}
\caption{Variation of QPO RMS with time for the LE (red), ME (green), and HE (blue) bands, respectively.}
\end{figure}

\begin{figure}[!h]
%\vspace{0.8cm}
  \centering
    \includegraphics[width=8.0cm]{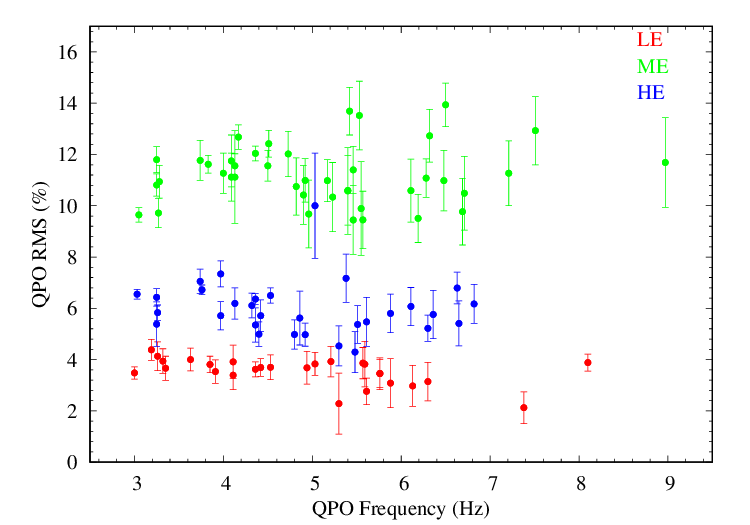}
\caption{Variation of QPO RMS with QPO frequency for the LE (red), ME (green), and HE (blue) bands, respectively.}
\end{figure}

\begin{figure*}[h]
%\vspace{-9cm}
\centering
\vbox{
\includegraphics[width=4.4truecm,angle=270]{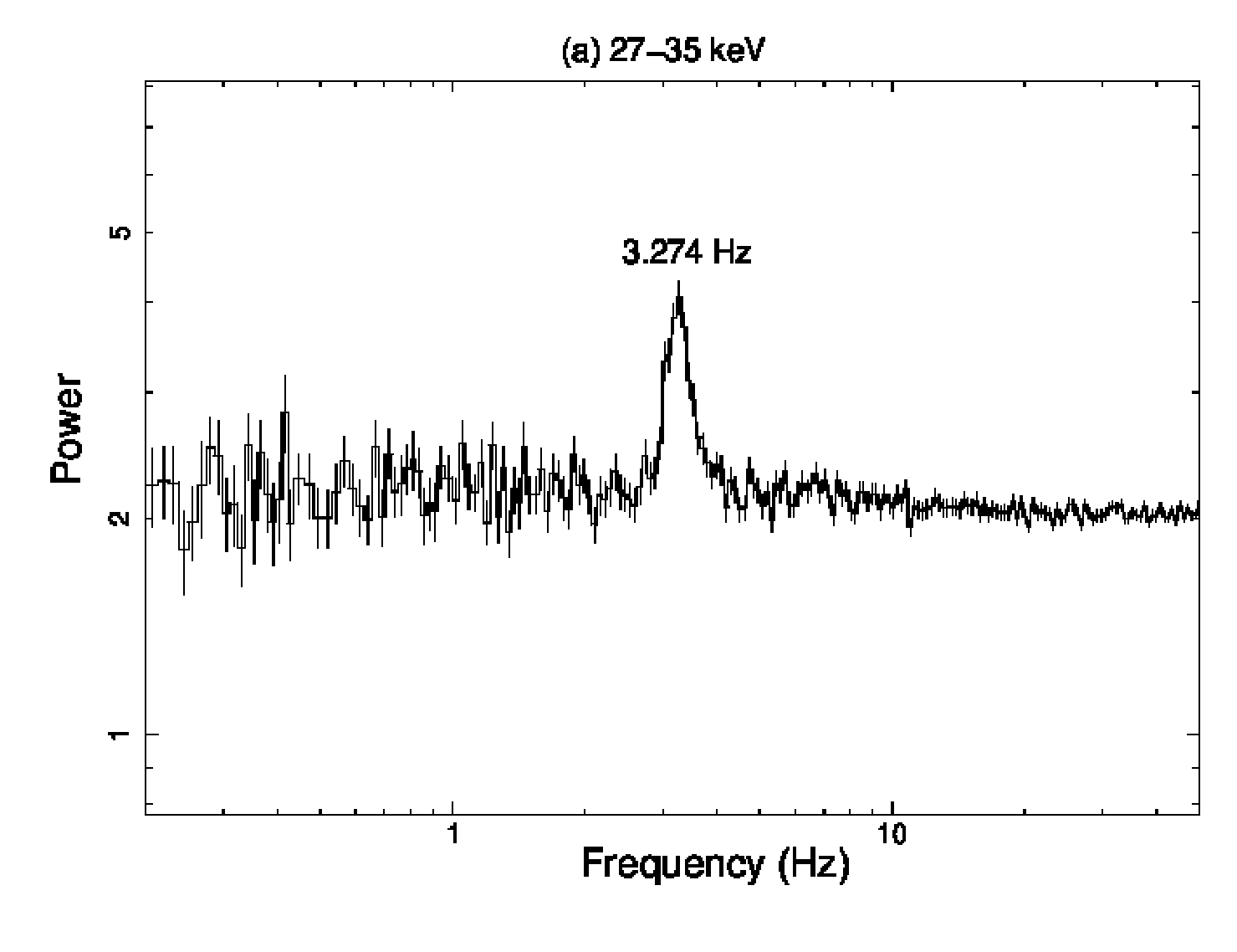}
\includegraphics[width=4.4truecm,angle=270]{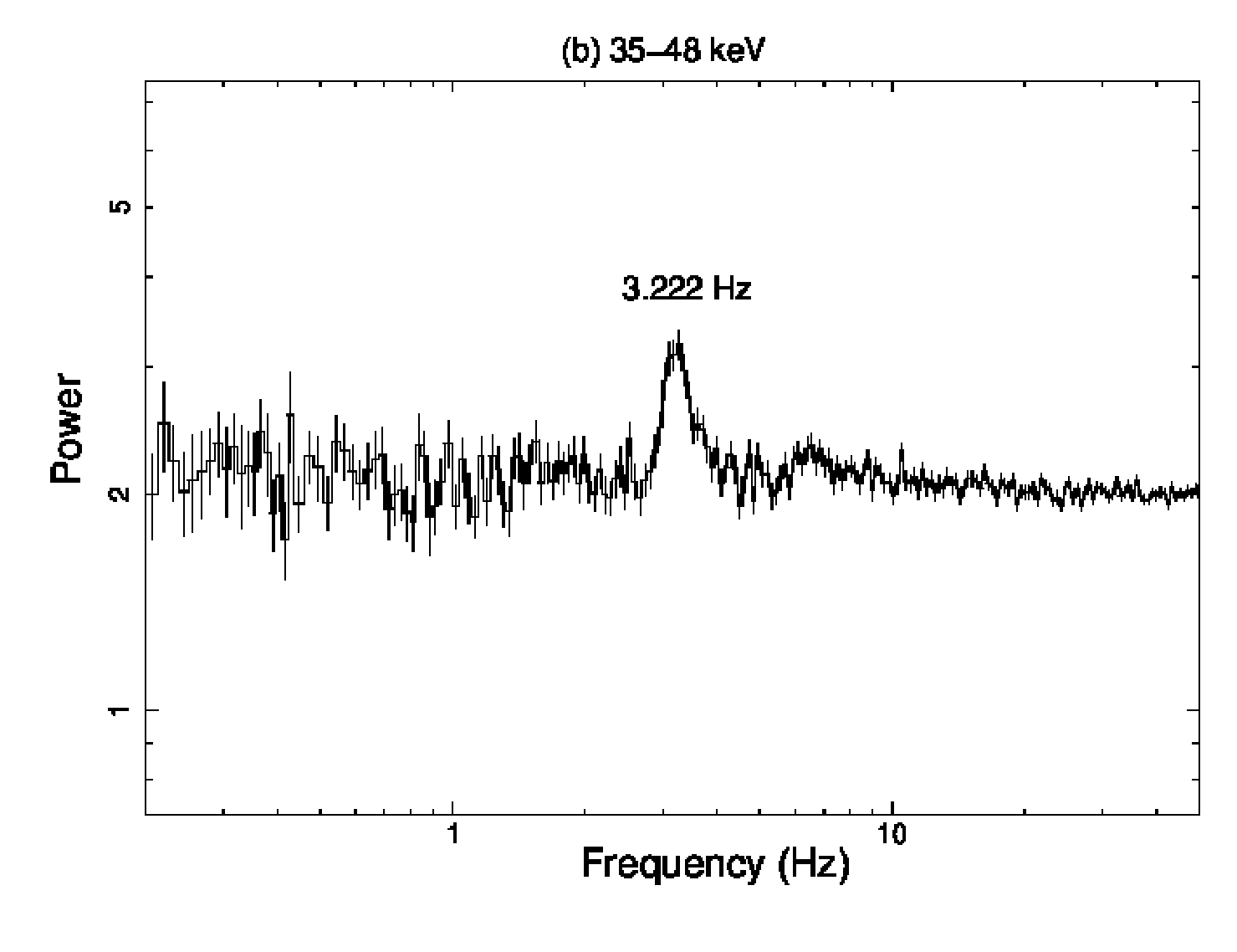}
\includegraphics[width=4.4truecm,angle=270]{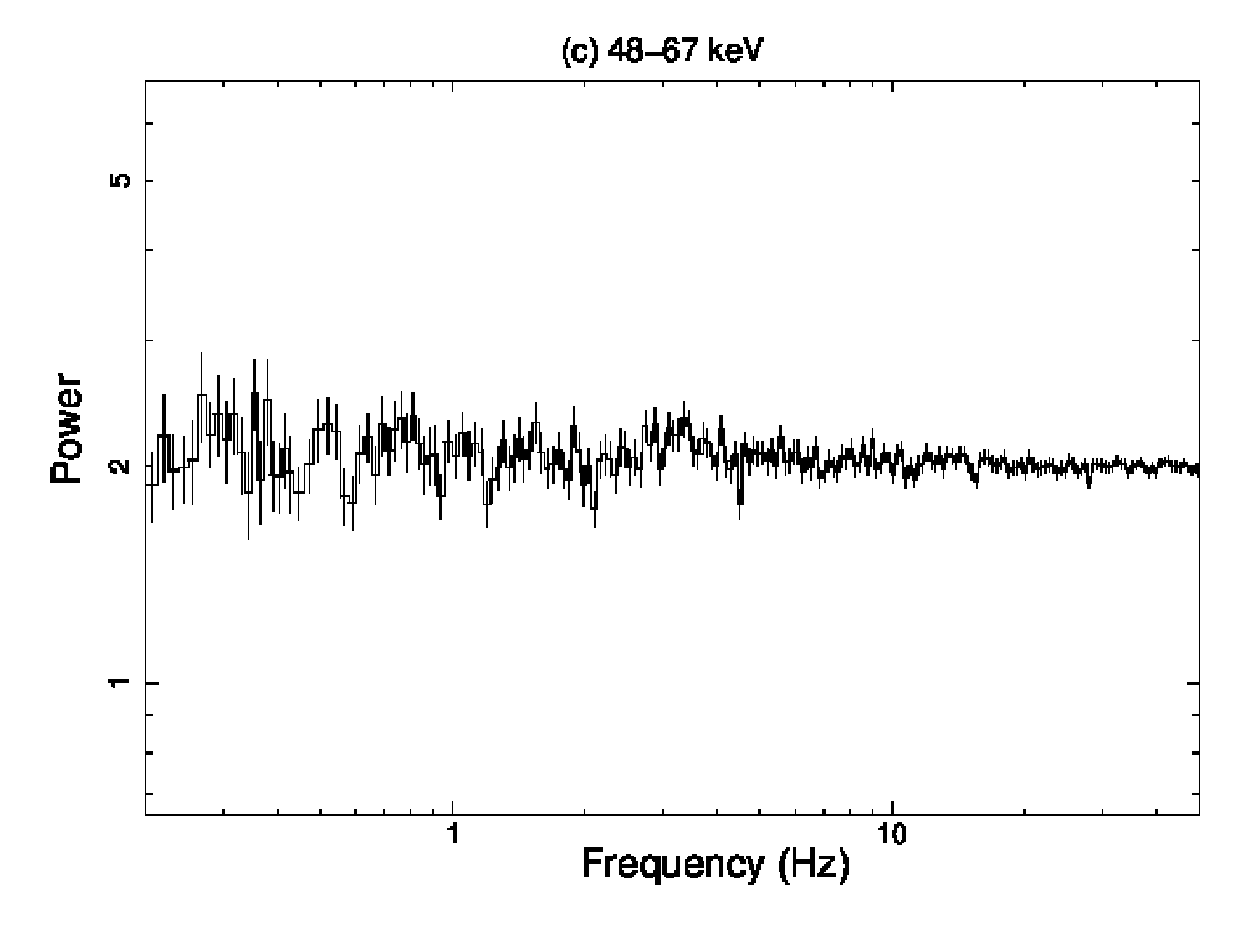}
\includegraphics[width=4.4truecm,angle=270]{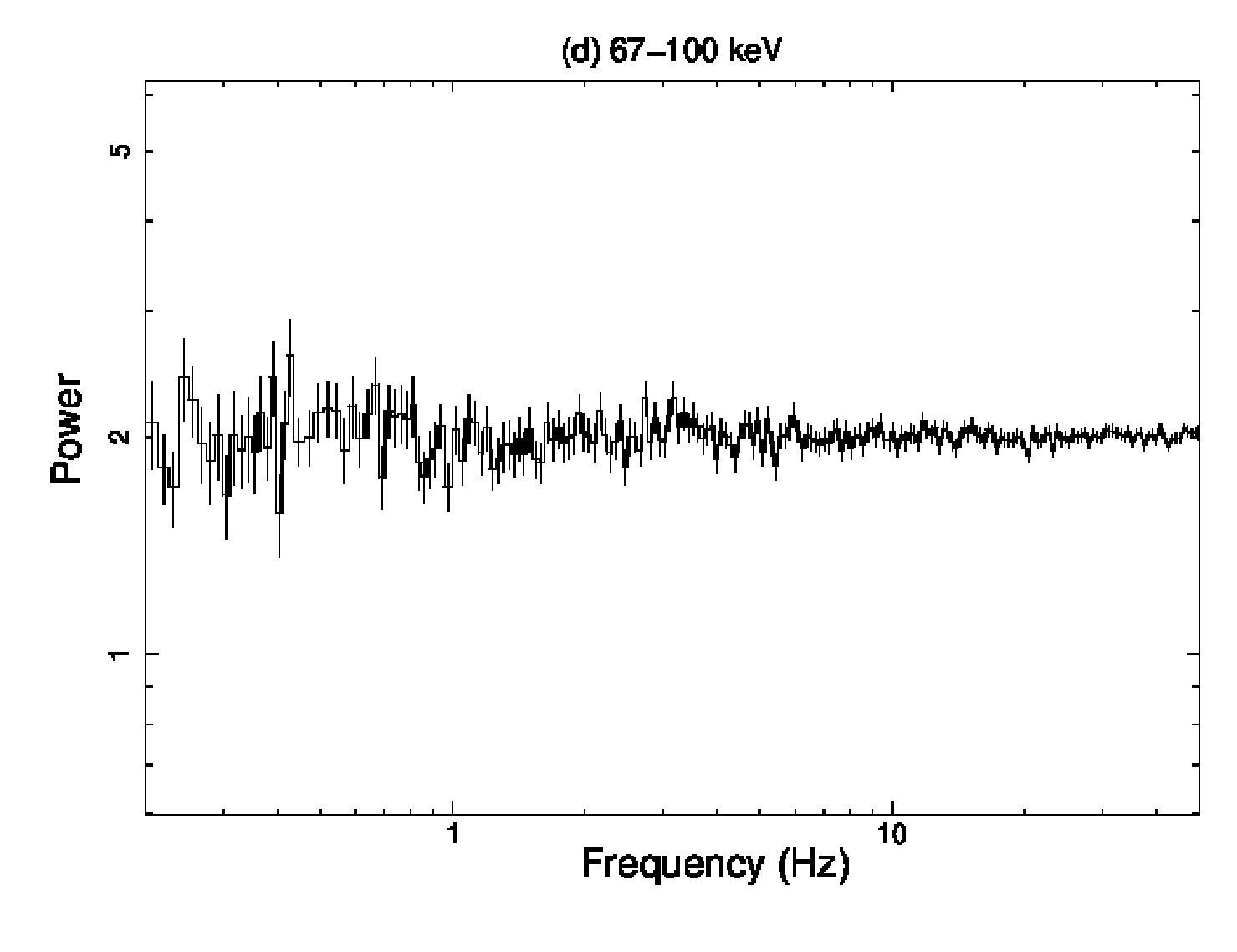}
\includegraphics[width=4.4truecm,angle=270]{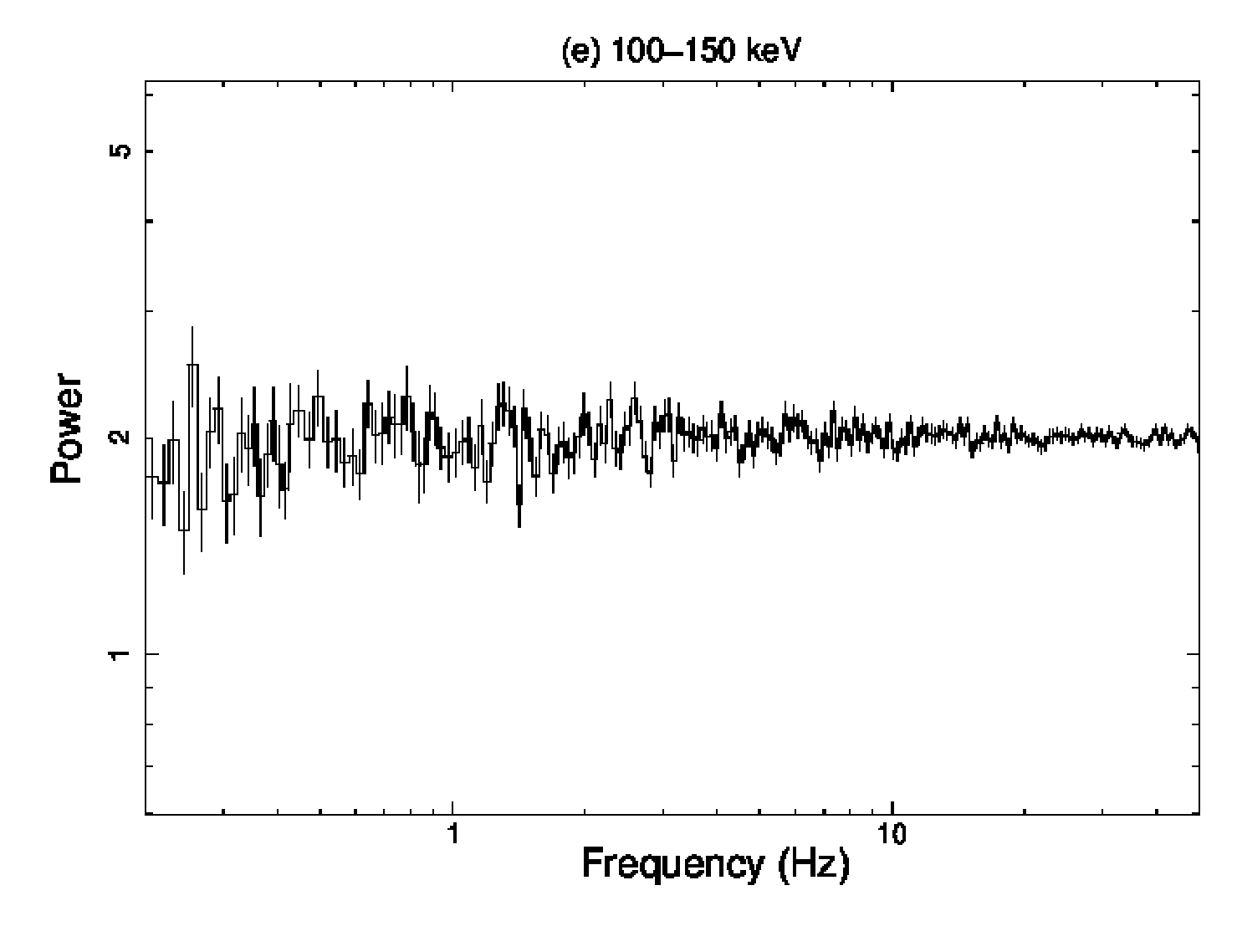}
\includegraphics[width=4.4truecm,angle=270]{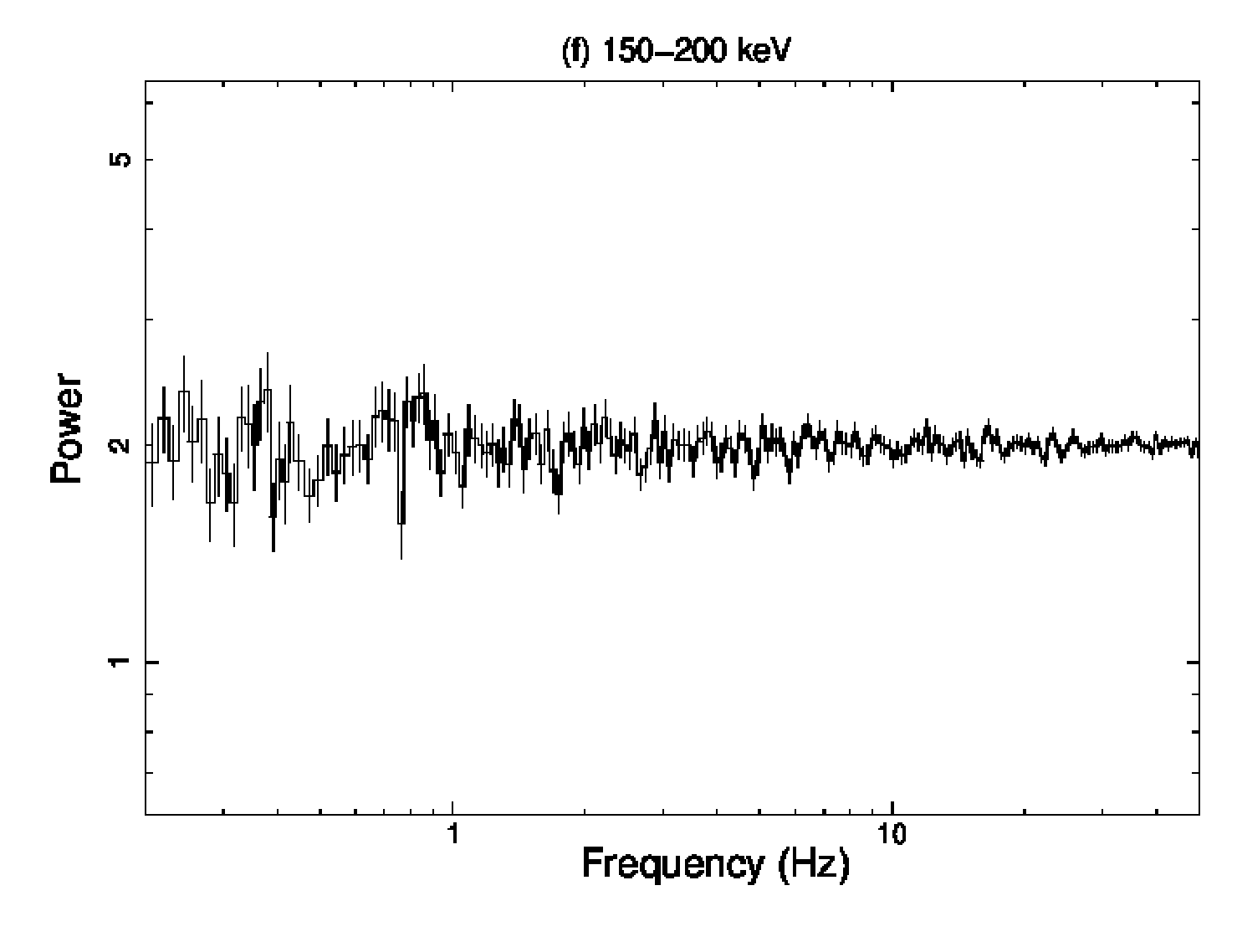}
\includegraphics[width=4.4truecm,angle=270]{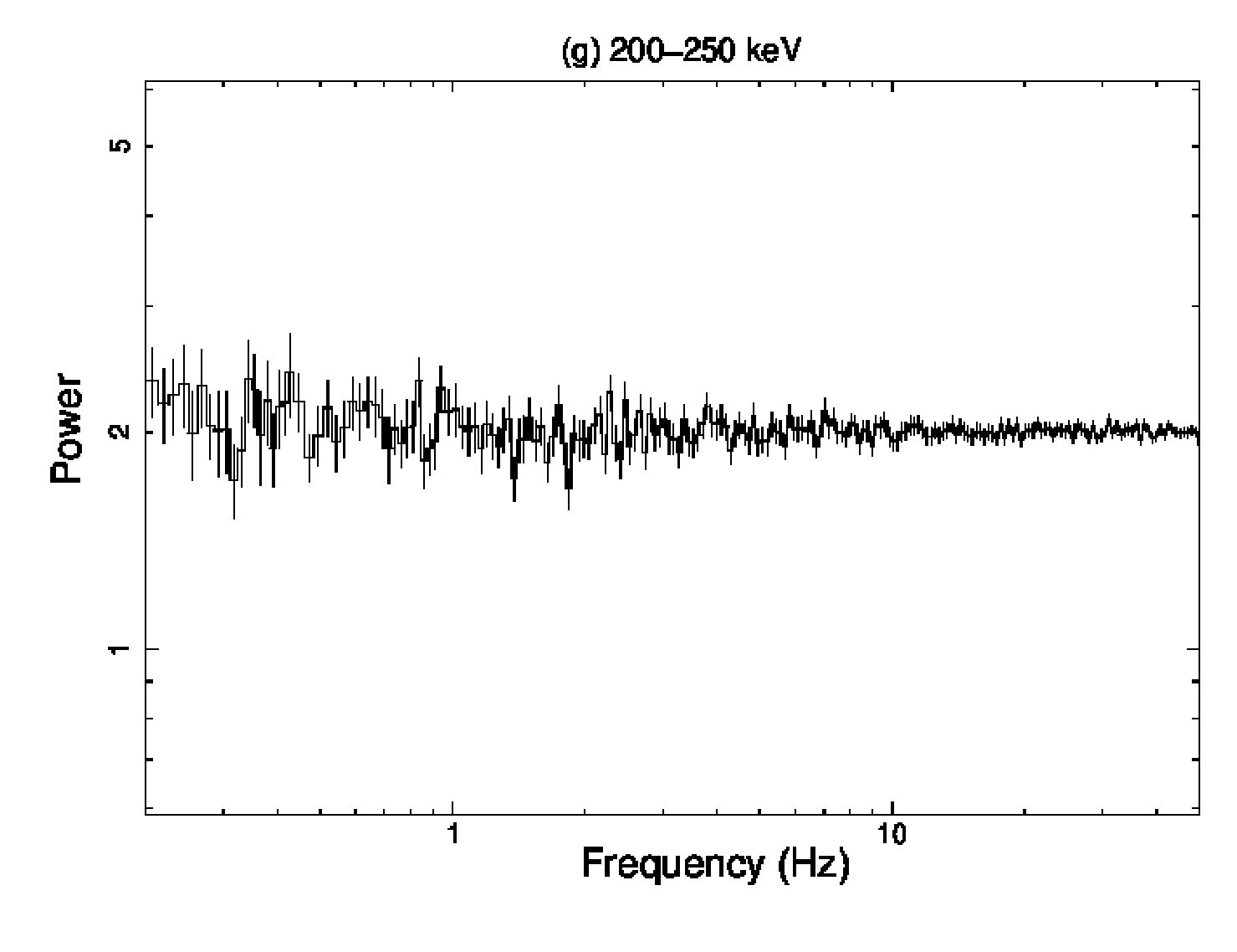}
}
\caption{Energy dependent PDS, produced in (a) 27-35, (b) 35-48, (c) 48-67, (d) 67-100, (e) 100-150, (f) 150-200, and (g) 200-250 keV energy bands using 0.01 s time-binned HE light curves. 
         This is for the observation ID P0614374001 (exposure ID: P061437400101-20240304-01-0).}
\end{figure*}

We were able to extract certain information about QPOs, such as full-width at half maximum (FWHM) and Normalization (LN) by the use of PDS fitting. For the exposures, we additionally retrieved 
the source and background count rates. Using the formula from Bu et al. (2015), we estimated the fractional RMS as $RMS = \sqrt{\frac{P}{S+B}} \frac{S+B}{S}$, which denotes the fractional 
variability in the PDS. Here, $S$, and $B$ represent the count rates of the source and the background, respectively. $P$ is the Leahy normalized power. We also estimated the $Q$-factor 
($\nu/\delta\nu$), which measures the sharpness of the QPO. Table 3 lists these values for LE, ME, and HE in columns 5--7 ($Q$-value) and 8--10 (RMS), respectively. This is shown in Fig. 5. 
The variations of the $Q$ factor were consistent in the three different bands. To check if there is any correlation between the QPO RMS and QPO frequency, we plotted those two properties against 
each other in Fig. 6. We have not found any correlation between them for this source in all three energy bands.

\begin{figure}[!h]
%\vspace{0.8cm}
  \centering
    \includegraphics[width=8.5cm]{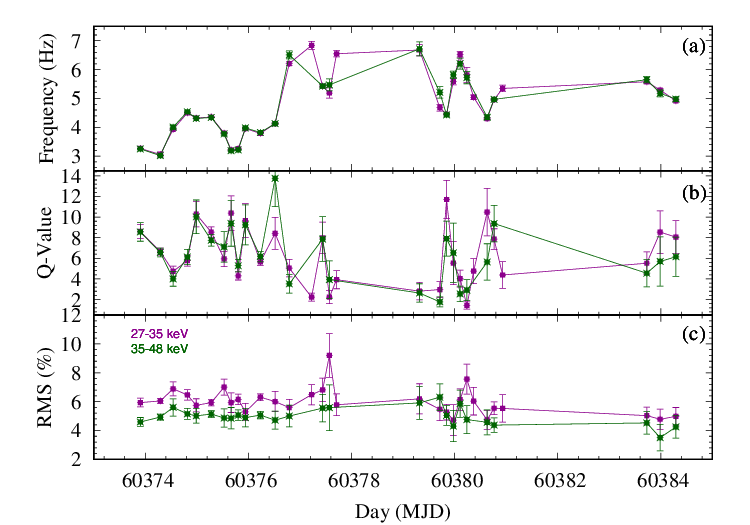}
    \caption{Variation of energy-dependent (a) QPO frequency (in Hz), (b) Q-factor, and (c) RMS (\%) with time. Here, the magenta-colored points represent the values for 27–35 keV, whereas the 
             green-colored points represent the values for 35–48 keV energy bands, respectively.}
\end{figure}

As explained before, we also checked the energy dependence of QPOs using the HE light curve in 7 different energy bands. These energy ranges were chosen to maintain similarity with Ma et al. 
(2023). The PDS continuums for the observation ID P0614374001 (exposure ID: P061437400101-20240304-01-01) are given in Fig. 7(a-g) for respective energy bands. For this exposure, we find that 
the fundamental QPO was prominently present at 3.274 Hz in the $27-35$ keV energy band, while it is also present in the $35-48$ keV with a little change of frequency of 3.222 Hz. However, the 
nature of QPO was not as strong as in the $27-35$ keV. Above 48 keV, we did not find any nature of fundamental QPO. We notice a sharp fall of QPO strength above 48 keV. Chatterjee et al. (2021) 
studied QPO energy dependence for the BHC GRS 1716-249 using AstroSat data. Although the fundamental QPO nature got weaker in high energies in that report, it did not show this type of sharp 
fall of QPO nature after some energy band. A possible weak harmonic nature was noticed in the $35-48$ keV band, which was not present in the $27-35$ keV. However, it looks very weak and we did 
not model it. Harmonic nature was also not observed above 48 keV. We checked this for all the 31 exposures for which fundamental LFQPO was present in the HE light curve. We find that for all of 
these exposures, QPO was absent above 48 keV. For some exposures, we found that LFQPO was absent in the 35-48 keV band, although it was present in the 27-35 keV band. 

Using the formulas, as mentioned above, we estimated $Q$-values and RMS (\%) for all these exposures in both these bands. In Fig. 8, we show the variation of QPO frequency, $Q$-value, and RMS 
(\%) for all these exposures with time. We notice that the $\nu_{qpo}$ varied in a very narrow range between these two energy bands, which is within the error range. The $Q$-value shows a random 
variation for both the bands, where it was sometimes higher for 27-35 keV bands and sometimes for 35-48 keV. The overall variation of RMS (\%) was higher in case of 27-35 keV band, compared to 
the higher band. The values of the variation of QPO properties in case of energy dependence is given in the Table 4.

\begin{figure}%[!h]
%\vspace{0.8cm}
  \centering
    \includegraphics[width=8.5cm]{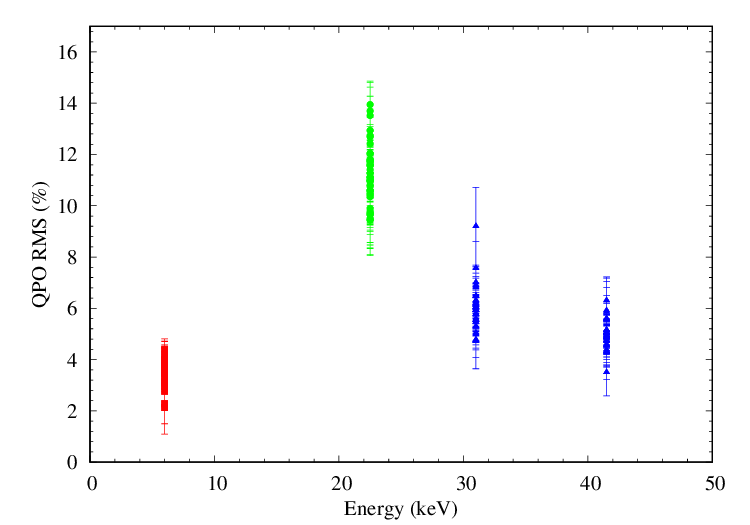}
    \caption{Variation of QPO RMS with energy. The red, green, and blue points represent data points using LE, ME, and HE bands light curves respectively for all the exposures.}
\end{figure}

We also show the variation of QPO RMS with energy in Fig. 9. We noticed that the QPO RMS was lowest in the LE band. It was the highest in the ME band. Then it started to decrease. Above 48 keV, 
we did not find the presence of any QPO.

\begin{figure}[!h]
%\vspace{0.8cm}
  \centering
    \includegraphics[width=6.5cm, angle=270]{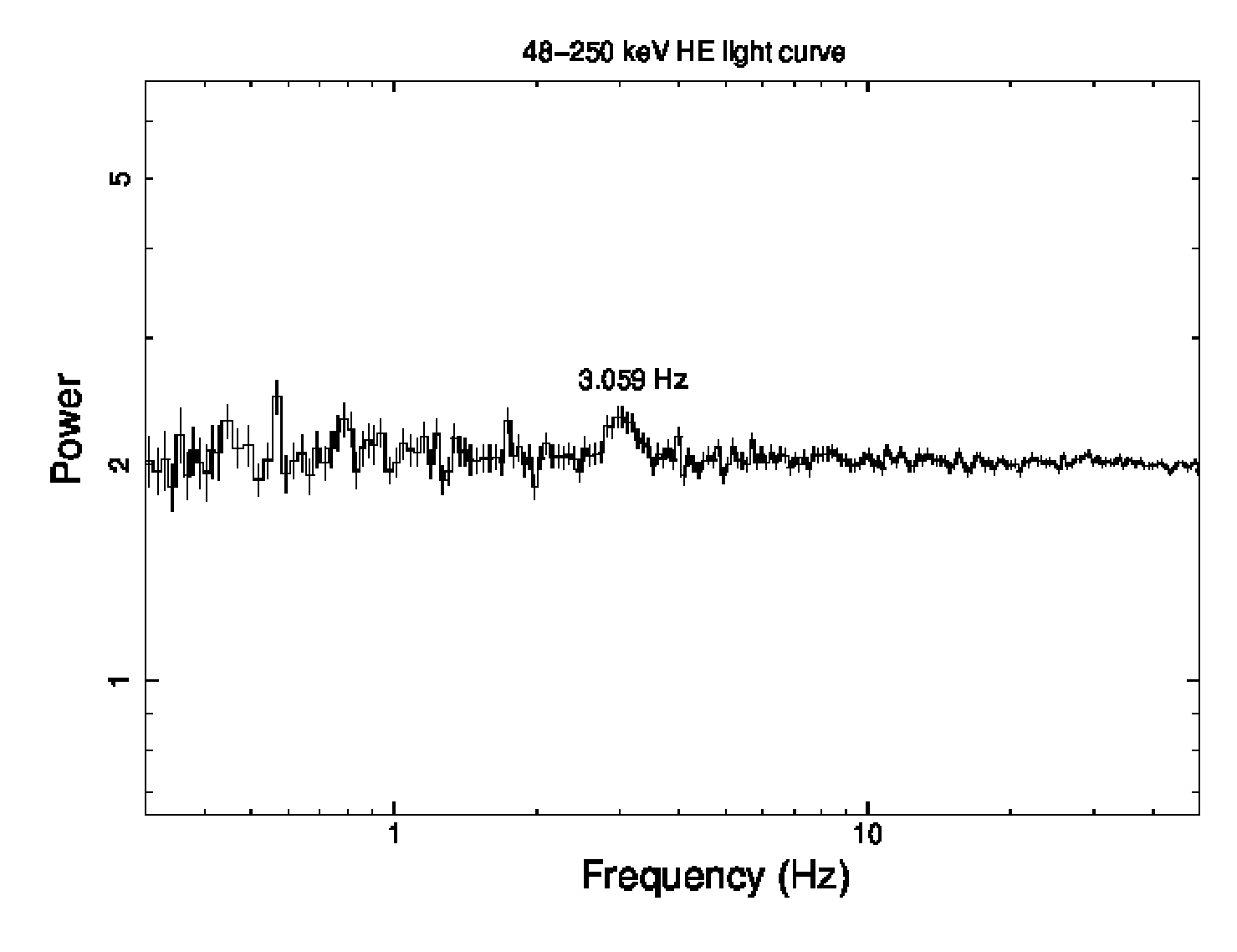}
    \caption{PDS in the $48-250$~keV energy band for the HE light curve of the exposure ID. P061437400103-20240305-02-01}
\end{figure}

Apart from this, we also searched for QPOs in the $48-250$~keV energy band light curve. As we move to the higher energies, the detector's effective area decreases. Thus, the number of 
photons also reduces. However, in high energies (e.g., $\sim 50$~keV), if the energy range is large, the PDS of the light curve may show the presence of QPOs. This is what we wanted to check. 
Since above $48$~keV, no QPO was found in the PDS, we wanted to make a further consistency check in a broader energy band, if it shows any QPOs or not. We could not find the presence of QPO 
in the higher energy bands in any of the exposures, except for the observation ID. P0614374001 with exposure ID. P061437400103-20240305-02-01. We found that there was the presence of a 
fundamental QPO in this exposure at $3.06 \pm 0.05$~Hz. This is shown in Fig. 10. For all the detected QPOs, we have estimated the significance ($\sigma$) using the relations given in 
Sreehari et al. (2019). The values of the significance are given in each table where QPO information is listed.

\subsubsection{High Frequency QPOs}

\begin{figure}[!h]
\centering
\vbox{
\includegraphics[width=5.0truecm,angle=270]{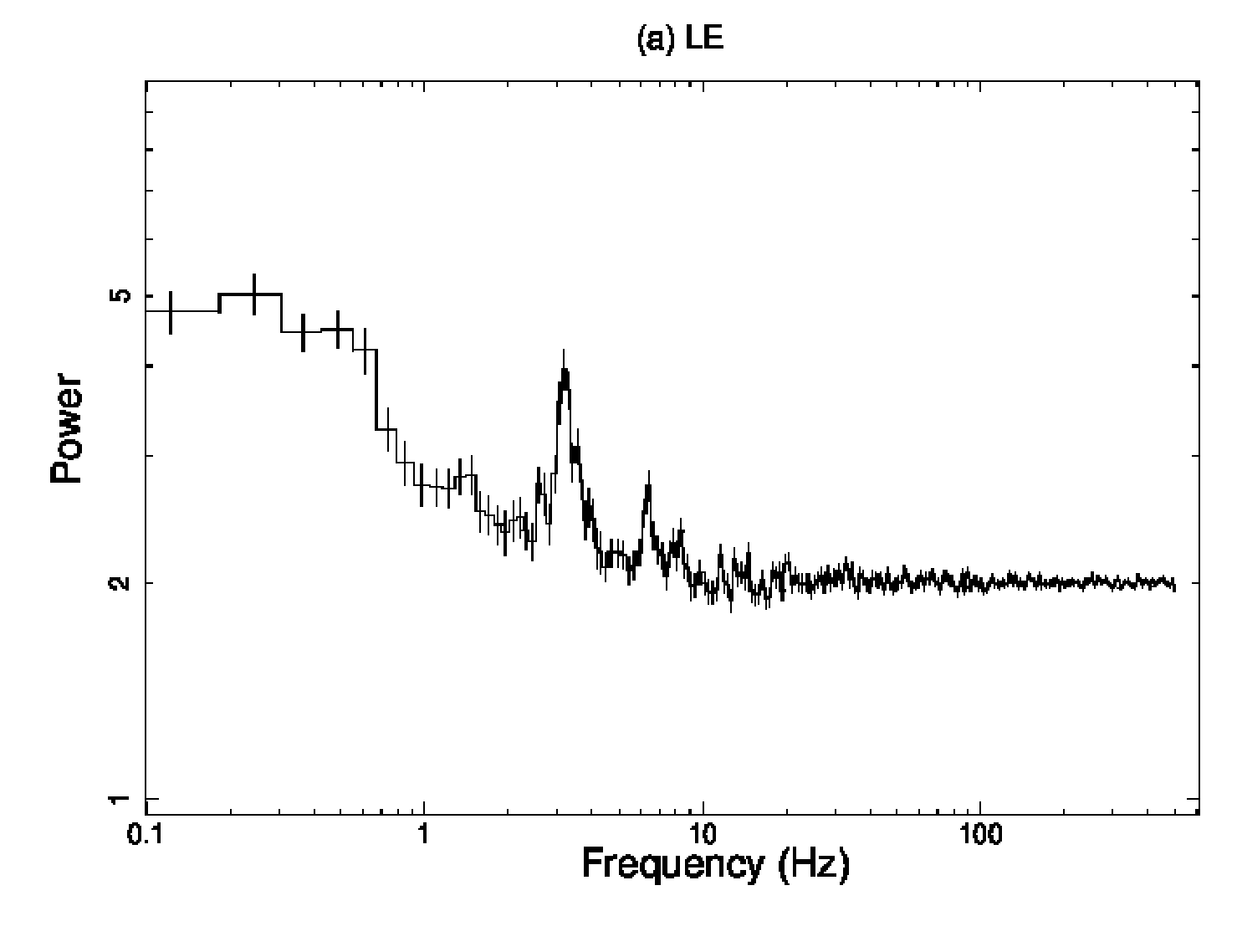}
\includegraphics[width=5.0truecm,angle=270]{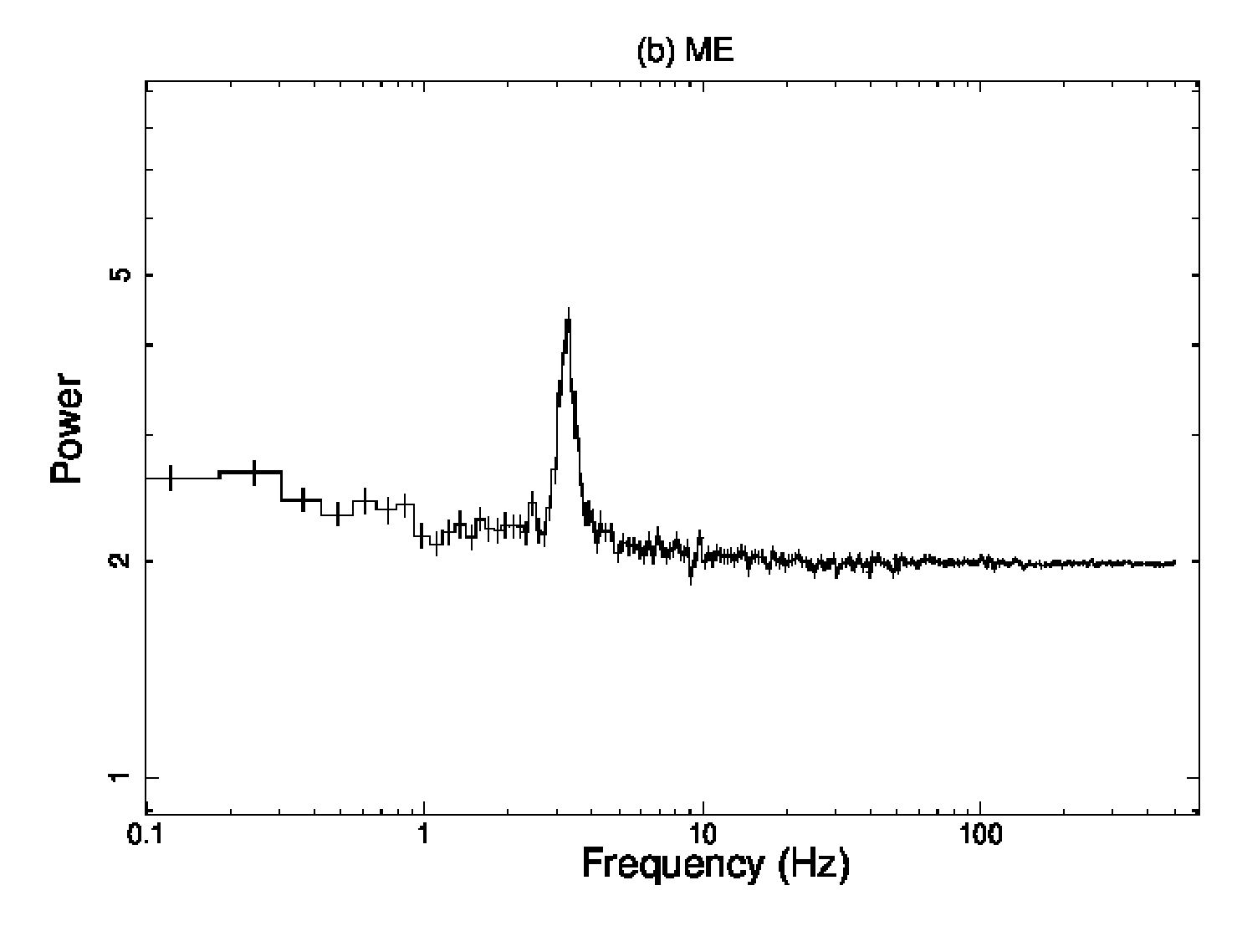}
\includegraphics[width=5.0truecm,angle=270]{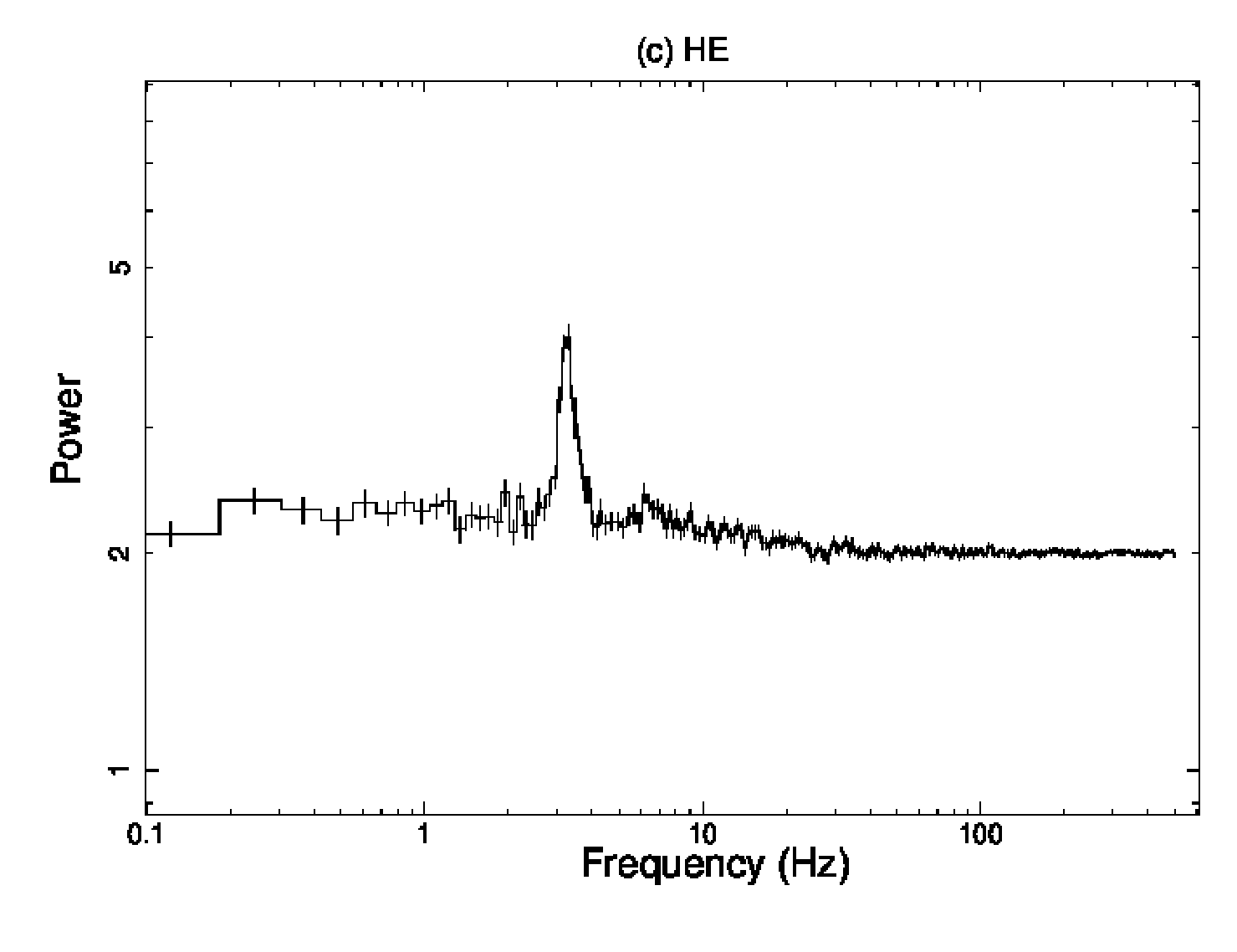}
}
\caption{PDS continuum in the $0.1-500$ Hz frequency range for (a) LE, (b) ME, and (c) HE bands. This is for the observation ID P0614374001 (exposure ID: P061437400101-20240304-01-0).}
\end{figure}

Apart from looking for low-frequency QPOs, we also searched for high-frequency QPOs (HFQPOs) in all the light curves for all three bands in all 62 exposures. In Fig. 11(a-c), we show the 
PDS continuum for $0.001$ sec time-binned (Nyquist frequency = 500 Hz) curve for (a) LE, (b) ME, and (c) HE. However, we did not find any signature of HFQPOs in any of our light curves. 
The frequency in the PDS in LE, ME, and HE in Fig. 11, are similar to those in Fig. 3. Those are the LFQPOs present in those light curves during that exposure.

\subsection{Spectral Properties}

Studying the spectral features sheds additional light on the nature of the outburst in addition to the temporal properties. We examined the source using the {\it Insight}-HXMT data that was 
available for 14 exposures in total. The exposure IDs in Table 2's first column have a `*' symbol next to them. We perform a thorough spectral study using HXMT data on this source for every 
consecutive days for the available data. Our spectrum investigation was initiated with MJD 60373.9. For spectral fitting, we have simultaneously analyzed {\fontfamily{pcr}\selectfont LE+ME+HE} 
in the $2-100$ keV energy band (LE in 2--10, ME in 10--35, and HE in 27--100 keV) for all of the chosen exposure IDs.

\begin{figure}[!h]
\vskip 0.2cm
\centering
\vbox{
\includegraphics[width=4.5truecm,angle=270]{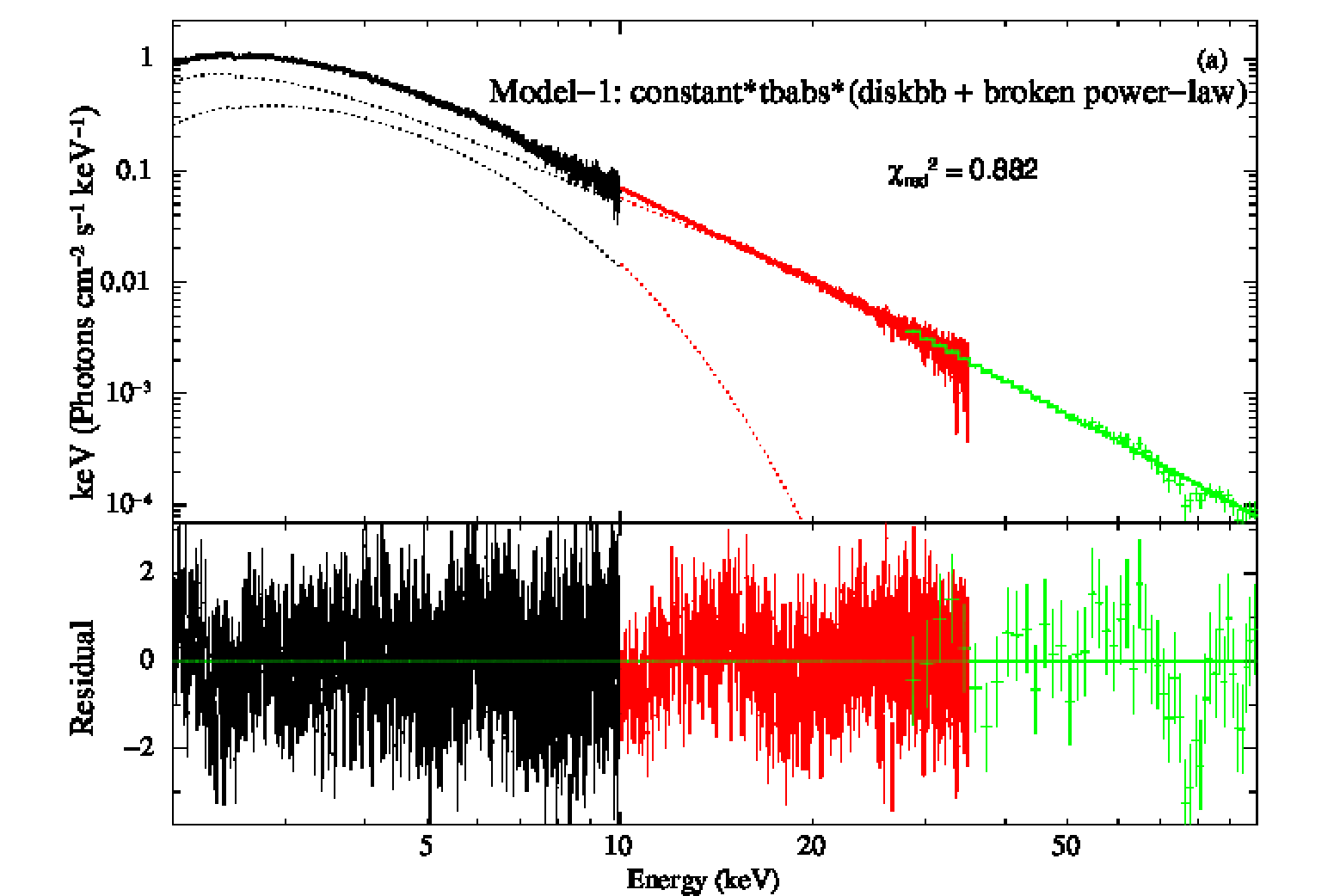}\hskip -0.5cm
\includegraphics[width=4.5truecm,angle=270]{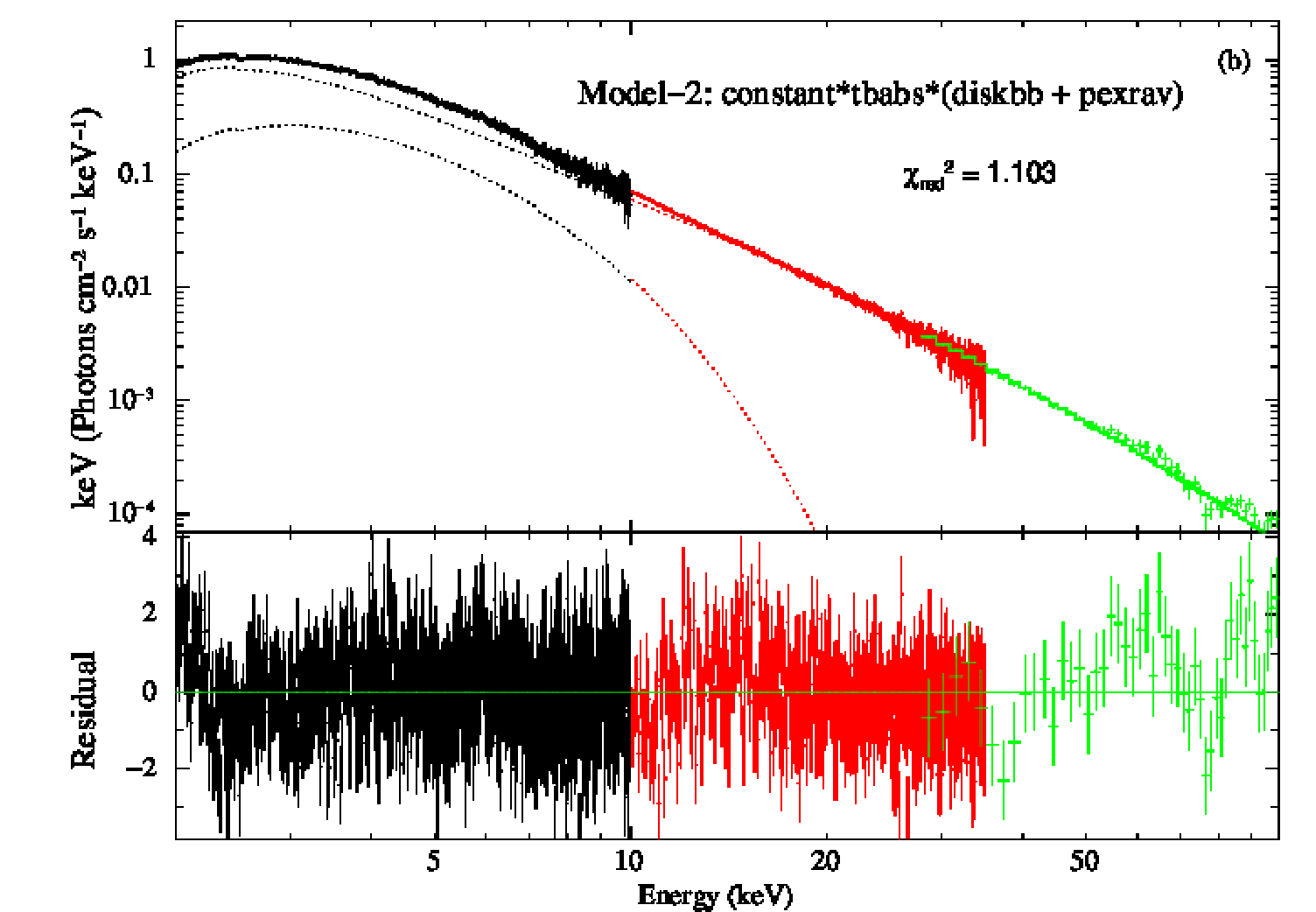}\hskip -0.5cm
\includegraphics[width=4.5truecm,angle=270]{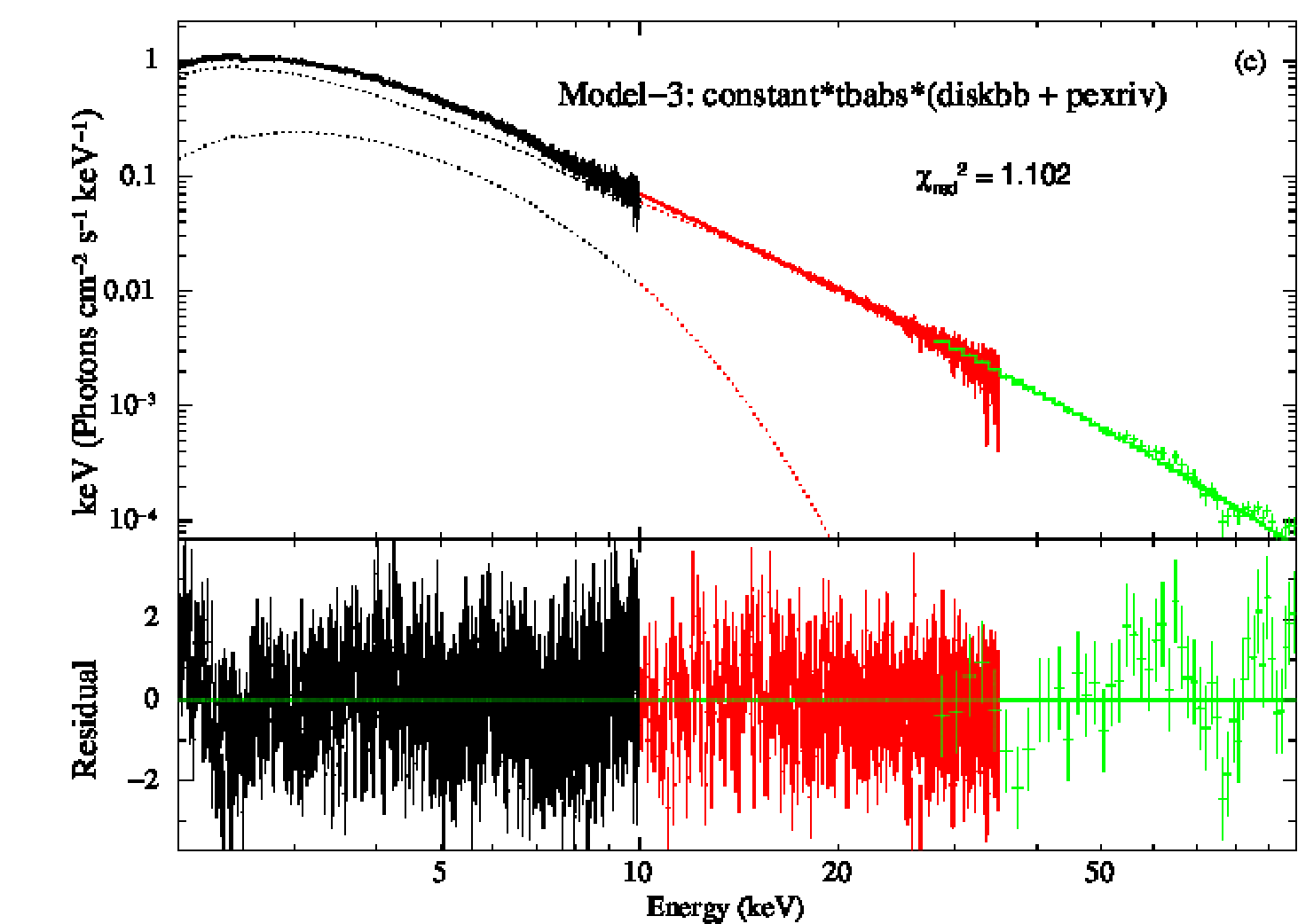}\hskip -0.5cm
\includegraphics[width=4.5truecm,angle=270]{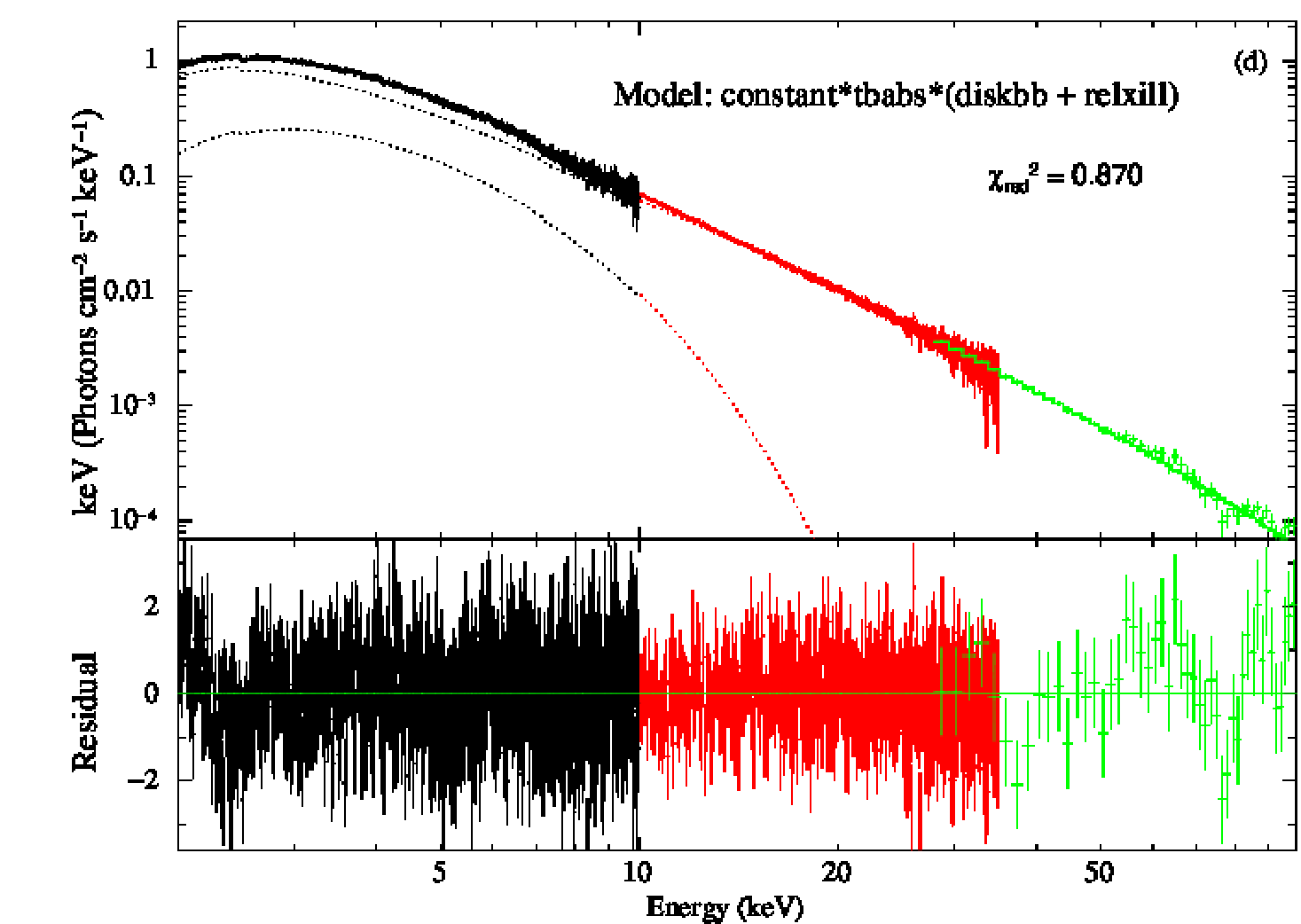}}
\caption{Best model fitted unfolded spectra for observation ID. P0614374001 (Exposure: P061437400101-20240304-01-01) using (a) Model-1, (b) Model-2, (c) Model-3, and (d) Model-4.}
\end{figure}

First, we tried to model the spectrum with simple additive models \href{https://heasarc.gsfc.nasa.gov/xanadu/xspec/manual/node166.html}{{\fontfamily{pcr}\selectfont diskbb}} 
and \href{https://heasarc.gsfc.nasa.gov/xanadu/xspec/manual/node221.html}{{\fontfamily{pcr}\selectfont power-law}}. We also used the multiplicative 
\href{https://heasarc.gsfc.nasa.gov/xanadu/xspec/manual/node273.html}{{\fontfamily{pcr}\selectfont tbabs}} (with \textit{wilm} abundance, Wilms et al. 2000) model to account for the 
interstellar absorption. The model fitted unfolded spectrum is given in the Appendix section in Fig. 15. Although the $\chi^2/DOF$ value was acceptable, we noticed that the spectrum 
changes its slope above $\sim 20$~keV. Thus, we replaced the {\fontfamily{pcr}\selectfont power-law} with the 
\href{https://heasarc.gsfc.nasa.gov/xanadu/xspec/manual/node141.html}{{\fontfamily{pcr}\selectfont broken power-law}} model, which accounts for the change of slope after certain energy, 
called break energy ($E_b$). We call this model as {\fontfamily{pcr}\selectfont Model-1}. The three distinct instruments (LE, ME, and HE) are normalized using the {\fontfamily{pcr}\selectfont 
constant}. Although this model fit was acceptable, there was a reflection nature in the spectrum. To account for that, we replaced the {\fontfamily{pcr}\selectfont broken power-law} 
model with the reflection model in neutral medium \href{https://heasarc.gsfc.nasa.gov/xanadu/xspec/manual/node214.html}{{\fontfamily{pcr}\selectfont pexrav}} (Model-2). With this Model-2, 
we also achieved the best-fit statistics. Then we checked the reflection component by using the reflection model {\fontfamily{pcr}\selectfont pexriv} (Model-3) which takes ionization 
into account. Using the Model-3, as mentioned before, we achieved the best fit statistics. Here, we like to point out our approach using the Model-3. Except for 2 parameters, all the 
parameters of this model are the same as the {\fontfamily{pcr}\selectfont pexrav} model. While fitting with this model, we set the cut-off energy value to the one obtained
from the fit with the {\fontfamily{pcr}\selectfont pexrav} model. Also, we found while fitting that the disk temperature parameter (in units of Kelvin) was always pegging to the highest 
value of $10^6$~Kelvin. 
Thus, for all the spectral fitting using this model, we freeze the value of this parameter to this highest value. The extra parameter that this model has over {\fontfamily{pcr}\selectfont 
pexrav} is the disk ionization parameter ($\xi$), which is given as $\xi = 4 \pi F_{ion}/n$, where $n$ is the density of the reflector (Done et al. 1992) and $F_{ion}$ is the irradiating 
flux in the 5~eV to 20~keV energy band. For the analysis with Model-2 and Model-3, we fixed the abundances to solar abundance and also varied the value of the inclination to a narrow 
range of around $30^\circ$ as reported by Mondal et al. (2024). 

For all the fitting processes mentioned so far, the results are achieved by keeping the $N_H$ free, and it varies in a broad range. Therefore, we take the average of $N_H$ from 
all three models which is $\sim 5.6 \times 10^{22} cm^{-2}$. Then we reanalyzed all the observations using the same models (Model-1 to 3) combinations by keeping $N_H$ fixed to the average 
value. The parameters of the fit do not change significantly by keeping $N_H$ fixed to the average value. We report both the results by keeping $N_H$ fixed and free in the next paragraph. 
All the best-fitted model parameters and statistics are given in Tables 5 to 10. 

Since none of these models takes into account the relativistic effects, we performed the same spectral analysis using the relativistic reflection model, 
\href{https://www.sternwarte.uni-erlangen.de/~dauser/research/relxill/}{\fontfamily{pcr}\selectfont relxill} to see the relativistic effects. We consider this as Model-4, as mentioned 
before. For this model fitting, we used the average $N_H$ value and froze it to $5.6 \times 10^{22}~cm^{-2}$ for all observations. The {\fontfamily{pcr}\selectfont relxill} model has 
a total of 14 parameters, of which several can be fixed to reasonable values in order to avoid degeneracy in the best fit parameters (Peng et al. 2024; Mondal et al. 2024). From this model 
fitting, we can get some valuable information about the source like its spin and disk inclination. To achieve the best fitting with Model-4, we fixed some of the model parameters. We fixed 
$R_{in}=1$ and $R_{out}=1000$, $index1 = index2 = 3$, $R_{br} = 15$. For being a Galactic source, the redshift ($z$) was always kept to 0. The best-fitted parameters and statistics for this 
model fitting are listed in Table 11. In Fig. 12(a-d), we show the model-fitted unfolded spectra using (a) Model-1, (b) Model-2, (c) Model-3, and (d) Model-4.

\begin{figure}[!h]
\vskip 0.2cm
\centering
\vbox{
\includegraphics[width=8.5cm]{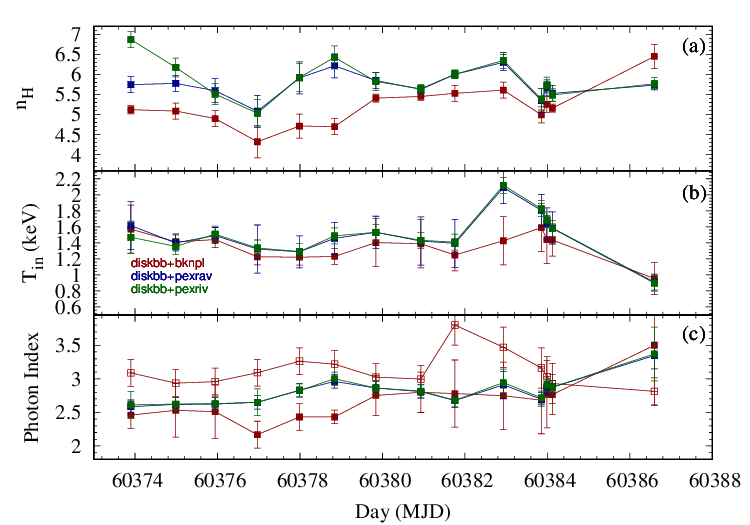}\hskip -0.5cm
\includegraphics[width=8.5cm]{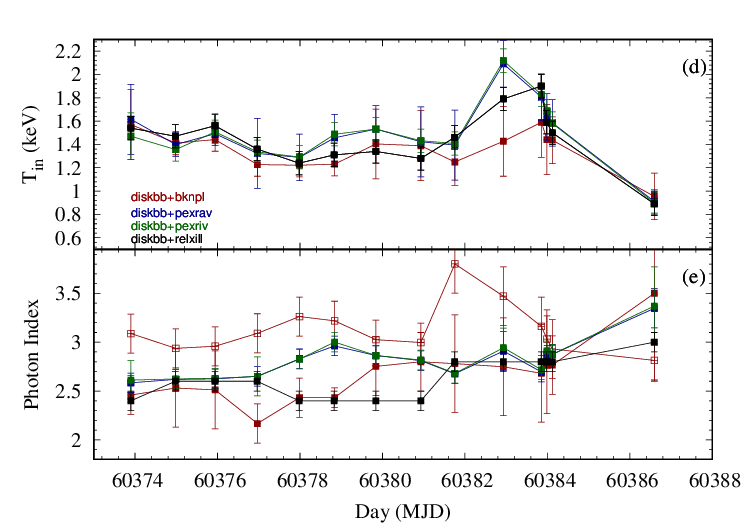}\hskip -0.5cm}
\caption{Best-fitted spectral model parameters are shown with time MJD. The panels (a), (b), and (c) show the variation of hydrogen column density (in 10$^{22}$ cm$^{-2}$ unit), inner-disk 
temperature ($T_{in}$ in keV), and photon index ($\Gamma$) for all the three models. The red, blue, and green colors represent the parameters for the \textit{Model-1}, \textit{Model-2}, and 
\textit{Model-3}, respectively. In the panel (c), we show the $\Gamma1$ and $\Gamma2$ of the \textit{Model-1} using red color filled and empty squares, respectively. The panels (d) and (e) 
represent the same parameters as in panels (b) and (c) but for the $N_H$ fixed to $5.6 \times 10^{22}$ cm$^{-2}$.}
\end{figure}

In Fig. 13, we show the variations of some of the spectral parameters for all four models. In the top figure, we show the variations of the parameters when the $N_H$ was free, whereas, in 
the bottom figure, we show the variations of the same parameters when $N_H$ was fixed to an average value. In panel (a), we show the variations of the $N_H$ for three different models (red 
filled-square for Model-1, blue filled-square for Model-2, and green filled-square for Model-3). We notice that they show consistent variations within the error range throughout. The $N_H$ 
for Model-1 varies between $(4.3-6.5) \times 10^{22}$ cm$^{-2}$, $(5.1-6.3) \times 10^{22}$ cm$^{-2}$ for Model-2 and $(5-6.9) \times 10^{22}$ for Model-3. In panel (d), we show the variations 
in the inner-disk temperature ($T_{in}$ in keV) for all four models for a fixed $N_H$. The $T_{in}$ shows variation in the range of $0.9-1.9$ keV for all four models. In panel (e), we show 
the variations of photon indices. The red filled-squares and empty squares represent $\Gamma1$ and $\Gamma2$ for the \textit{broken power-law} model, where the blue, green, and black filled 
squares represent the $\Gamma$ of the \textit{pexrav}, \textit{pexriv}, and \textit{relxill} models. We notice that the photon index was high if we take into the consideration of the presence 
of type-C QPOs. We also note that both the $T_{in}$ and $Norm_{diskbb}$ values did not show typical variations as was observed for other BHCs (Remillard \& McClintock 2006). When an outburst 
progresses, the normalization generally decreases from a high value. In this case, we have not seen that trend, rather it varied randomly. In general, $T_{in}$ increases as an outburst 
progresses, reflecting the movement of the disk inwards. Here, for all the model combinations, we did not notice that increasing profile. The {\fontfamily{pcr}\selectfont pexriv} model 
fitted ionization parameter is low here in the range of $2.4 \times 10^{-13} - 3.8 \times 10^{-8}$, the reason of which is not clear. Since the ionization parameter is low, the estimated
irradiation flux ($F_{ion}$) is also very low, which can be estimated using the relation $\xi = 4 \pi F_{ion}/n$, where $n$ is the density of the reflector (Done et al. 1992) and $F_{ion}$ 
is the irradiating flux in the 5~eV to 20~keV energy band.

The {\fontfamily{pcr}\selectfont relxill} model fits provided a broad range for both the spin and inclination parameters of the source to be $0.5-0.81$ and $10-41$ (in degrees). This also could 
be due to the degeneracy in model parameters and significant changes in $Norm_{diskbb}$. Specifically on the last exposure Id, we found that the $Norm_{diskbb}$ increased to a very high value, 
compared to other Ids and disk temperature decreased to 0.9, whereas the photon index increased to 3.3. However, the ranges of the spin and inclination parameters cover the estimations in Mondal 
et al. (2024). Since the last exposure Id returned a very high photon index, we conclude the spectral state as SS, as was reported earlier by Mondal et al. (2024). The absence of QPO also supports 
the same spectral state. We found that the ionization parameter in {\fontfamily{pcr}\selectfont relxill} model fitting was high ($>3$) and the iron abundance ($AFe$) varied from $1.1-4.9$. The 
reflection fraction varied in a range of $0.12-2.29$. The $E_{cut}$ also showed random variations. We discuss the possible reasons in the next section.

\subsection{Evolution of the Shock}

In the TCAF scenario, the oscillation of the shock produces QPOs. According to this model, the QPO frequency can be written as (Molteni et al. 1996, Chakrabarti et al. 2005),

$$\nu_{qpo} = \frac{c^3}{2GM_{BH}} \frac{1}{RX_s (X_s -1)^{1/2}} ~\text{Hz}, \eqno{(1)}$$

where the following are represented, respectively: $c$, $G$, $M_{BH}$, $X_s$, and $R$; these are the speed of light, the gravitational constant, the mass of the BH, the shock 
location, and the ratio of matter densities in post-shock to pre-shock regions ($\rho_+/\rho_-$). 

\begin{figure}[!h]
\vspace{0.8cm}
  \centering
    \includegraphics[width=8.5cm]{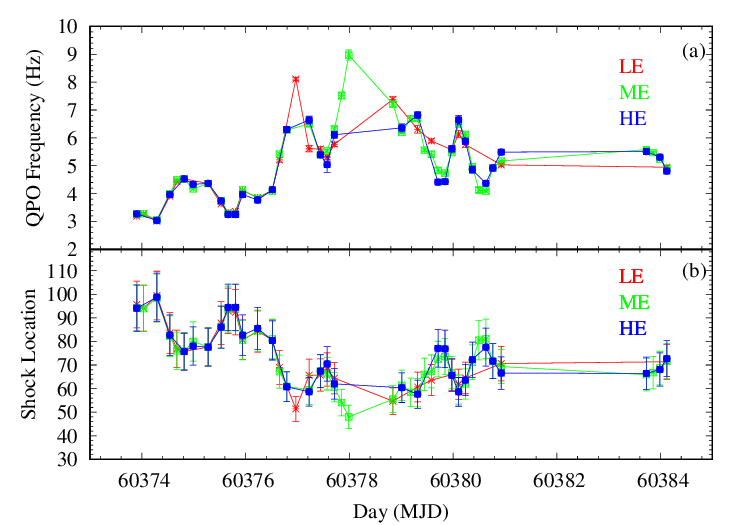}
    \caption{Evolution of the location of the shock, estimated from observed $\nu_{qpo}$. The panel (a) shows the variation of $\nu_{qpo}$ with time and (b) shows the variation of $X_s$ 
    with time.}
\end{figure}

The QPO frequencies ($\nu_{qpo}$) are estimated from timing analysis as discussed earlier and the shock locations during the outburst are estimated using the above relation in Eq. 1. We 
found that at the beginning of the outburst, the shock was located at a distance of $\sim 100~r_s$ from the BH (see Fig. 14b). Later, as the $\nu_{qpo}$ increased, the shock moved inwards,
decreasing the size of the corona. After a few days, the shock became stable. Table 3 provides the values for the shock locations (columns 11–13).

\section{Discussions}

The Galactic black hole Swift J151857.0-572147 started an outburst recently in March 2024. We have used {\it Insight}-HXMT data for our both timing and spectral studies from 2024 March 
04 to 2024 March 17. Using the $0.01$~sec time-binned light curves from the three instruments of HXMT (LE, ME, and HE), we studied the source's timing properties. We also searched for 
the energy dependence of LFQPOs by producing light curves in 7 different energy ranges within the HE band. Along with these, we searched for HFQPOs in all the PDS from LE, ME, and HE 
using $1$~ms time-binned light curves. We then examined the combined LE+ME+HE spectra in the $2-100$~keV broad energy band to learn more about the spectral characteristics of this source 
using the spectra files from these three instruments.

For stellar-mass black holes, quasi-periodic oscillation is one of the most significant and frequent occurrences. We examined 186 exposures in total for this recently found source (62 
for each of LE, ME, and HE). Nevertheless, incorrect light curve production occurred in 2 LE exposures. A total of 184 light curves for LE, ME, and HE were obtained. The details are 
listed in Table 2. We discovered that QPO was not present in each of these light curves. The details on QPO properties are listed in Table 3. Over the analysis period of $\sim 
13$ days, the QPO frequency has rapidly changed. In our analysis period, the $\nu_{qpo}$ showed variation approximately from 3.3 to 7.4, 3.2 to 9, and 3.2 to 7 Hz in LE, ME, and HE bands, 
respectively. Even in a single day, there was a change in the QPO frequency ($\nu_{qpo}$). The results section contains a general discussion on the evolution of QPO frequencies. Type-C 
QPO nature is identified from the fluctuation of QPO frequency, (\%)RMS, and $Q$-factor. One thing we would like to discuss here that the difference in RMS value (in Table 3) for LE, ME, 
and HE is due to the large variation in background counts in these three bands. As we can notice in Table 2, the background count in the HE band is almost equivalent to the source count 
in the HE band for this source. The high background count rate in the HE band could be due to the contribution of the close proximity of the other source Cir X-1 and the high 
effective area of the detector in this band. However, it was previously reported by Peng et al. (2024) that Cir X-1 was mainly present in the soft spectral state during the outburst of 
Swift J151857.0-572147. Thus, it could only have contributed to the LE band. Thereby the reason for this high background rate can not be firmly concluded. 

Even though the QPOs have been thoroughly examined in the literature using observations from other sources, further modeling is necessary to understand their origin and connection with 
the spectral properties. Here, we want to concentrate on the physical scenario that explains how shock instabilities in advective flows near black holes (BHs) give rise to QPOs. It 
is already explained in the introduction how the shock oscillations in the TCAF model explains the origin of the QPOs. This shock may not be stable at the outer edge over time. There 
could be oscillations in the CENBOL boundary, which can be caused due to either of two reasons: 

(i) According to Chakrabarti (1989), the satisfaction of the Rankine-Hugoniot condition makes the boundary of the shock stable and steady. However, if this condition is not satisfied 
(Ryu et al. 1997), the shock could oscillate at the outer boundary. It could produce variabilities in the light curves.

(ii) Molteni et al. (1996) stated that the presence of cooling may cause the shock to oscillate. QPOs emerge during the oscillation when the compressional heating timescale and the 
cooling timescale due to inverse Comptonization process match (see Chakrabarti et al. 2015).

Depending on the flow parameters causing shocks, $X_s$ can be anywhere over $10~r_s$. When the spectral nature of an outburst is hard, the shock forms far away $\sim 1000 ~r_s$, and it 
gradually becomes small in the following days as cooling increases (Mondal et al. 2015). For this outburst, the shock was far from the BH at the start of our analysis period. As the 
spectral nature of the outburst was softening, the shock moved inwards, suggesting cooling was in progress. As the shock moved inwards, the QPO frequency increased.  

Although MAXI/GSC observed the source, it was not identified as a new source due to the proximity of another source, called Cir X-1. Thus, from the HXMT extracted light curves 
and HR variations, we may say that the source transitioned past its HS at the start of our analysis period and was already moved to the intermediate state. The variation of the 
photon index supports the above spectral state. At the start of our analysis period, the shock was at a distance, which suggests that the source had already completed its HS and was in 
the intermediate state. Later, the shock moved inward, suggesting the source is making a transition towards the soft state. The non-identification of type-B QPO does not help to designate 
a transition between the HIMS and SIMS, and thus we consider this overall observation period belonging to the intermediate state. On several exposure Ids, we did not notice any QPO signature 
from the start of our analysis to the end in any of the three bands (Table 2 \& 3 for correspondence). This could be because of the mismatch of the heating and cooling timescales at the 
shock. For several exposure Ids, QPO is only present in one of the three bands, whereas, for some of them it was present in two bands (Table 2 \& 3 for correspondence). We find 
that on the last day of our analysis period, the $\Gamma$ became high ($> 3.3$ from all the three models). These values suggest that the source transitioned into the SS that day 
(Remillard \& McClintock 2006). Thus, we did not find QPO in any band on that day. The absence of QPOs and high $\Gamma$ agrees with the SS as inferred in Mondal et al. (2024) 
using joint {\it IXPE} and {\it NuSTAR} observations of the source. As mentioned previously in the result section, we have not noticed typical variations of the $\Gamma$, 
$Norm_{diskbb}$, $T_{in}$ and we have also found a broad range of the spin and inclination from the spectral analysis. We speculate that this is due to the contribution from the Cir X-1 
source, albeit it was in the low energy band.

We did not find the presence of any high-frequency QPO during the entire analysis period of the outburst. The HFQPO phenomenon is not very common. To date, HFQPO has been observed 
in a few sources only, e.g., GRO J1655-40 (Remillard et al. 1999; Strohmayer 2001b; Remillard et al. 2006), H1743-322 (Homan et al. 2005; Remillard et al. 2006), XTE J1550-564 (Homan et al. 
2001; Miller et al. 2001; Remillard et al. 2002a), and GRS 1915 + 105 (Morgan, Remillard \& Greiner 1997; Strohmayer 2001b; Belloni et al. 2006). This suggests that this is not a very common 
phenomenon, like LFQPO in BHXRBs. Its absence could be because the disk did not proceed very close to the compact object to produce variabilities with high frequency. Detection of
HFQPOs requires high timing resolution, large effective area, both of which are synonymous with the {\it Insight}-HXMT satellite. However, the photon statistics should be very strong to 
detect HFQPOs that requires very high signal-to-noise (S/N) ratio, especially in soft energy band, as it is generally thought to be produced by the disk when it is very close to the black 
hole. Considering this source was at proximity of another source, this may not have satisfied. These could be plausible causes for the non-detection of HFQPOs for this outburst. We also 
studied the energy dependence of LFQPOs in the HE band for those exposures in which LFQPO was present in the full energy band. The energy 
dependence of QPOs could give valuable insight into the origin of the QPO. Ma et al. (2023) reported that LFQPO was present till very high energy, which suggested that the origin of the QPO 
could be from the precession of the jet. Examining all 31 exposures, we find that LFQPO was present till $48$~keV, above which there is no prominent or weak QPO nature, either. In the $27-35$
~keV band, the nature of LFQPOs was stronger than in the $35-48$~keV band. Such observational findings can be explained from TCAF model scenario, where in the intermediate states CENBOL shrinks, 
due to the increase in cooling effects (see, Mondal et al. 2015), therefore the spectral break energy permissible for inverse-Compton scattering also decreases, which is the case for the 
present source. Thereby, much higher energy photons possibly could not contribute to the observed QPOs. There could be another explanation for the absence of QPOs at high energies. 
We have noticed that the background count rates are higher for the HE band and are very comparable to the source count. Due to the high background, there could be less contribution from the 
source at higher energy bands. This could also explain why we do not observe QPOs at these HE bands. Since the source is already in the SIMS, the RMS amplitude is generally lower than in the 
HS. Therefore, the upper limit on the RMS amplitude can not be constrained from the present data sets, requiring a detailed study with the complete cycle of an outburst.

The source Swift J151857.0-572147 has shown a very high value of $N_H$, using all three combinations of models. The $N_H$ varied in the range $(4.3-6.5) \times 10^{22}$
~cm$^{-2}$, $(5.1-6.3) \times 10^{22}$~cm$^{-2}$, and $(5-6.9) \times 10^{22}$~cm$^{-2}$ for Model-1, Model-2, and Model-3, respectively. The average column density was $\sim 5.6 \times 
10^{22}~cm^{-2}$. This value is significantly high, compared to other Galactic black holes. For example, some BHCs Swift J1727.8-1613, MAXI J1803-298, GX 339-4, and Swift J1357.2-0933 
the $N_H$ varied in the range $(0.1-0.5) \times 10^{22}~cm^{-2}$ (Liu et al. 2024; Debnath et al. 2024; Chatterjee et al. 2024), $(0.2-0.5) \times 10^{22}~cm^{-2}$ (Adegoke et al. 2024; 
Jana et al. 2022), $0.5 \times 10^{22}~cm^{-2}$ (Motta et al. 2009), and $0.13 \times 10^{22}~cm^{-2}$ (Mondal \& Chakrabarti 2019). For these sources, there was no local absorption and 
thus the values were close to the Galactic hydrogen column density (HI4PI Collaboration 2016). This has also been observed for other BHXRBs for which there was no local absorption to the 
source. This indicates some absorption local to the source, which could be due to the outflows from the disk or the presence of some blobs along the line of sight (see Neilsen \& Homan, 
2012; Mondal \& Jithesh, 2023). To confirm this, we need a detailed study of the outflow/jet properties of the source.

\section{Summary and Conclusions}

We have studied the timing and spectral properties of the very first outburst of the BHC Swift J151857.0-572147 in 2024. Using \textit{Insight}-HXMT LE, ME, and HE exposure 
average light curve data, we present the evolution of the light curve and its hardness ratio across our full analysis period from March 04 2024 (MJD 60373) to March 17 2024 (MJD 60386). 
For our investigation, we selected the 7 observation IDs using the {\it Insight}-HXMT data, publicly available during the analysis. For timing analysis, we employed all of the exposures 
from those observation IDs, and for spectrum analysis, we employed selective exposures, respectively. We produced a power density spectrum and used $0.01 ~s$ time-binned light curves from 
the three HXMT instruments, i.e., LE, ME, and HE, to study the QPO properties. We used the {\fontfamily{pcr}\selectfont \textit{Lorentzian}} model to obtain the QPO properties. We also studied
energy-dependent QPO by producing HE light curves in seven different energy bands. We extracted the energy-dependent QPO properties in the same way we did for the LE, ME, and HE light 
curves in the full band. Apart from these, we also produced $0.001$~s time-binned light curve to search for high-frequency QPOs. We employ {\fontfamily{pcr}\selectfont LE + ME + HE} 
spectrum files in the broad $2-100$~keV energy band for spectral analysis. We found that the models i) {\fontfamily{pcr}\selectfont constant*tbabs*(diskbb + broken power-law)} and ii) 
{\fontfamily{pcr}\selectfont constant*tbabs*(diskbb + pexrav)} fit the spectra for the best statistics. Based on our investigation, we conclude that:

i) The source was present in the intermediate state at the start of our analysis period and proceeded toward the soft state as the outburst progressed.

ii) It was in the soft state at the last observation ID of our analysis period.

iii) Type-C QPO was present in the intermediate state, which could be produced by the shock instability in the transonic accretion flow. 

iv) As the source transited to the soft state, we did not find any QPOs.

v) LFQPOs were present up to 48 keV, above which we did not find the presence of LFQPO for any of the exposures. 

vi) HFQPOs were absent during this analysis period.

vii) As the shock was of intermediate strength, it could not produce variabilities up to very high energies. Thus, we only found QPOs up to 48 keV. 

viii) The average hydrogen column density was high with $N_H \sim 5.6 \times 10^{22}$~cm$^{-2}$ in accordance with the estimation by Mondal et al. (2024), Peng et al. (2024). 
This could be due to the presence of outflows from the disk or some blobs along the line of sight, which prompts a further detailed study.

\section{Data Availability}

This work has made use of public data from several satellite/instrument archives and has made use of software from the HEASARC, which is developed and monitored by the Astrophysics 
Science Division at NASA/GSFC and the High Energy Astrophysics Division of the Smithsonian Astrophysical Observatory. This work made use of the data from the {\it Insight}-HXMT mission, 
a project funded by the China National Space Administration (CNSA) and the Chinese Academy of Sciences (CAS).

\section{Acknowledgements}

We thank Dr. Lian Tao of the Institute of High Energy Physics (IHEP), Chinese Academy of Sciences (CAS) for publicizing the {\it Insight}-HXMT data by request. We also thank Dr. Mutsumi 
Sugizaki of the Kanazawa University, Japan for providing fruitful information about the absence of MAXI/GSC daily average light curve on the source. We sincerely thank Prof. John A. Tomsick 
of the Space Sciences Lab, University of California, Berkeley, USA for providing suggestions on the draft. CBS is supported by the National Natural Science Foundation of China under grant 
no. 12073021. SPS and SM acknowledge the Ramanujan Fellowship (RJF/2020/000113) by SERB/ANRF-DST, Govt. of India for this research.  KC acknowledges support from the SWIFAR postdoctoral 
fund of Yunnan University.

\clearpage

\begin{appendix}

\begin{center}
\begin{figure*}[!h]
\centering
    \includegraphics[width=5.5cm, angle=270]{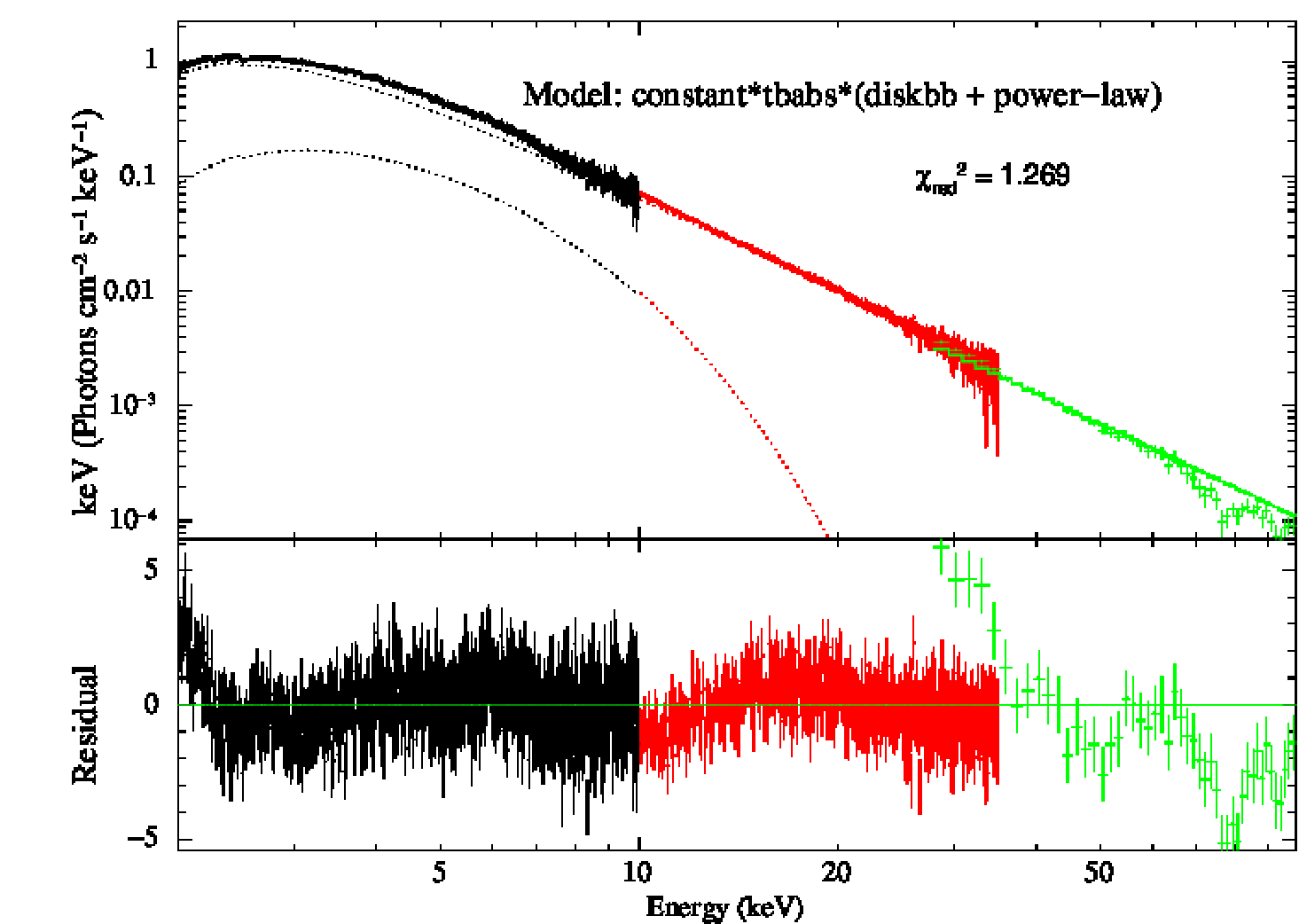}
    \caption{Model fitted unfolded spectrum for the combination of \textit{tbabs}, \textit{diskbb}, and \textit{power-law} models. This is for the observation ID. P0614374001 (exposure 
             ID: P061437400101-20240304-01-01).}
%\label{figa1}
\end{figure*}
\end{center}

Before fitting the data with the combination of phenomenological \textit{broken power-law} or physical \textit{pexrav} models with the combination of \textit{diskbb} model, we tried to 
perform the spectral analysis using only the combination of \textit{diskbb} and \textit{power-law} models with the interstellar absorption model \textit{tbabs}. The model combination reads 
as:\textit{constant*tbabs(diskbb + power-law)}. However, from Fig. 15, we notice that at the high energy end, after 20 keV, there is the presence of high residuals, which could be due to 
the presence of reflection radiation. Thus, we modeled the data using those above mentioned models to better fit the data, which we achieved.

%\newpage
\addtolength{\tabcolsep}{-1.5pt}
\hspace{-50.0cm}
\begin{longtable}{|c|c|c|c|c|c|c|c|c|c|}
\caption{Time and Count Rates of all the HXMT Exposures.}\label{tab:long} \\
 \hline
     Exposure                 &            \multicolumn{3}{|c|}{MJD}        & \multicolumn{3}{|c|}{Source Count Rate}  &  \multicolumn{3}{|c|}{Background Count Rate}    \\
\hline
        ID                    &      Start   &     Stop     &     Average   &     LE       &      ME     &      HE     &      LE       &    ME      &       HE           \\
        (1)                   &       (2)    &      (3)     &        (4)    &     (5)      &     (6)     &      (7)    &      (8)      &    (9)     &      (10)          \\
\hline
\endfirsthead
\multicolumn{10}{c}
{{\bfseries \tablename\ \thetable{} -- continued from previous page}} \\
\hline
     Exposure                 &            \multicolumn{3}{|c|}{MJD}        & \multicolumn{3}{|c|}{Source Count Rate}  &  \multicolumn{3}{|c|}{Background Count Rate}    \\
\hline
        ID                    &      Start   &     Stop     &     Average   &     LE       &      ME     &      HE     &      LE       &    ME      &       HE           \\
        (1)                   &       (2)    &      (3)     &        (4)    &     (5)      &     (6)     &      (7)    &      (8)      &    (9)     &      (10)          \\
\hline                                                                                                                                                                                                           
\endhead                                                                                                                                                                                                         
\hline \multicolumn{10}{|r|}{Continued on next page} \\ \hline                                                                                                                                                   
\endfoot
\endlastfoot
P061437400101-20240304-01-01* &  60373.83	 &  60373.96  &   60373.90  &    536.58    &    138.35   &    570.52   &      10.86    &   23.95    &     404.71         \\
P061437400102-20240304-01-01  &  60373.96	 &  60374.10  &   60374.03  &    580.17    &    156.85   &    624.12   &      10.96    &   24.83    &     523.82         \\
P061437400103-20240305-02-01  &  60374.10	 &  60374.47  &   60374.29  &    623.90    &    159.03   &    567.25   &      11.07    &   23.00    &     375.57         \\
P061437400104-20240305-02-01  &  60374.47	 &  60374.60  &   60374.54  &    637.14    &    140.78   &    522.86   &      10.54    &   22.75    &     357.61         \\
P061437400105-20240305-02-01  &  60374.60	 &  60374.74  &   60374.67  &    662.66    &    140.50   &    585.67   &      10.31    &   25.13    &     442.21         \\
P061437400106-20240305-02-01  &  60374.74	 &  60374.88  &   60374.81  &    644.42    &    132.14   &    541.91   &      10.74    &   21.73    &     404.61         \\
P061437400107-20240305-02-01* &  60374.88	 &  60375.09  &   60374.98  &    632.28    &    130.30   &    552.83   &      10.73    &   24.15    &     397.43         \\
P061437400201-20240306-01-01  &  60375.09	 &  60375.46  &   60375.27  &    622.66    &    126.47   &    523.01   &      10.95    &   22.51    &     380.94         \\
P061437400202-20240306-01-01  &  60375.46	 &  60375.59  &   60375.52  &    595.72    &    130.94   &    535.99   &      10.65    &   22.87    &     382.38         \\
P061437400203-20240306-01-01  &  60375.59	 &  60375.73  &   60375.66  &    600.56    &    139.07   &    621.86   &      10.96    &   25.96    &     423.93         \\
P061437400204-20240306-01-01  &  60375.73	 &  60375.87  &   60375.80  &    599.74    &    139.67   &    575.39   &      10.65    &   22.71    &     397.88         \\
P061437400205-20240306-01-01* &  60375.87	 &  60376.01  &   60375.94  &    595.50    &    131.44   &    555.10   &      11.40    &   24.49    &     403.45         \\
P061437400206-20240307-02-01  &  60376.01	 &  60376.45  &   60376.23  &    608.29    &    133.40   &    535.65   &      11.06    &   22.87    &     377.09         \\
P061437400207-20240307-02-01  &  60376.45	 &  60376.58  &   60376.51  &    621.09    &    129.42   &    522.51   &      11.19    &   22.66    &     366.92         \\
P061437400208-20240307-02-01  &  60376.58	 &  60376.72  &   60376.65  &    604.54    &    113.97   &    550.91   &      10.61    &   24.82    &     436.03         \\
P061437400209-20240307-02-01  &  60376.72	 &  60376.86  &   60376.79  &    638.58    &    108.06   &    500.53   &      10.37    &   22.94    &     398.94         \\
P061437400210-20240307-02-01* &  60376.86	 &  60377.07  &   60376.96  &    666.77    &    98.46    &    493.26   &      11.88    &   23.53    &     425.96         \\
P061437400301-20240308-01-01  &  60377.07	 &  60377.37  &   60377.22  &    633.28    &    103.44   &    473.36   &      10.53    &   21.98    &     375.15         \\
P061437400302-20240308-01-01  &  60377.37	 &  60377.50  &   60377.43  &    669.41    &    117.44   &    481.87   &      11.93    &   23.36    &     374.84         \\
P061437400303-20240308-01-01  &  60377.50	 &  60377.63  &   60377.56  &    625.12    &    113.95   &    518.43   &      11.07    &   27.79    &     396.08         \\
P061437400304-20240308-01-01  &  60377.63	 &  60377.78  &   60377.70  &    605.70    &    106.46   &    510.03   &      10.22    &   22.89    &     410.33         \\
P061437400305-20240308-01-01  &  60377.78	 &  60377.91  &   60377.85  &    600.03    &    92.76    &    473.50   &      10.47    &   22.67    &     391.08         \\
P061437400306-20240308-01-01* &  60377.91	 &  60378.05  &   60377.98  &    627.36    &    92.00    &    472.99   &      10.61    &   23.77    &     419.09         \\
P061437400307-20240309-02-01  &  60378.05	 &  60378.36  &   60378.21  &    651.44    &    82.46    &    416.10   &      10.34    &   21.32    &     372.58         \\
P061437400308-20240309-02-01  &  60378.36	 &  60378.49  &   60378.42  &      -       &    77.33    &    409.98   &        -      &   24.98    &     355.93         \\
P061437400309-20240309-02-01  &  60378.49	 &  60378.62  &   60378.55  &    598.49    &    79.97    &    446.97   &      11.36    &   25.21    &     366.72         \\
P061437400310-20240309-02-01  &  60378.62	 &  60378.77  &   60378.69  &    642.67    &    87.64    &    460.32   &      10.32    &   23.54    &     421.21         \\
P061437400311-20240309-02-01* &  60378.77	 &  60378.90  &   60378.83  &    613.15    &    98.77    &    473.44   &      10.27    &   23.16    &     387.30         \\
P061437400312-20240309-02-01  &  60378.90	 &  60379.11  &   60379.01  &    600.63    &    105.11   &    489.30   &      10.49    &   23.10    &     402.28         \\
P061437400401-20240310-01-01  &  60379.11	 &  60379.25  &   60379.18  &    602.94    &    96.22    &    455.29   &      10.20    &   22.20    &     375.02         \\
P061437400402-20240310-01-01  &  60379.25	 &  60379.38  &   60379.31  &    565.28    &    93.30    &    451.50   &      10.00    &   22.68    &     357.27         \\
P061437400403-20240310-01-01  &  60379.38	 &  60379.51  &   60379.44  &      -       &    108.09   &    467.63   &        -      &   26.14    &     357.47         \\
P061437400404-20240310-01-01  &  60379.51	 &  60379.64  &   60379.58  &    555.07    &    115.94   &    515.31   &      11.10    &   27.98    &     416.68         \\
P061437400405-20240310-01-01  &  60379.64	 &  60379.77  &   60379.71  &    537.59    &    113.94   &    521.13   &      10.30    &   23.27    &     415.00         \\
P061437400406-20240310-01-01* &  60379.77	 &  60379.90  &   60379.84  &    509.38    &    104.94   &    503.53   &      10.25    &   24.60    &     378.08         \\
P061437400407-20240310-01-01  &  60379.90	 &  60380.04  &   60379.97  &    517.66    &    97.23    &    502.49   &      10.53    &   23.50    &     405.01         \\
P061437400408-20240311-02-01  &  60380.04	 &  60380.17  &   60380.10  &    527.40    &    90.35    &    465.21   &      10.32    &   21.46    &     376.86         \\
P061437400409-20240311-02-01  &  60380.17	 &  60380.30  &   60380.23  &    571.07    &    106.01   &    462.14   &      9.93     &   22.41    &     360.21         \\
P061437400410-20240311-02-01  &  60380.30	 &  60380.43  &   60380.36  &    548.30    &    115.14   &    474.22   &      10.07    &   22.07    &     373.48         \\
P061437400411-20240311-02-01  &  60380.43	 &  60380.56  &   60380.50  &    580.27    &    134.88   &    525.68   &      11.17    &   29.07    &     386.17         \\
P061437400412-20240311-02-01  &  60380.56	 &  60380.69  &   60380.63  &    462.08    &    106.48   &    537.55   &      9.94     &   25.56    &     420.90         \\
P061437400413-20240311-02-01  &  60380.69	 &  60380.82  &   60380.76  &    477.94    &    97.35    &    493.75   &      10.20    &   22.13    &     396.88         \\
P061437400414-20240311-02-01* &  60380.82	 &  60381.03  &   60380.93  &    474.07    &    96.15    &    498.83   &      10.35    &   23.93    &     404.79         \\
P061437400501-20240312-01-01  &  60381.42	 &  60381.55  &   60381.49  &    483.34    &    68.63    &    433.63   &      10.76    &   26.38    &     405.46         \\
P061437400502-20240312-01-01  &  60381.55	 &  60381.68  &   60381.61  &    482.73    &    64.80    &    445.23   &      10.16    &   24.11    &     424.39         \\
P061437400503-20240312-01-01* &  60381.68	 &  60381.82  &   60381.75  &    537.66    &    80.44    &    417.55   &      10.13    &   21.57    &     402.64         \\
P061437400601-20240313-01-01  &  60382.41	 &  60382.58  &   60382.49  &    510.02    &    79.82    &    443.41   &      10.98    &   26.42    &     397.53         \\
P061437400602-20240313-01-01  &  60382.58	 &  60382.72  &   60382.65  &    516.28    &    82.24    &    437.49   &      9.78     &   23.70    &     412.44         \\
P061437400603-20240313-01-01  &  60382.72	 &  60382.86  &   60382.79  &    506.55    &    88.41    &    435.97   &      9.95     &   23.59    &     391.56         \\
P061437400604-20240313-01-01* &  60382.86	 &  60383.00  &   60382.93  &    501.54    &    92.93    &    460.78   &      10.00    &   24.10    &     401.87         \\
P061437400605-20240314-02-01  &  60383.00	 &  60383.13  &   60383.06  &    446.42    &    67.57    &    428.66   &      10.19    &   21.78    &     380.04         \\
P061437400606-20240314-02-01  &  60383.13	 &  60383.26  &   60383.19  &    441.60    &    69.75    &    419.85   &      9.98     &   21.97    &     368.76         \\
P061437400607-20240314-02-01  &  60383.26	 &  60383.39  &   60383.33  &    430.38    &    75.95    &    434.04   &      9.76     &   23.01    &     355.99         \\
P061437400608-20240314-02-01  &  60383.39	 &  60383.52  &   60383.46  &    494.91    &    100.00   &    482.17   &      11.36    &   28.54    &     445.78         \\
P061437400609-20240314-02-01  &  60383.52	 &  60383.66  &   60383.59  &    473.40    &    97.84    &    479.17   &      9.87     &   24.27    &     428.51         \\
P061437400610-20240314-02-01  &  60383.66	 &  60383.79  &   60383.72  &    470.76    &    107.58   &    473.73   &      9.83     &   22.89    &     400.49         \\
P061437400611-20240314-02-01* &  60383.79	 &  60383.92  &   60383.85  &    442.65    &    94.15    &    520.03   &      10.68    &   24.74    &     411.57         \\
P061437400612-20240314-02-01* &  60383.92	 &  60384.05  &   60383.98  &    405.67    &    84.88    &    471.21   &      10.46    &   22.74    &     395.89         \\
P061437400613-20240315-03-01* &  60384.05	 &  60384.18  &   60384.12  &    385.96    &    84.14    &    459.38   &      10.01    &   22.73    &     376.74         \\
P061437400614-20240315-03-01  &  60384.18	 &  60384.39  &   60384.29  &    408.51    &    80.65    &    438.74   &      9.89     &   22.18    &     366.30         \\
P061437400801-20240317-01-01* &  60386.50	 &  60386.67  &   60386.59  &    265.65    &    36.99    &    421.11   &      9.68     &   22.76    &     404.34         \\
P061437400802-20240317-01-01  &  60386.67	 &  60386.91  &   60386.79  &    256.75    &    43.10    &    440.27   &      9.88     &   24.65    &     422.77         \\
\hline
\end{longtable}
\vspace{-0.4cm}
 \leftline{Column 1 represents the Exposure IDs, taken for this complete analysis.} 
 \leftline{Column 2 and 3 represent the start and end MJDs of those  exposures respectively.} 
 \leftline{Column 4 represents the average MJD for those exposure IDs.} 
 \leftline{Columns 5, 6, \& 7 represent source count rate in LE, ME, and HE bands.} 
 \leftline{Columns 8, 9, \& 10 represent background count rate in LE, ME, and HE bands in `\textit{counts/sec}' unit.}

\vskip 0.5cm
%\newpage
\addtolength{\tabcolsep}{-2.075pt}
\hspace{-110cm}
\begin{longtable}{|c|c|c|c|c|c|c|c|c|c|c|c|c|c|c|c|}
\caption{Results from timing analysis of LE, ME, and HE light curves.}\label{tab:long} \\
\hline
Time       &              \multicolumn{3}{|c|}{QPO Frequency (Hz)}               &                      \multicolumn{3}{|c|}{Q-Value}                     &                             \multicolumn{3}{|c|}{RMS (\%)}           &   \multicolumn{3}{|c|}{Significance}   &      \multicolumn{3}{|c|}{Shock Location ($X_s$)}    \\
\hline  
 (MJD)     &         LE           &            ME        &            HE         &          LE          &            ME          &           HE           &           LE          &            ME         &          HE           &     LE     &    ME     &     HE       &             LE       &    ME      &     HE           \\   
\hline                                                                                                                                          
	(1)      &        (2)           &           (3)        &           (4)         &          (5)         &            (6)         &           (7)          &           (8)         &           (9)         &         (10)          &    (11)    &   (12)    &    (13)      &       (14)     &    (15)    &    (16)           \\  
  \hline
\endfirsthead
\multicolumn{16}{c}
{{\bfseries \tablename\ \thetable{} -- continued from previous page}} \\
\hline 
Time       &              \multicolumn{3}{|c|}{QPO Frequency (Hz)}               &                      \multicolumn{3}{|c|}{Q-Value}                     &                             \multicolumn{3}{|c|}{RMS (\%)}           &   \multicolumn{3}{|c|}{Significance}   &      \multicolumn{3}{|c|}{Shock Location ($X_s$)}    \\
\hline  
 (MJD)     &         LE           &            ME        &            HE         &          LE          &            ME          &           HE           &           LE          &            ME         &          HE           &     LE     &    ME     &     HE       &             LE       &    ME      &     HE           \\   
\hline                                                                                                                                          
	(1)      &        (2)           &           (3)        &           (4)         &          (5)         &            (6)         &           (7)          &           (8)         &           (9)         &         (10)          &    (11)    &   (12)    &    (13)      &       (14)     &    (15)    &    (16)           \\  
  \hline
\endhead
\hline \multicolumn{16}{|r|}{{Continued on next page}} \\ \hline
\endfoot
\hline %\hline
\endlastfoot
 3.90   &  $3.19\pm{0.02}$  &  $3.25\pm{0.08}$  &  $3.26\pm{0.01}$  &  $9.1 \pm{1.3 }$  &  $9.0 \pm{0.5 }$  &  $9.0 \pm{0.7 }$  &  $4.3 \pm{0.4 }$  &  $10.8 \pm{0.4 }$  &  $5.8 \pm{0.3 }$   &    8.0    &   17.4    &   15.4    &  $95.6 \pm{9.9 }$  &  $94.2 \pm{9.7 }$  &  $94.0 \pm{9.7 }$   \\ 
 4.03   &  $3.26\pm{0.04}$  &  $3.27\pm{0.02}$  &        -          &  $7.1 \pm{1.5 }$  &  $7.1 \pm{0.6 }$  &        -          &  $4.1 \pm{0.5 }$  &  $9.7  \pm{0.5 }$  &          -         &    6.1    &   12.6    &      -    &  $94.0 \pm{9.7 }$  &  $93.9 \pm{9.7 }$  &          -          \\
 4.28   &  $3.00\pm{0.01}$  &  $3.05\pm{0.09}$  &  $3.03\pm{0.07}$  &  $8.1 \pm{0.9 }$  &  $6.5 \pm{0.3 }$  &  $7.0 \pm{0.3 }$  &  $3.4 \pm{0.2 }$  &  $9.6  \pm{0.2 }$  &  $6.5 \pm{0.1 }$   &   10.6    &   25.0    &   27.1    &  $99.4 \pm{9.3 }$  &  $98.3 \pm{9.4 }$  &  $98.7 \pm{9.5 }$   \\   
 4.53   &  $3.91\pm{0.04}$  &  $4.00\pm{0.03}$  &  $3.97\pm{0.04}$  &  $6.5 \pm{1.4 }$  &  $4.6 \pm{0.5 }$  &  $4.1 \pm{0.4 }$  &  $3.5 \pm{0.4 }$  &  $11.2 \pm{0.7 }$  &  $7.3 \pm{0.5 }$   &    6.9    &   12.6    &   12.3    &  $83.5 \pm{8.6 }$  &  $82.2 \pm{8.5 }$  &  $82.6 \pm{8.5 }$   \\ 
 4.67   &  $4.42\pm{0.03}$  &  $4.50\pm{0.02}$  &        -          &  $8.5 \pm{1.3 }$  &  $6.1 \pm{0.5 }$  &        -          &  $3.6 \pm{0.3 }$  &  $11.5 \pm{0.6 }$  &          -         &    8.3    &   15.4    &      -    &  $76.8 \pm{7.9 }$  &  $76.0 \pm{7.8 }$  &          -          \\
 4.81   &  $4.53\pm{0.04}$  &  $4.51\pm{0.02}$  &  $4.53\pm{0.02}$  &  $7.9 \pm{1.7 }$  &  $5.7 \pm{0.3 }$  &  $6.6 \pm{0.4 }$  &  $3.7 \pm{0.4 }$  &  $12.4 \pm{0.5 }$  &  $6.5 \pm{0.3 }$   &    6.4    &   17.8    &   17.5    &  $75.7 \pm{7.8 }$  &  $75.8 \pm{7.8 }$  &  $75.7 \pm{7.8 }$   \\
 4.98   &        -          &  $4.17\pm{0.02}$  &  $4.32\pm{0.02}$  &        -          &  $6.2 \pm{0.3 }$  &  $8.8 \pm{1.0 }$  &        -          &  $12.6 \pm{0.4 }$  &  $6.1 \pm{0.4 }$   &      -    &   21.0    &    9.7    &          -         &  $80.0 \pm{8.3 }$  &  $78.0 \pm{8.0 }$   \\
 5.27   &  $4.36\pm{0.02}$  &  $4.36\pm{0.08}$  &  $4.36\pm{0.09}$  &  $9.4 \pm{1.2 }$  &  $7.6 \pm{0.3 }$  &  $7.6 \pm{0.4 }$  &  $3.6 \pm{0.2 }$  &  $12.0 \pm{0.2 }$  &  $6.3 \pm{0.2 }$   &   10.0    &   29.0    &   22.8    &  $77.6 \pm{8.0 }$  &  $77.6 \pm{8.0 }$  &  $77.5 \pm{8.0 }$   \\
 5.52   &  $3.63\pm{0.02}$  &  $3.74\pm{0.02}$  &  $3.74\pm{0.02}$  &  $10.6\pm{1.9 }$  &  $8.6 \pm{0.8 }$  &  $7.4 \pm{0.7 }$  &  $4.0 \pm{0.4 }$  &  $11.7 \pm{0.7 }$  &  $7.0 \pm{0.4 }$   &    7.6    &   10.4    &   10.4    &  $87.7 \pm{9.1 }$  &  $86.0 \pm{8.9 }$  &  $86.0 \pm{8.9 }$   \\
 5.66   &  $3.32\pm{0.03}$  &  $3.28\pm{0.02}$  &  $3.25\pm{0.02}$  &  $6.6 \pm{1.3 }$  &  $7.4 \pm{0.6 }$  &  $12. \pm{2.8 }$  &  $3.9 \pm{0.4 }$  &  $10.9 \pm{0.6 }$  &  $5.3 \pm{0.8 }$   &    6.8    &   13.5    &    4.3    &  $92.9 \pm{9.6 }$  &  $93.7 \pm{9.7 }$  &  $94.4 \pm{9.8 }$   \\
 5.80   &  $3.35\pm{0.03}$  &  $3.25\pm{0.02}$  &  $3.25\pm{0.02}$  &  $7.0 \pm{1.4 }$  &  $4.4 \pm{0.2 }$  &  $4.6 \pm{0.3 }$  &  $3.6 \pm{0.4 }$  &  $11.8 \pm{0.5 }$  &  $6.4 \pm{0.3 }$   &    6.5    &   18.0    &   16.0    &  $92.4 \pm{9.5 }$  &  $94.2 \pm{9.7 }$  &  $94.4 \pm{9.8 }$   \\
 5.94   &  $4.11\pm{0.03}$  &  $4.13\pm{0.02}$  &  $3.97\pm{0.02}$  &  $8.9 \pm{2.3 }$  &  $7.0 \pm{0.2 }$  &  $11.3\pm{1.6 }$  &  $3.9 \pm{0.6 }$  &  $11.5 \pm{0.4 }$  &  $5.7 \pm{0.5 }$   &    4.7    &   13.3    &    7.6    &  $80.7 \pm{8.3 }$  &  $80.5 \pm{8.3 }$  &  $82.6 \pm{8.5 }$   \\
 6.23   &  $3.85\pm{0.02}$  &  $3.83\pm{0.02}$  &  $3.76\pm{0.02}$  &  $6.7 \pm{0.9 }$  &  $6.0 \pm{0.2 }$  &  $5.6 \pm{0.2 }$  &  $3.8 \pm{0.3 }$  &  $11.6 \pm{0.3 }$  &  $6.7 \pm{0.1 }$   &   10.7    &   25.5    &   27.0    &  $84.2 \pm{8.7 }$  &  $84.5 \pm{8.7 }$  &  $85.5 \pm{8.8 }$   \\
 6.51   &  $4.11\pm{0.02}$  &  $4.09\pm{0.02}$  &  $4.13\pm{0.02}$  &  $16.4\pm{3.9 }$  &  $12.7\pm{1.5 }$  &  $9.0 \pm{1.3 }$  &  $3.3 \pm{0.5 }$  &  $11.1 \pm{0.9 }$  &  $6.1 \pm{0.6 }$   &    4.3    &    8.7    &    7.7    &  $80.7 \pm{8.3 }$  &  $81.0 \pm{8.4 }$  &  $80.4 \pm{8.3 }$   \\ 
 6.65   &  $5.21\pm{0.08}$  &  $5.42\pm{0.03}$  &        -          &  $5.0 \pm{1.3 }$  &  $4.2 \pm{0.4 }$  &        -          &  $3.9 \pm{0.5 }$  &  $13.6 \pm{0.9 }$  &          -         &    6.2    &   12.5    &      -    &  $68.9 \pm{7.1 }$  &  $67.1 \pm{6.9 }$  &          -          \\
 6.79   &        -          &  $6.28\pm{0.02}$  &  $6.30\pm{0.05}$  &        -          &  $7.5 \pm{0.8 }$  &  $6.2 \pm{1.0 }$  &        -          &  $11.0 \pm{0.7 }$  &  $5.2 \pm{0.5 }$   &      -    &   12.0    &    9.6    &          -         &  $60.9 \pm{6.3 }$  &  $60.7 \pm{6.3 }$   \\
 6.96   &  $8.10\pm{0.06}$  &        -          &        -          &  $5.9 \pm{0.9 }$  &        -          &        -          &  $3.8 \pm{0.3 }$  &          -         &          -         &   11.5    &      -    &      -    &  $51.3 \pm{5.3 }$  &          -         &          -          \\
 7.22   &  $5.61\pm{0.15}$  &  $6.50\pm{0.09}$  &  $6.63\pm{0.15}$  &  $4.6 \pm{1.5 }$  &  $2.3 \pm{0.2 }$  &  $1.9 \pm{0.3 }$  &  $2.7 \pm{0.5 }$  &  $13.9 \pm{0.8 }$  &  $6.7 \pm{0.6 }$   &    5.0    &   19.0    &   11.5    &  $65.6 \pm{6.8 }$  &  $59.5 \pm{6.1 }$  &  $58.6 \pm{6.0 }$   \\
 7.44   &  $5.59\pm{0.09}$  &  $5.40\pm{0.03}$  &  $5.38\pm{0.08}$  &  $6.5 \pm{2.5 }$  &  $9.3 \pm{1.7 }$  &  $4.8 \pm{1.0 }$  &  $3.8 \pm{0.8 }$  &  $10.6 \pm{1.3 }$  &  $7.1 \pm{0.9 }$   &    3.6    &    5.7    &    6.6    &  $65.7 \pm{6.8 }$  &  $67.3 \pm{6.9 }$  &  $67.4 \pm{6.9 }$   \\
 7.57   &  $5.30\pm{0.09}$  &  $5.53\pm{0.06}$  &  $5.03\pm{0.21}$  &  $17.6\pm{1.7 }$  &  $4.4 \pm{0.7 }$  &  $1.4 \pm{0.5 }$  &  $2.2 \pm{1.1 }$  &  $13.5 \pm{1.3 }$  &  $10.0\pm{2.0 }$   &    3.4    &    8.6    &    5.5    &  $68.1 \pm{7.0 }$  &  $66.2 \pm{6.8 }$  &  $70.5 \pm{7.3 }$   \\
 7.71   &  $5.76\pm{0.08}$  &  $6.32\pm{0.06}$  &  $6.11\pm{0.14}$  &  $6.1 \pm{1.6 }$  &  $3.6 \pm{0.5 }$  &  $3.1 \pm{0.6 }$  &  $3.4 \pm{0.5 }$  &  $12.7 \pm{1.0 }$  &  $6.0 \pm{0.7 }$   &    6.0    &   13.0    &    8.5    &  $64.4 \pm{6.6 }$  &  $60.6 \pm{6.2 }$  &  $62.0 \pm{6.4 }$   \\
 7.85   &        -          &  $7.51\pm{0.16}$  &        -          &        -          &  $3.3 \pm{0.5 }$  &        -          &        -          &  $12.9 \pm{1.3 }$  &          -         &      -    &    8.7    &      -    &          -         &  $54.0 \pm{5.6 }$  &          -          \\
 7.98   &        -          &  $8.97\pm{0.18}$  &        -          &        -          &  $3.6 \pm{0.9 }$  &        -          &        -          &  $11.6 \pm{1.7 }$  &          -         &      -    &    6.5    &      -    &          -         &  $48.0 \pm{4.9 }$  &          -          \\
 8.21   &        -          &        -          &        -          &        -          &        -          &        -          &        -          &          -         &          -         &      -    &      -    &      -    &          -         &          -         &          -          \\
 8.42   &        -          &        -          &        -          &        -          &        -          &        -          &        -          &          -         &          -         &      -    &      -    &      -    &          -         &          -         &          -          \\
 8.55   &        -          &        -          &        -          &        -          &        -          &        -          &        -          &          -         &          -         &      -    &      -    &      -    &          -         &          -         &          -          \\
 8.69   &        -          &        -          &        -          &        -          &        -          &        -          &        -          &          -         &          -         &      -    &      -    &      -    &          -         &          -         &          -          \\ 
 8.83   &  $7.38\pm{0.09}$  &  $7.21\pm{0.09}$  &        -          &  $13.6\pm{6.8 }$  &  $4.0 \pm{0.7 }$  &        -          &  $2.1 \pm{0.6 }$  &  $11.2 \pm{1.2 }$  &          -         &    3.2    &    9.0    &      -    &  $54.6 \pm{5.6 }$  &  $55.5 \pm{5.7 }$  &          -          \\
 9.01   &        -          &  $6.19\pm{0.03}$  &  $6.36\pm{0.15}$  &        -          &  $7.7 \pm{1.2 }$  &  $3.8 \pm{1.0 }$  &        -          &  $9.5  \pm{0.9 }$  &  $5.7 \pm{0.9 }$   &      -    &   8.85    &    5.6    &          -         &  $61.4 \pm{6.3 }$  &  $60.3 \pm{6.2 }$   \\
 9.18   &        -          &  $6.69\pm{0.09}$  &        -          &        -          &  $5.0 \pm{1.1 }$  &        -          &        -          &  $9.7  \pm{1.3 }$  &          -         &      -    &    7.2    &      -    &          -         &  $58.3 \pm{6.0 }$  &          -          \\
 9.31   &  $6.30\pm{0.13}$  &  $6.71\pm{0.07}$  &  $6.82\pm{0.13}$  &  $5.6 \pm{2.3 }$  &  $5.9 \pm{1.3 }$  &  $4.5 \pm{0.9 }$  &  $3.1 \pm{0.7 }$  &  $10.4 \pm{1.4 }$  &  $6.1 \pm{0.7 }$   &    3.8    &    6.7    &    8.2    &  $60.7 \pm{6.2 }$  &  $58.2 \pm{6.0 }$  &  $57.6 \pm{5.9 }$   \\
 9.44   &        -          &  $5.55\pm{0.04}$  &        -          &        -          &  $11.5\pm{3.3 }$  &        -          &        -          &  $9.8  \pm{1.8 }$  &          -         &      -    &    4.3    &      -    &          -         &  $66.0 \pm{6.8 }$  &          -          \\
 9.58   &  $5.88\pm{0.07}$  &  $5.40\pm{0.13}$  &        -          &  $13.4\pm{6.9 }$  &  $4.5 \pm{1.2 }$  &        -          &  $3.0 \pm{0.9 }$  &  $10.5 \pm{1.6 }$  &          -         &    2.8    &    6.2    &      -    &  $63.5 \pm{6.5 }$  &  $67.2 \pm{6.9 }$  &          -          \\
 9.71   &        -          &  $4.82\pm{0.09}$  &  $4.40\pm{0.03}$  &        -          &  $2.9 \pm{0.5 }$  &  $7.7 \pm{1.2 }$  &        -          &  $10.7 \pm{1.1 }$  &  $4.9 \pm{0.4 }$   &      -    &   10.5    &    9.0    &          -         &  $72.5 \pm{7.5 }$  &  $77.0 \pm{7.9 }$   \\
 9.84   &        -          &  $4.73\pm{0.03}$  &  $4.42\pm{0.03}$  &        -          &  $5.2 \pm{0.6 }$  &  $7.6 \pm{1.3 }$  &        -          &  $12.0 \pm{0.8 }$  &  $5.7 \pm{0.6 }$   &      -    &   12.4    &    7.3    &          -         &  $73.5 \pm{7.6 }$  &  $76.8 \pm{7.9 }$   \\
 9.97   &  $5.57\pm{0.06}$  &  $5.46\pm{0.02}$  &  $5.61\pm{0.15}$  &  $8.0 \pm{2.1 }$  &  $7.9 \pm{1.0 }$  &  $4.7 \pm{1.3 }$  &  $3.8 \pm{0.6 }$  &  $11.4 \pm{0.9 }$  &  $5.4 \pm{0.9 }$   &    5.3    &   10.4    &    5.0    &  $65.9 \pm{6.8 }$  &  $66.7 \pm{6.9 }$  &  $65.6 \pm{6.8 }$   \\ 
10.10   &  $6.13\pm{0.12}$  &  $6.48\pm{0.06}$  &  $6.65\pm{0.16}$  &  $7.3 \pm{3.3 }$  &  $5.8 \pm{1.0 }$  &  $3.4 \pm{0.9 }$  &  $2.9 \pm{0.8 }$  &  $10.9 \pm{1.1 }$  &  $5.4 \pm{0.8 }$   &    3.1    &    8.5    &    6.2    &  $61.8 \pm{6.4 }$  &  $59.5 \pm{6.1 }$  &  $58.6 \pm{6.0 }$   \\
10.23   &  $5.76\pm{0.10}$  &  $6.11\pm{0.08}$  &  $5.88\pm{0.09}$  &  $5.1 \pm{1.6 }$  &  $4.3 \pm{0.8 }$  &  $3.7 \pm{0.8 }$  &  $3.4 \pm{0.6 }$  &  $10.5 \pm{1.2 }$  &  $5.8 \pm{0.7 }$   &    5.4    &    8.8    &    7.2    &  $64.5 \pm{6.6 }$  &  $61.9 \pm{6.4 }$  &  $63.6 \pm{6.6 }$   \\
10.36   &        -          &  $4.96\pm{0.04}$  &  $4.86\pm{0.14}$  &        -          &  $8.2 \pm{1.7 }$  &  $4.1 \pm{1.3 }$  &        -          &  $9.6  \pm{1.3 }$  &  $5.6 \pm{1.0 }$   &      -    &    6.0    &    5.1    &          -         &  $71.1 \pm{7.3 }$  &  $72.2 \pm{7.4 }$   \\
10.50   &        -          &  $4.13\pm{0.08}$  &        -          &        -          &  $3.8 \pm{1.0 }$  &        -          &        -          &  $11.1 \pm{1.8 }$  &          -         &      -    &    5.4    &      -    &          -         &  $80.5 \pm{8.3 }$  &          -          \\
10.63   &        -          &  $4.09\pm{0.03}$  &  $4.36\pm{0.03}$  &        -          &  $4.4 \pm{0.6 }$  &  $9.2 \pm{1.7 }$  &        -          &  $11.7 \pm{1.0 }$  &  $5.3 \pm{0.6 }$   &      -    &   10.3    &    6.1    &          -         &  $81.0 \pm{8.4 }$  &  $77.5 \pm{8.0 }$   \\
10.76   &        -          &  $4.92\pm{0.02}$  &  $4.92\pm{0.02}$  &        -          &  $8.9 \pm{1.1 }$  &  $10.2\pm{1.4 }$  &        -          &  $10.9 \pm{0.8 }$  &  $4.9 \pm{0.4 }$   &      -    &   11.1    &    9.0    &          -         &  $71.6 \pm{7.4 }$  &  $71.6 \pm{7.4 }$   \\
10.92   &  $5.03\pm{0.04}$  &  $5.17\pm{0.02}$  &  $5.48\pm{0.07}$  &  $7.9 \pm{1.5 }$  &  $8.9 \pm{1.0 }$  &  $8.3 \pm{2.5 }$  &  $3.8 \pm{0.4 }$  &  $10.9 \pm{0.8 }$  &  $4.2 \pm{0.8 }$   &    6.9    &   11.3    &    4.4    &  $70.6 \pm{7.3 }$  &  $69.3 \pm{7.1 }$  &  $66.6 \pm{6.9 }$   \\
11.49   &        -          &        -          &        -          &        -          &        -          &        -          &        -          &          -         &          -         &      -    &      -    &      -    &          -         &          -         &          -          \\
11.61   &        -          &        -          &        -          &        -          &        -          &        -          &        -          &          -         &          -         &      -    &      -    &      -    &          -         &          -         &          -          \\
11.75   &        -          &        -          &        -          &        -          &        -          &        -          &        -          &          -         &          -         &      -    &      -    &      -    &          -         &          -         &          -          \\
12.49   &        -          &        -          &        -          &        -          &        -          &        -          &        -          &          -         &          -         &      -    &      -    &      -    &          -         &          -         &          -          \\ 
12.65   &        -          &        -          &        -          &        -          &        -          &        -          &        -          &          -         &          -         &      -    &      -    &      -    &          -         &          -         &          -          \\
12.79   &        -          &        -          &        -          &        -          &        -          &        -          &        -          &          -         &          -         &      -    &      -    &      -    &          -         &          -         &          -          \\
12.93   &        -          &        -          &        -          &        -          &        -          &        -          &        -          &          -         &          -         &      -    &      -    &      -    &          -         &          -         &          -          \\
13.06   &        -          &        -          &        -          &        -          &        -          &        -          &        -          &          -         &          -         &      -    &      -    &      -    &          -         &          -         &          -          \\
13.19   &        -          &        -          &        -          &        -          &        -          &        -          &        -          &          -         &          -         &      -    &      -    &      -    &          -         &          -         &          -          \\
13.33   &        -          &        -          &        -          &        -          &        -          &        -          &        -          &          -         &          -         &      -    &      -    &      -    &          -         &          -         &          -          \\ 
13.46   &        -          &        -          &        -          &        -          &        -          &        -          &        -          &          -         &          -         &      -    &      -    &      -    &          -         &          -         &          -          \\
13.59   &        -          &        -          &        -          &        -          &        -          &        -          &        -          &          -         &          -         &      -    &      -    &      -    &          -         &          -         &          -          \\
13.72   &        -          &  $5.57\pm{0.06}$  &  $5.51\pm{0.13}$  &        -          &  $4.9 \pm{0.9 }$  &  $3.8 \pm{0.8 }$  &        -          &  $9.4  \pm{1.1 }$  &  $5.3 \pm{0.7 }$   &      -    &    7.5    &    6.6    &          -         &  $65.8 \pm{6.8 }$  &  $66.4 \pm{6.8 }$   \\ 
13.85   &        -          &  $5.46\pm{0.04}$  &        -          &        -          &  $7.1 \pm{1.6 }$  &        -          &        -          &  $9.4  \pm{1.3 }$  &          -         &      -    &    6.1    &      -    &          -         &  $66.7 \pm{6.9 }$  &          -          \\
13.98   &        -          &  $5.23\pm{0.06}$  &  $5.30\pm{0.07}$  &        -          &  $6.3 \pm{1.3 }$  &  $6.6 \pm{1.9 }$  &        -          &  $10.3 \pm{1.3 }$  &  $4.5 \pm{0.7 }$   &      -    &    6.8    &    5.2    &          -         &  $68.7 \pm{7.1 }$  &  $68.1 \pm{7.0 }$   \\
14.12   &  $4.94\pm{0.05}$  &  $4.90\pm{0.02}$  &  $4.80\pm{0.03}$  &  $8.3 \pm{2.4 }$  &  $8.4 \pm{1.6 }$  &  $8.4 \pm{1.6 }$  &  $3.6 \pm{0.6 }$  &  $10.4 \pm{1.1 }$  &  $4.9 \pm{0.5 }$   &    5.0    &    8.7    &    7.7    &  $71.3 \pm{7.4 }$  &  $71.7 \pm{7.4 }$  &  $72.7 \pm{7.5 }$   \\
14.29   &        -          &        -          &        -          &        -          &        -          &        -          &        -          &          -         &          -         &      -    &      -    &      -    &          -         &          -         &          -          \\
16.59   &        -          &        -          &        -          &        -          &        -          &        -          &        -          &          -         &          -         &      -    &      -    &      -    &          -         &          -         &          -          \\
16.79   &        -          &        -          &        -          &        -          &        -          &        -          &        -          &          -         &          -         &      -    &      -    &      -    &          -         &          -         &          -          \\
\hline
\end{longtable}
\vspace{-0.4cm}
\leftline{In column 1, we have listed the MJD-60370 (to save space) of the exposure IDs we used.} 
\leftline{Columns 2, 3, \& 4 represent the QPO frequency (in $Hz$ unit) in LE, ME, and HE energy bands respectively.} 
\leftline{Columns 5, 6, \& 7 represent the Q-values of QPOs in LE, ME, and HE energy bands respectively.} 
\leftline{Columns 8, 9, \& 10 represent the QPO RMS (\%) in LE, ME, and HE energy bands respectively.} 
\leftline{Columns 11, 12, \& 13 represent the significance of QPO in LE, ME, and HE energy bands respectively.}
\leftline{Columns 14, 15, \& 16 represent the shock location in LE, ME, and HE energy bands respectively. $X_s$ is in the units of Schwarzschild} \leftline{radius ($r_s = 2GM_{BH}/c^2$).}

\vskip 0.5cm
%\newpage
\addtolength{\tabcolsep}{1.1pt}
\hspace{-50.0cm}
\begin{longtable}{|c|c|c|c|c|c|c|c|c|}
  \caption{Results of timing analysis from energy dependent HE light curves}\label{tab:long} \\
 \hline
  Time       &                   \multicolumn{4}{|c|}{27-35 keV}                                       &                                \multicolumn{4}{|c|}{35-48 keV}                               \\
  \hline
 (MJD)       &   Frequency (Hz)     &     $Q$-Value         &          RMS (\%)     &  Significance    &   Frequency (Hz)      &        $Q$-Value         &          RMS (\%)    &     Significance   \\
 (1)         &         (2)          &        (3)            &            (4)        &       (5)        &        (6)            &            (7)           &            (8)       &          (9)       \\
\hline
\endfirsthead
\multicolumn{7}{c}
{{\bfseries \tablename\ \thetable{} -- continued from previous page}} \\
\hline
  Time       &                   \multicolumn{4}{|c|}{27-35 keV}                                       &                                \multicolumn{4}{|c|}{35-48 keV}                               \\
  \hline
 (MJD)       &   Frequency (Hz)     &     $Q$-Value         &          RMS (\%)     &  Significance    &   Frequency (Hz)      &        $Q$-Value         &          RMS (\%)    &     Significance   \\
 (1)         &         (2)          &        (3)            &            (4)        &       (5)        &        (6)            &            (7)           &            (8)       &          (9)       \\
\hline
\endhead

\hline \multicolumn{7}{|r|}{Continued on next page} \\ \hline
\endfoot
\hline %\hline
\endlastfoot
60373.90     &  $ 3.27 \pm 0.01 $   &   $  8.6  \pm 0.7 $   &   $ 5.9  \pm 0.3 $    &     15.1     &   $ 3.25 \pm 0.01 $   &   $  8.6  \pm 0.9 $   &   $  4.6  \pm 0.3 $  &    11.8      \\
60374.29     &  $ 3.06 \pm 0.09 $   &   $  6.6  \pm 0.3 $   &   $ 6.1  \pm 0.2 $    &     24.7     &   $ 3.02 \pm 0.01 $   &   $  6.6  \pm 0.4 $   &   $  4.9  \pm 0.2 $  &    19.2      \\
60374.54     &  $ 3.93 \pm 0.03 $   &   $  4.7  \pm 0.5 $   &   $ 6.9  \pm 0.5 $    &     11.3     &   $ 4.01 \pm 0.06 $   &   $  4.1  \pm 0.7 $   &   $  5.6  \pm 0.6 $  &     8.8      \\
60374.81     &  $ 4.50 \pm 0.02 $   &   $  5.8  \pm 0.5 $   &   $ 6.5  \pm 0.4 $    &     15.1     &   $ 4.54 \pm 0.03 $   &   $  6.1  \pm 0.7 $   &   $  5.2  \pm 0.4 $  &    11.8      \\
60374.98     &  $ 4.32 \pm 0.01 $   &   $ 10.2  \pm 1.2 $   &   $ 5.7  \pm 0.4 $    &      9.4     &   $ 4.31 \pm 0.02 $   &   $ 10.0  \pm 1.6 $   &   $  5.0  \pm 0.5 $  &     8.0      \\
60375.27     &  $ 4.36 \pm 0.01 $   &   $  8.5  \pm 0.5 $   &   $ 5.9  \pm 0.2 $    &     22.1     &   $ 4.34 \pm 0.01 $   &   $  7.8  \pm 0.6 $   &   $  5.1  \pm 0.2 $  &    18.2      \\
60375.52     &  $ 3.80 \pm 0.02 $   &   $  5.9  \pm 0.7 $   &   $ 7.1  \pm 0.6 $    &     10.2     &   $ 3.77 \pm 0.04 $   &   $  7.1  \pm 1.5 $   &   $  4.9  \pm 0.6 $  &     6.4      \\
60375.66     &  $ 3.22 \pm 0.02 $   &   $  10.  \pm 1.7 $   &   $ 5.9  \pm 0.7 $    &      6.1     &   $ 3.19 \pm 0.03 $   &   $  9.4  \pm 2.2 $   &   $  4.8  \pm 0.7 $  &     5.1      \\
60375.80     &  $ 3.26 \pm 0.02 $   &   $  4.3  \pm 0.4 $   &   $ 6.2  \pm 0.4 $    &     13.5     &   $ 3.22 \pm 0.02 $   &   $  5.3  \pm 0.6 $   &   $  5.1  \pm 0.4 $  &    11.4      \\
60375.94     &  $ 3.96 \pm 0.02 $   &   $  9.6  \pm 1.6 $   &   $ 5.3  \pm 0.6 $    &      7.0     &   $ 3.98 \pm 0.03 $   &   $  9.2  \pm 1.9 $   &   $  4.9  \pm 0.6 $  &     6.3      \\
60376.23     &  $ 3.79 \pm 0.01 $   &   $  5.6  \pm 0.3 $   &   $ 6.3  \pm 0.2 $    &     23.8     &   $ 3.82 \pm 0.01 $   &   $  6.2  \pm 0.5 $   &   $  5.1  \pm 0.2 $  &    16.6      \\
60376.51     &  $ 4.13 \pm 0.03 $   &   $  8.4  \pm 1.5 $   &   $ 6.1  \pm 0.7 $    &      6.8     &   $ 4.13 \pm 0.02 $   &   $ 13.7  \pm 2.8 $   &   $  4.7  \pm 0.6 $  &     5.8      \\
60376.79     &  $ 6.21 \pm 0.06 $   &   $  5.0  \pm 0.9 $   &   $ 5.6  \pm 0.6 $    &      9.0     &   $ 6.51 \pm 0.14 $   &   $  3.5  \pm 0.9 $   &   $  5.1  \pm 0.8 $  &     6.0      \\
60377.22     &  $ 6.84 \pm 0.14 $   &   $  2.2  \pm 0.4 $   &   $ 6.5  \pm 0.7 $    &      7.7     &   $       -       $   &   $        -      $   &   $        -      $  &      -       \\
60377.43     &  $ 5.43 \pm 0.04 $   &   $  7.9  \pm 1.5 $   &   $ 6.8  \pm 0.8 $    &      6.9     &   $ 5.43 \pm 0.06 $   &   $  7.9  \pm 2.2 $   &   $  5.5  \pm 0.9 $  &     4.8      \\
60377.56     &  $ 5.19 \pm 0.18 $   &   $  2.2  \pm 0.6 $   &   $ 9.2  \pm 1.5 $    &      5.7     &   $ 5.47 \pm 0.21 $   &   $  3.9  \pm 1.8 $   &   $  5.6  \pm 1.5 $  &     3.0      \\
60377.70     &  $ 6.55 \pm 0.11 $   &   $  3.9  \pm 0.9 $   &   $ 5.8  \pm 0.8 $    &      7.2     &   $       -       $   &   $        -      $   &   $        -      $  &      -       \\
60379.31     &  $ 6.68 \pm 0.20 $   &   $  2.8  \pm 0.8 $   &   $ 6.2  \pm 1.1 $    &      6.5     &   $ 6.72 \pm 0.24 $   &   $  2.6  \pm 0.9 $   &   $  5.9  \pm 1.1 $  &     5.5      \\
60379.71     &  $ 4.68 \pm 0.12 $   &   $  2.9  \pm 0.8 $   &   $ 5.5  \pm 0.8 $    &      8.8     &   $ 5.21 \pm 0.20 $   &   $  1.8  \pm 0.5 $   &   $  6.3  \pm 0.9 $  &     8.3      \\
60379.84     &  $ 4.45 \pm 0.02 $   &   $  11.  \pm 1.8 $   &   $ 5.3  \pm 0.5 $    &      7.4     &   $ 4.43 \pm 0.04 $   &   $  7.9  \pm 1.7 $   &   $  5.0  \pm 0.7 $  &     5.9      \\
60379.97     &  $ 5.59 \pm 0.12 $   &   $  5.5  \pm 2.1 $   &   $ 4.7  \pm 1.1 $    &      3.9     &   $ 5.82 \pm 0.13 $   &   $  6.5  \pm 2.9 $   &   $  4.3  \pm 1.0 $  &     4.1      \\
60380.10     &  $ 6.53 \pm 0.10 $   &   $  4.0  \pm 0.8 $   &   $ 6.1  \pm 0.7 $    &      7.6     &   $ 6.21 \pm 0.19 $   &   $  2.5  \pm 0.8 $   &   $  5.9  \pm 0.9 $  &     7.7      \\
60380.23     &  $ 5.82 \pm 0.26 $   &   $  1.4  \pm 0.4 $   &   $ 7.6  \pm 1.1 $    &      7.7     &   $ 5.72 \pm 0.21 $   &   $  2.9  \pm 1.1 $   &   $  4.8  \pm 0.9 $  &     4.8      \\
60380.36     &  $ 5.04 \pm 0.08 $   &   $  4.8  \pm 1.2 $   &   $ 6.1  \pm 0.9 $    &      5.3     &   $       -       $   &   $        -      $   &   $        -      $  &      -       \\
60380.63     &  $ 4.30 \pm 0.03 $   &   $ 10.4  \pm 2.3 $   &   $ 4.7  \pm 0.6 $    &      5.8     &   $ 4.35 \pm 0.08 $   &   $  5.6  \pm 1.8 $   &   $  4.6  \pm 0.9 $  &     4.7      \\
60380.76     &  $ 4.96 \pm 0.03 $   &   $  7.9  \pm 1.1 $   &   $ 5.5  \pm 0.4 $    &     10.6     &   $ 4.97 \pm 0.03 $   &   $  9.4  \pm 1.8 $   &   $  4.4  \pm 0.5 $  &     7.9      \\
60380.93     &  $ 5.35 \pm 0.11 $   &   $  4.4  \pm 1.3 $   &   $ 5.5  \pm 0.9 $    &      5.5     &   $       -       $   &   $        -      $   &   $        -      $  &      -       \\
60383.72     &  $ 5.58 \pm 0.07 $   &   $  5.5  \pm 1.1 $   &   $ 5.1  \pm 0.6 $    &      8.2     &   $ 5.65 \pm 0.11 $   &   $  4.6  \pm 1.3 $   &   $  4.5  \pm 0.8 $  &     5.4      \\
60383.98     &  $ 5.29 \pm 0.05 $   &   $  8.5  \pm 2.1 $   &   $ 4.8  \pm 0.7 $    &      6.0     &   $ 5.18 \pm 0.12 $   &   $  5.7  \pm 2.4 $   &   $  3.5  \pm 0.9 $  &     3.1      \\
60384.29     &  $ 4.92 \pm 0.04 $   &   $  8.1  \pm 1.6 $   &   $ 4.9  \pm 0.6 $    &      6.9     &   $ 4.98 \pm 0.08 $   &   $  6.1  \pm 1.9 $   &   $  4.2  \pm 0.8 $  &     4.8      \\
\hline 
\end{longtable}
\vspace{-0.4cm}
\leftline{In column 1, we have listed the MJD of the exposure IDs we used.} 
\leftline{Columns 2, 3, 4, \& 5 represent the QPO frequency (in $Hz$), $Q$-value, RMS (\%), and significance in 27-35 keV energy band.} 
\leftline{Columns 6, 7, 8, \& 9 represent the QPO frequency (in $Hz$), $Q$-value, RMS (\%), and significance in 35-48 keV energy band.}
%\clearpage

\vskip 0.5cm
%\newpage
\begin{table*}[!h]
%\scriptsize
\addtolength{\tabcolsep}{3.0pt}
 \centering
 \caption{Properties from spectral analysis using Model-1. Column 1 represents the MJD of those respective Exposure IDs for which we have performed spectral analysis. Column 2 gives the 
 values of the hydrogen column densities ($N_H$) of those analyzed exposures. Columns 3 \& 4 give the values of the parameters from the \textit{diskbb} model. Columns 5-8 give  the values 
 of the parameters from the \textit{broken power-law} model. Columns 9 \& 10 give the values of the constants needed to achieve simultaneous broadband fitting. Column  11 gives the values 
 of the $\chi^2/DOF$ for each fitting. The errors are estimated with 90\% confidence interval, that corresponds to 1.645$\sigma$ in XSPEC.}
 \label{tab:table5}
 \begin{tabular}{|c|c|c|c|c|c|c|c|c|c|c|}
 \hline
 Time      &      TBabs       &     \multicolumn{2}{|c|}{diskbb}        &                    \multicolumn{4}{|c|}{broken power-law}                       & \multicolumn{2}{|c|}{Fitting constants}  &     Fitting Stat  \\
\hline 
 MJD       &      $N_H$       & $T_{in}$ (keV)     &        Norm        &      $\Gamma1$      &   $E_b$ (keV)    &     $\Gamma2$       &        Norm      &     Constant1       &    Constant2       &     $\chi^2/DOF$  \\
\hline 
 (1)       &       (2)        &        (3)         &         (4)        &         (5)         &        (6)       &        (7)          &        (8)       &        (9)          &       (10)         &        (11)       \\ 
\hline 
60373.90   &   $5.1 \pm 0.1$  &   $1.6 \pm 0.3$    &   $145 \pm   8 $   &    $2.4 \pm 0.2$    &  $19.2 \pm 0.3$  &    $3.1 \pm 0.2$    &  $15.4 \pm 1.1$  &   $1.10 \pm 0.01$   &  $1.15 \pm 0.02$   &  $1242.96/1410$   \\
60374.98   &   $5.1 \pm 0.2$  &   $1.4 \pm 0.1$    &   $278 \pm  23 $   &    $2.5 \pm 0.4$    &  $15.1 \pm 0.4$  &    $2.9 \pm 0.2$    &  $19.8 \pm 2.5$  &   $1.01 \pm 0.01$   &  $0.97 \pm 0.02$   &  $1238.56/1410$   \\
60375.94   &   $4.9 \pm 0.2$  &   $1.4 \pm 0.1$    &   $222 \pm  23 $   &    $2.5 \pm 0.4$    &  $14.4 \pm 0.5$  &    $2.9 \pm 0.2$    &  $18.5 \pm 2.3$  &   $1.04 \pm 0.01$   &  $1.04 \pm 0.02$   &  $1188.51/1410$   \\
60376.96   &   $4.3 \pm 0.4$  &   $1.2 \pm 0.1$    &   $844 \pm 103 $   &    $2.2 \pm 0.2$    &  $10.7 \pm 0.2$  &    $3.1 \pm 0.2$    &  $ 7.2 \pm 4.9$  &   $0.96 \pm 0.02$   &  $0.91 \pm 0.03$   &  $1186.14/1410$   \\
60377.98   &   $4.7 \pm 0.3$  &   $1.2 \pm 0.1$    &   $755 \pm  75 $   &    $2.4 \pm 0.2$    &  $11.4 \pm 0.2$  &    $3.3 \pm 0.2$    &  $11.8 \pm 4.4$  &   $1.00 \pm 0.02$   &  $0.89 \pm 0.04$   &  $1204.31/1410$   \\
60378.83   &   $4.7 \pm 0.2$  &   $1.2 \pm 0.1$    &   $590 \pm  61 $   &    $2.4 \pm 0.1$    &  $11.0 \pm 0.2$  &    $3.2 \pm 0.2$    &  $14.2 \pm 3.4$  &   $0.95 \pm 0.01$   &  $1.06 \pm 0.03$   &  $1266.20/1410$   \\
60379.84   &   $5.4 \pm 0.1$  &   $1.4 \pm 0.3$    &   $136 \pm  20 $   &    $2.8 \pm 0.3$    &  $14.5 \pm 0.8$  &    $3.1 \pm 0.2$    &  $27.9 \pm 2.0$  &   $1.01 \pm 0.01$   &  $1.15 \pm 0.03$   &  $1234.25/1410$   \\
60380.93   &   $5.4 \pm 0.1$  &   $1.4 \pm 0.3$    &   $129 \pm  20 $   &    $2.8 \pm 0.3$    &  $14.2 \pm 1.1$  &    $2.9 \pm 0.2$    &  $27.9 \pm 2.0$  &   $1.03 \pm 0.01$   &  $0.96 \pm 0.03$   &  $1184.85/1410$   \\
60381.75   &   $5.5 \pm 0.2$  &   $1.2 \pm 0.2$    &   $339 \pm  49 $   &    $2.8 \pm 0.5$    &  $11.8 \pm 0.1$  &    $3.8 \pm 0.3$    &  $27.5 \pm 3.4$  &   $0.96 \pm 0.01$   &  $0.93 \pm 0.04$   &  $1270.61/1410$   \\
60382.93   &   $5.6 \pm 0.2$  &   $1.4 \pm 0.3$    &   $102 \pm  34 $   &    $2.7 \pm 0.5$    &  $12.0 \pm 0.3$  &    $3.5 \pm 0.3$    &  $29.2 \pm 3.2$  &   $0.94 \pm 0.01$   &  $1.06 \pm 0.04$   &  $1243.06/1410$   \\
60383.85   &   $5.0 \pm 0.2$  &   $1.6 \pm 0.3$    &   $ 69 \pm  19 $   &    $2.7 \pm 0.5$    &  $13.1 \pm 0.7$  &    $3.2 \pm 0.3$    &  $21.0 \pm 2.4$  &   $0.99 \pm 0.02$   &  $1.32 \pm 0.09$   &  $1299.35/1410$   \\
60383.98   &   $5.2 \pm 0.2$  &   $1.4 \pm 0.3$    &   $ 80 \pm  25 $   &    $2.8 \pm 0.5$    &  $13.2 \pm 1.2$  &    $3.1 \pm 0.3$    &  $23.2 \pm 2.6$  &   $1.02 \pm 0.02$   &  $1.03 \pm 0.03$   &  $1312.79/1410$   \\
60384.12   &   $5.2 \pm 0.1$  &   $1.4 \pm 0.2$    &   $ 75 \pm  18 $   &    $2.8 \pm 0.3$    &  $14.4 \pm 1.6$  &    $2.9 \pm 0.3$    &  $21.9 \pm 1.8$  &   $1.04 \pm 0.02$   &  $0.99 \pm 0.03$   &  $1188.29/1410$   \\
60386.59   &   $6.4 \pm 0.3$  &   $0.9 \pm 0.2$    &   $758 \pm 117 $   &    $3.5 \pm 0.9$    &  $ 8.5 \pm 0.3$  &    $2.8 \pm 0.2$    &  $34.5 \pm 6.8$  &   $1.05 \pm 0.02$   &  $1.07 \pm 0.05$   &  $1344.08/1410$   \\
\hline 
 \end{tabular}
\end{table*}

\vskip 0.5cm
%\newpage
\begin{table*}[!h]
%\scriptsize
\addtolength{\tabcolsep}{3.0pt}
 \centering
\caption{Properties from spectral analysis using Model-1 by freezing the column density ($N_H$) to $5.6 \times 10^{22}~cm^{-2}$. Column 1 represents the MJD of those respective Exposure IDs for 
which we have performed spectral analysis. Column 2 \& 3 gives the values of the parameters from the \textit{diskbb} model. Columns 4-7 give the values of the parameters from the \textit{broken 
power-law} model. Columns 8 \& 9 give the values of the constants needed to achieve simultaneous broadband fitting. Column 10 gives the values of the $\chi^2/DOF$ for each fitting. The errors are 
estimated with 90\% confidence interval, that corresponds to 1.645$\sigma$ in XSPEC.}
 \label{tab:table5}
 \begin{tabular}{|c|c|c|c|c|c|c|c|c|c|}
 \hline
 Time      &     \multicolumn{2}{|c|}{diskbb}      &                    \multicolumn{4}{|c|}{broken power-law}                       & \multicolumn{2}{|c|}{Fitting constants}  &     Fitting Stat  \\
\hline 
 MJD       & $T_{in}$ (keV)   &        Norm        &      $\Gamma1$      &   $E_b$ (keV)    &     $\Gamma2$       &        Norm      &     Constant1       &    Constant2       &     $\chi^2/DOF$  \\
\hline 
 (1)       &       (2)        &        (3)         &          (4)        &        (5)       &         (6)         &        (7)       &        (8)          &        (9)         &         (10)      \\ 
\hline 
60373.90   &   $1.6 \pm0.1 $    &   $ 116 \pm  7  $   &    $2.6 \pm0.3 $    &  $20.5 \pm1.2 $  &    $3.1 \pm0.2 $    &  $20.6 \pm1.2 $  &   $1.08 \pm0.02 $   &  $1.14 \pm0.02 $   &  $1261.70/1411$   \\
60374.98   &   $1.4 \pm0.1 $    &   $ 221 \pm 27  $   &    $2.6 \pm0.4 $    &  $16.1 \pm0.8 $  &    $2.9 \pm0.1 $    &  $27.6 \pm2.3 $  &   $1.00 \pm0.02 $   &  $0.97 \pm0.02 $   &  $1246.82/1411$   \\
60375.94   &   $1.4 \pm0.2 $    &   $ 141 \pm 25  $   &    $2.7 \pm0.4 $    &  $16.4 \pm0.6 $  &    $2.9 \pm0.3 $    &  $28.8 \pm3.2 $  &   $1.04 \pm0.03 $   &  $1.07 \pm0.02 $   &  $1204.47/1411$   \\
60376.96   &   $1.2 \pm0.1 $    &   $ 581 \pm 93  $   &    $2.7 \pm0.2 $    &  $ 6.9 \pm0.5 $  &    $3.1 \pm0.4 $    &  $28.9 \pm3.9 $  &   $1.05 \pm0.02 $   &  $1.01 \pm0.03 $   &  $1211.79/1328$   \\
60377.98   &   $1.2 \pm0.3 $    &   $ 537 \pm 68  $   &    $2.8 \pm0.2 $    &  $12.0 \pm0.7 $  &    $3.3 \pm0.4 $    &  $28.3 \pm3.4 $  &   $1.03 \pm0.02 $   &  $0.92 \pm0.02 $   &  $1210.63/1411$   \\
60378.83   &   $1.3 \pm0.1 $    &   $ 357 \pm 45  $   &    $2.8 \pm0.2 $    &  $11.5 \pm0.5 $  &    $3.2 \pm0.2 $    &  $30.7 \pm3.3 $  &   $0.98 \pm0.02 $   &  $1.09 \pm0.02 $   &  $1283.22/1411$   \\
60379.84   &   $1.4 \pm0.2 $    &   $ 110 \pm 15  $   &    $2.8 \pm0.3 $    &  $15.2 \pm1.2 $  &    $3.0 \pm0.2 $    &  $31.1 \pm2.3 $  &   $1.02 \pm0.04 $   &  $1.16 \pm0.02 $   &  $1236.82/1411$   \\
60380.93   &   $1.4 \pm0.1 $    &   $ 109 \pm 17  $   &    $2.8 \pm0.4 $    &  $14.9 \pm1.2 $  &    $2.9 \pm0.2 $    &  $30.3 \pm3.0 $  &   $1.03 \pm0.02 $   &  $0.96 \pm0.02 $   &  $1186.33/1411$   \\
60381.75   &   $1.2 \pm0.2 $    &   $ 320 \pm 27  $   &    $2.8 \pm0.4 $    &  $11.9 \pm1.1 $  &    $3.8 \pm0.2 $    &  $28.9 \pm3.2 $  &   $0.96 \pm0.02 $   &  $0.93 \pm0.04 $   &  $1270.78/1411$   \\
60382.93   &   $1.4 \pm0.2 $    &   $ 103 \pm 37  $   &    $2.7 \pm0.3 $    &  $12.1 \pm1.2 $  &    $3.5 \pm0.2 $    &  $29.1 \pm3.1 $  &   $0.94 \pm0.02 $   &  $1.06 \pm0.02 $   &  $1243.06/1411$   \\
60383.85   &   $1.8 \pm0.1 $    &   $  30 \pm  6  $   &    $2.8 \pm0.2 $    &  $14.9 \pm0.5 $  &    $3.1 \pm0.2 $    &  $28.9 \pm2.6 $  &   $1.00 \pm0.03 $   &  $1.33 \pm0.02 $   &  $1310.91/1411$   \\
60383.98   &   $1.5 \pm0.1 $    &   $  48 \pm  7  $   &    $2.8 \pm0.4 $    &  $14.1 \pm1.1 $  &    $3.0 \pm0.2 $    &  $27.8 \pm2.5 $  &   $1.02 \pm0.02 $   &  $1.03 \pm0.02 $   &  $1317.29/1411$   \\
60384.12   &   $1.5 \pm0.2 $    &   $  40 \pm  8  $   &    $2.8 \pm0.3 $    &  $15.4 \pm1.3 $  &    $2.9 \pm0.2 $    &  $27.3 \pm1.9 $  &   $1.03 \pm0.02 $   &  $0.98 \pm0.03 $   &  $1201.22/1411$   \\
60386.59   &   $0.9 \pm0.2 $    &   $1029 \pm108  $   &    $3.3 \pm0.8 $    &  $ 8.6 \pm0.6 $  &    $2.8 \pm0.3 $    &  $17.6 \pm5.9 $  &   $1.06 \pm0.02 $   &  $1.06 \pm0.05 $   &  $1355.04/1411$   \\
\hline 
 \end{tabular}
\end{table*}

%\newpage
\begin{table*}
%\scriptsize
 \addtolength{\tabcolsep}{1.5pt}
 \centering
 \caption{Properties from spectral analysis using Model-2. Column 1 represents the MJD of those respective Exposure IDs for which we have performed spectral analysis. Column 2 gives the values of 
 hydrogen column densities ($N_H$) of those analyzed exposures. Columns 3 \& 4 give the values of the parameters from the \textit{diskbb} model. Columns 5-8 give the values of the parameters from 
 the \textit{pexrav} model. Columns 9 \& 10 give the values of the constants needed to achieve simultaneous broadband fitting. Column 11 gives the values of the $\chi^2/DOF$ for each fitting. The 
 errors are estimated with 90\% confidence interval, that corresponds to 1.645$\sigma$ in XSPEC.} 
 \label{tab:table6}
% \resizebox{1 \textwidth}{!}{
 \begin{tabular}{|c|c|c|c|c|c|c|c|c|c|c|}
 \hline
 Time       &        TBabs       &         \multicolumn{2}{|c|}{diskbb}         &                               \multicolumn{4}{|c|}{pexrav}                                     &    \multicolumn{2}{|c|}{Fitting constants}    &    Fitting Stat  \\
\hline                                                                                                                                                                                                                           
 MJD        &        $N_H$       &    $T_{in}$ (keV)     &        Norm          &        $\Gamma$       &       $E_{cut}$       &      $rel_{frac}$       &         Norm         &       Constant1       &      Constant2        &    $\chi^2/DOF$  \\
\hline                                                                                                                                                                                                                                                                       
 (1)        &         (2)        &           (3)         &         (4)          &           (5)         &          (6)          &           (7)           &          (8)         &          (9)          &         (10)          &        (11)      \\ 
\hline
60373.90    &   $ 5.7 \pm 0.2 $  &    $ 1.6 \pm 0.3 $    &    $  98   \pm 2 $   &    $ 2.6 \pm 0.1 $    &     $  82 \pm 3  $    &    $ 0.32 \pm 0.01  $   &   $ 22.9 \pm 2.4 $   &   $ 1.07 \pm 0.01 $   &   $ 1.07 \pm 0.02 $   &   1554.44/1409   \\
60374.98    &   $ 5.8 \pm 0.2 $  &    $ 1.4 \pm 0.1 $    &    $ 198   \pm 5 $   &    $ 2.6 \pm 0.1 $    &     $  82 \pm 1  $    &    $ 0.01 \pm 0.01  $   &   $ 30.1 \pm 2.6 $   &   $ 0.98 \pm 0.02 $   &   $ 0.98 \pm 0.02 $   &   1451.32/1409   \\
60375.94    &   $ 5.6 \pm 0.3 $  &    $ 1.5 \pm 0.1 $    &    $ 130   \pm 3 $   &    $ 2.6 \pm 0.1 $    &     $  76 \pm 1  $    &    $ 0.05 \pm 0.01  $   &   $ 28.6 \pm 2.6 $   &   $ 1.04 \pm 0.02 $   &   $ 1.08 \pm 0.03 $   &   1397.76/1409   \\
60376.96    &   $ 5.1 \pm 0.4 $  &    $ 1.3 \pm 0.3 $    &    $ 477   \pm 8 $   &    $ 2.6 \pm 0.1 $    &     $  45 \pm 2  $    &    $ 0.10 \pm 0.01  $   &   $ 22.2 \pm 2.6 $   &   $ 1.09 \pm 0.03 $   &   $ 1.17 \pm 0.05 $   &   1522.15/1409   \\
60377.98    &   $ 5.9 \pm 0.4 $  &    $ 1.3 \pm 0.2 $    &    $ 404   \pm 8 $   &    $ 2.8 \pm 0.1 $    &     $  42 \pm 3  $    &    $ 0.05 \pm 0.01  $   &   $ 35.6 \pm 2.4 $   &   $ 1.07 \pm 0.02 $   &   $ 1.10 \pm 0.08 $   &   1400.06/1409   \\
60378.83    &   $ 6.2 \pm 0.3 $  &    $ 1.4 \pm 0.2 $    &    $ 166   \pm 5 $   &    $ 2.9 \pm 0.1 $    &     $ 106 \pm 4  $    &    $ 0.05 \pm 0.01  $   &   $ 46.9 \pm 2.4 $   &   $ 1.04 \pm 0.01 $   &   $ 1.15 \pm 0.04 $   &   1620.60/1409   \\
60379.84    &   $ 5.8 \pm 0.2 $  &    $ 1.5 \pm 0.2 $    &    $  74   \pm 3 $   &    $ 2.9 \pm 0.1 $    &     $ 179 \pm 2  $    &    $ 0.14 \pm 0.01  $   &   $ 35.9 \pm 2.4 $   &   $ 1.03 \pm 0.02 $   &   $ 1.17 \pm 0.04 $   &   1474.45/1409   \\
60380.93    &   $ 5.6 \pm 0.1 $  &    $ 1.4 \pm 0.3 $    &    $ 100   \pm 4 $   &    $ 2.8 \pm 0.1 $    &     $ 124 \pm 2  $    &    $ 0.05 \pm 0.01  $   &   $ 30.6 \pm 2.8 $   &   $ 1.04 \pm 0.02 $   &   $ 1.00 \pm 0.04 $   &   1409.79/1409   \\
60381.75    &   $ 6.1 \pm 0.1 $  &    $ 1.4 \pm 0.3 $    &    $ 120   \pm 3 $   &    $ 2.7 \pm 0.1 $    &     $  16 \pm 1  $    &    $ 0.05 \pm 0.01  $   &   $ 35.9 \pm 2.8 $   &   $ 1.04 \pm 0.02 $   &   $ 1.33 \pm 0.07 $   &   1632.36/1409   \\
60382.93    &   $ 6.3 \pm 0.2 $  &    $ 2.1 \pm 0.2 $    &    $  16   \pm 1 $   &    $ 2.9 \pm 0.2 $    &     $  53 \pm 2  $    &    $ 0.05 \pm 0.01  $   &   $ 42.6 \pm 2.8 $   &   $ 0.99 \pm 0.01 $   &   $ 1.13 \pm 0.06 $   &   1431.01/1409   \\
60383.85    &   $ 5.4 \pm 0.3 $  &    $ 1.8 \pm 0.2 $    &    $  31   \pm 1 $   &    $ 2.7 \pm 0.1 $    &     $  44 \pm 1  $    &    $ 0.15 \pm 0.01  $   &   $ 25.3 \pm 2.8 $   &   $ 1.01 \pm 0.02 $   &   $ 1.48 \pm 0.05 $   &   1433.76/1409   \\
60383.98    &   $ 5.7 \pm 0.2 $  &    $ 1.6 \pm 0.2 $    &    $  36   \pm 2 $   &    $ 2.9 \pm 0.1 $    &     $ 211 \pm 2  $    &    $ 0.05 \pm 0.01  $   &   $ 29.2 \pm 2.8 $   &   $ 1.04 \pm 0.02 $   &   $ 1.03 \pm 0.04 $   &   1440.43/1409   \\
60384.12    &   $ 5.5 \pm 0.2 $  &    $ 1.6 \pm 0.2 $    &    $  41   \pm 2 $   &    $ 2.9 \pm 0.1 $    &     $ 373 \pm 5  $    &    $ 0.11 \pm 0.01  $   &   $ 27.2 \pm 2.8 $   &   $ 1.05 \pm 0.01 $   &   $ 1.02 \pm 0.04 $   &   1321.88/1409   \\
60386.59    &   $ 5.7 \pm 0.1 $  &    $ 0.9 \pm 0.1 $    &    $1320   \pm72 $   &    $ 3.3 \pm 0.2 $    &     $     374    $    &    $ 1.95 \pm 0.19  $   &   $ 18.5 \pm 1.9 $   &   $ 1.05 \pm 0.02 $   &   $ 1.64 \pm 0.11 $   &   1420.99/1410   \\
\hline 
 \end{tabular}
\end{table*}

%\newpage
\begin{table*}
%\scriptsize
 \addtolength{\tabcolsep}{1.5pt}
 \centering
 \caption{Properties from spectral analysis using Model-2 by freezing the column density ($N_H$) to $5.6 \times 10^{22}~cm^{-2}$. Column 1 represents the MJD of those respective Exposure IDs for 
 which we have performed spectral analysis. Columns 2 \& 3 give the values of the parameters from the \textit{diskbb} model. Columns 4-7 give the values of the parameters from the \textit{pexrav} 
 model. Columns 8 \& 9 give the values of the constants needed to achieve simultaneous broadband fitting. Column 10 gives the values of the $\chi^2/DOF$ for each fitting. The errors are estimated 
 with 90\% confidence interval, that corresponds to 1.645$\sigma$ in XSPEC.} 
 \label{tab:table6}
% \resizebox{1 \textwidth}{!}{
 \begin{tabular}{|c|c|c|c|c|c|c|c|c|c|}
 \hline
 Time       &       \multicolumn{2}{|c|}{diskbb}         &                               \multicolumn{4}{|c|}{pexrav}                                     &    \multicolumn{2}{|c|}{Fitting constants}    &    Fitting Stat  \\
\hline                                                                                                                                                                                                    
 MJD        &  $T_{in}$ (keV)     &        Norm          &        $\Gamma$       &       $E_{cut}$       &      $rel_{frac}$       &         Norm         &       Constant1       &      Constant2        &    $\chi^2/DOF$  \\
\hline                                                                                                                                                                                                                                                
 (1)        &         (2)        &           (3)         &         (4)          &           (5)         &          (6)          &           (7)           &          (8)         &          (9)          &         (10)      \\ 
\hline
60373.90    &     $1.6  \pm 0.2  $    &    $ 109  \pm 2 $   &    $2.5  \pm 0.1  $    &     $  70 \pm  3 $    &    $ 0.22 \pm 0.01  $  &   $ 20.7 \pm  2.2$   &   $ 1.07 \pm 0.01 $   &   $ 1.06 \pm 0.04 $    &  1555.59/1410  \\
60374.98    &     $1.4  \pm 0.2  $    &    $ 212  \pm 4 $   &    $2.6  \pm 0.1  $    &     $  73 \pm  4 $    &    $ 0.05 \pm 0.01  $  &   $ 27.3 \pm  2.4$   &   $ 0.99 \pm 0.02 $   &   $ 0.98 \pm 0.03 $    &  1453.48/1410  \\
60375.94    &     $1.5  \pm 0.3  $    &    $ 130  \pm 4 $   &    $2.6  \pm 0.1  $    &     $  76 \pm  4 $    &    $ 0.05 \pm 0.01  $  &   $ 28.5 \pm  2.4$   &   $ 1.04 \pm 0.04 $   &   $ 1.08 \pm 0.02 $    &  1397.81/1410  \\
60376.96    &     $1.4  \pm 0.4  $    &    $ 563  \pm18 $   &    $2.6  \pm 0.1  $    &     $  35 \pm  2 $    &    $ 0.10 \pm 0.01  $  &   $ 25.8 \pm  2.1$   &   $ 1.05 \pm 0.03 $   &   $ 1.26 \pm 0.04 $    &  1594.50/1410  \\
60377.98    &     $1.4  \pm 0.4  $    &    $ 457  \pm12 $   &    $2.7  \pm 0.1  $    &     $  33 \pm  3 $    &    $ 0.05 \pm 0.01  $  &   $ 28.7 \pm  2.1$   &   $ 1.07 \pm 0.02 $   &   $ 1.14 \pm 0.06 $    &  1403.88/1410  \\
60378.83    &     $1.4  \pm 0.3  $    &    $ 241  \pm 9 $   &    $2.8  \pm 0.1  $    &     $  53 \pm  4 $    &    $ 0.05 \pm 0.01  $  &   $ 33.1 \pm  2.3$   &   $ 1.05 \pm 0.01 $   &   $ 1.23 \pm 0.02 $    &  1657.64/1410  \\
60379.84    &     $1.4  \pm 0.2  $    &    $  98  \pm 3 $   &    $2.8  \pm 0.1  $    &     $ 117 \pm  9 $    &    $ 0.05 \pm 0.01  $  &   $ 31.1 \pm  1.9$   &   $ 1.03 \pm 0.02 $   &   $ 1.16 \pm 0.05 $    &  1477.93/1410  \\
60380.93    &     $1.4  \pm 0.1  $    &    $ 102  \pm 3 $   &    $2.8  \pm 0.1  $    &     $ 123 \pm  8 $    &    $ 0.05 \pm 0.01  $  &   $ 30.2 \pm  2.7$   &   $ 1.04 \pm 0.03 $   &   $ 1.00 \pm 0.03 $    &  1409.89/1410  \\
60381.75    &     $1.3  \pm 0.1  $    &    $ 200  \pm 3 $   &    $2.5  \pm 0.1  $    &     $  13 \pm  1 $    &    $ 0.05 \pm 0.01  $  &   $ 27.7 \pm  2.4$   &   $ 1.03 \pm 0.02 $   &   $ 1.42 \pm 0.06 $    &  1641.72/1410  \\
60382.93    &     $1.7  \pm 0.2  $    &    $  42  \pm 3 $   &    $2.7  \pm 0.2  $    &     $  26 \pm  2 $    &    $ 0.05 \pm 0.01  $  &   $ 29.8 \pm  2.3$   &   $ 1.01 \pm 0.01 $   &   $ 1.31 \pm 0.03 $    &  1470.61/1410  \\
60383.85    &     $1.9  \pm 0.3  $    &    $  24  \pm 1 $   &    $2.8  \pm 0.1  $    &     $  60 \pm  3 $    &    $ 0.25 \pm 0.01  $  &   $ 29.2 \pm  2.3$   &   $ 1.01 \pm 0.02 $   &   $ 1.45 \pm 0.04 $    &  1434.88/1410  \\
60383.98    &     $1.6  \pm 0.2  $    &    $  39  \pm 2 $   &    $2.8  \pm 0.1  $    &     $ 197 \pm  4 $    &    $ 0.06 \pm 0.01  $  &   $ 28.4 \pm  2.1$   &   $ 1.05 \pm 0.02 $   &   $ 1.03 \pm 0.03 $    &  1440.85/1410  \\
60384.12    &     $1.6  \pm 0.2  $    &    $  37  \pm 2 $   &    $2.9  \pm 0.1  $    &     $ 529 \pm  6 $    &    $ 0.22 \pm 0.01  $  &   $ 28.3 \pm  1.8$   &   $ 1.05 \pm 0.02 $   &   $ 1.02 \pm 0.06 $    &  1322.09/1410  \\
60386.59    &     $0.9  \pm 0.1  $    &    $1305  \pm97 $   &    $3.3  \pm 0.2  $    &     $     373    $    &    $ 1.68 \pm 0.15  $  &   $ 16.3 \pm  3.9$   &   $ 1.04 \pm 0.06 $   &   $ 1.54 \pm 0.09 $    &  1419.64/1411  \\
\hline 
 \end{tabular}
\end{table*}

%\newpage
\begin{table*}
\scriptsize
 \addtolength{\tabcolsep}{3.5pt}
 \centering
 \caption{Properties from spectral analysis using Model-3. Column 1 represents the MJD of those respective Exposure IDs for which we have performed spectral analysis. Column 2 gives the values of 
 hydrogen column densities ($N_H$) of those analyzed exposures. Columns 3 \& 4 give the values of the parameters from the \textit{diskbb} model. Columns 5-8 give the values of the parameters from 
 the \textit{pexriv} model. Columns 9 \& 10 give the values of the constants needed to achieve simultaneous broadband fitting. Column 11 gives the values of the $\chi^2/DOF$ for each fitting. For 
 this model fitting, we have fixed the $E_{cut}$ of this model to the $E_{cut}$ of the \textit{pexrav} model. Also, the disk temperature was set to $10^6$~Kelvin. Since, the ionization  parameter 
 $\xi$ has so small value, we did not estimate the error for this parameter. The errors are estimated with 90\% confidence interval, that corresponds to 1.645$\sigma$ in XSPEC.}
 \label{tab:table6}
% \resizebox{1 \textwidth}{!}{
 \begin{tabular}{|c|c|c|c|c|c|c|c|c|c|c|}
 \hline
 Time       &        TBabs       &         \multicolumn{2}{|c|}{diskbb}         &                               \multicolumn{4}{|c|}{pexriv}                                     &    \multicolumn{2}{|c|}{Fitting constants}    &    Fitting Stat  \\
\hline                                                                                                                                                                                                                           
 MJD        &        $N_H$       &    $T_{in}$ (keV)     &        Norm          &        $\Gamma$       &      $rel_{frac}$     &         $\xi$           &         Norm         &       Constant1       &      Constant2        &    $\chi^2/DOF$  \\
\hline                                                                                                                                                                                                                                                                       
 (1)        &         (2)        &           (3)         &         (4)          &           (5)         &          (6)          &           (7)           &          (8)         &          (9)          &         (10)          &        (11)      \\ 
\hline
60373.90    &   $6.9 \pm 0.2$    &    $1.5 \pm 0.2$      &    $121  \pm  2 $    &     $2.6 \pm 0.2 $    &   $0.13   \pm  0.02$  &         9.2E-09         &    $27.9 \pm 2.2$    &    $1.06 \pm 0.01$    &    $1.07 \pm 0.02$    &  1201.95/1409    \\
60374.98    &   $6.2 \pm 0.2$    &    $1.4 \pm 0.1$      &    $248  \pm  4 $    &     $2.6 \pm 0.1 $    &   $0.01   \pm  0.01$  &         4.4E-13         &    $30.1 \pm 2.4$    &    $1.00 \pm 0.01$    &    $1.00 \pm 0.03$    &  1210.39/1409    \\
60375.94    &   $5.5 \pm 0.3$    &    $1.5 \pm 0.1$      &    $136  \pm  3 $    &     $2.6 \pm 0.1 $    &   $0.01   \pm  0.01$  &         1.6E-10         &    $27.6 \pm 2.4$    &    $1.05 \pm 0.01$    &    $1.10 \pm 0.03$    &  1230.26/1409    \\
60376.96    &   $5.0 \pm 0.3$    &    $1.3 \pm 0.1$      &    $447  \pm  8 $    &     $2.6 \pm 0.2 $    &   $0.10   \pm  0.01$  &         1.0E-08         &    $22.5 \pm 2.4$    &    $1.07 \pm 0.02$    &    $1.16 \pm 0.05$    &  1279.92/1409    \\
60377.98    &   $5.9 \pm 0.4$    &    $1.3 \pm 0.1$      &    $389  \pm  8 $    &     $2.8 \pm 0.1 $    &   $0.01   \pm  0.01$  &         2.4E-13         &    $36.1 \pm 2.2$    &    $1.06 \pm 0.02$    &    $1.09 \pm 0.07$    &  1253.45/1409    \\
60378.83    &   $6.4 \pm 0.3$    &    $1.5 \pm 0.1$      &    $135  \pm  5 $    &     $3.0 \pm 0.1 $    &   $0.05   \pm  0.01$  &         2.9E-09         &    $52.4 \pm 2.2$    &    $1.03 \pm 0.01$    &    $1.19 \pm 0.04$    &  1392.68/1409    \\
60379.84    &   $5.8 \pm 0.2$    &    $1.5 \pm 0.1$      &    $76   \pm  4 $    &     $2.9 \pm 0.1 $    &   $0.15   \pm  0.05$  &         1.6E-10         &    $35.8 \pm 2.2$    &    $1.03 \pm 0.01$    &    $1.17 \pm 0.03$    &  1258.84/1409    \\
60380.93    &   $5.6 \pm 0.1$    &    $1.4 \pm 0.1$      &    $96   \pm  3 $    &     $2.8 \pm 0.1 $    &   $0.06   \pm  0.01$  &         6.6E-09         &    $31.0 \pm 2.6$    &    $1.04 \pm 0.01$    &    $1.00 \pm 0.04$    &  1195.57/1409    \\
60381.75    &   $6.0 \pm 0.1$    &    $1.4 \pm 0.1$      &    $112  \pm  2 $    &     $2.7 \pm 0.1 $    &   $0.05   \pm  0.01$  &         5.8E-12         &    $36.3 \pm 2.5$    &    $1.03 \pm 0.01$    &    $1.33 \pm 0.06$    &  1397.17/1409    \\
60382.93    &   $6.4 \pm 0.2$    &    $2.1 \pm 0.1$      &    $16   \pm  1 $    &     $2.9 \pm 0.2 $    &   $0.14   \pm  0.01$  &         1.5E-09         &    $44.2 \pm 2.5$    &    $0.99 \pm 0.01$    &    $1.17 \pm 0.06$    &  1285.72/1409    \\
60383.85    &   $5.4 \pm 0.3$    &    $1.8 \pm 0.1$      &    $30   \pm  1 $    &     $2.7 \pm 0.1 $    &   $0.18   \pm  0.01$  &         2.6E-09         &    $25.9 \pm 2.6$    &    $1.01 \pm 0.02$    &    $1.51 \pm 0.15$    &  1329.20/1409    \\
60383.98    &   $5.7 \pm 0.2$    &    $1.7 \pm 0.1$      &    $33   \pm  2 $    &     $2.9 \pm 0.1 $    &   $0.17   \pm  0.01$  &         2.2E-11         &    $30.8 \pm 2.6$    &    $1.04 \pm 0.02$    &    $1.07 \pm 0.04$    &  1326.96/1409    \\
60384.12    &   $5.5 \pm 0.2$    &    $1.6 \pm 0.1$      &    $44   \pm  2 $    &     $2.9 \pm 0.1 $    &   $0.18   \pm  0.01$  &         6.5E-11         &    $26.9 \pm 2.6$    &    $1.05 \pm 0.02$    &    $1.03 \pm 0.03$    &  1185.50/1409    \\
60386.59    &   $5.8 \pm 0.2$    &    $0.9 \pm 0.1$      &    $1437 \pm 11 $    &     $3.4 \pm 0.4 $    &   $2.24   \pm  0.15$  &         3.8E-08         &    $18.4 \pm 1.3$    &    $1.00 \pm 0.02$    &    $1.64 \pm 0.14$    &  1277.63/1409    \\
\hline 
\end{tabular}
\end{table*}
%\clearpage

\begin{table*}
\scriptsize
 \addtolength{\tabcolsep}{3.5pt}
 \centering
 \caption{Properties from spectral analysis using Model-3 by freezing the column density ($N_H$) to $5.6 \times 10^{22}~cm^{-2}$. Column 1 represents the MJD of those respective Exposure IDs for 
 which we have performed spectral analysis. Columns 2 \& 3 give the values of the parameters from the \textit{diskbb} model. Columns 4-7 give the values of the parameters from the \textit{pexriv} 
 model. Columns 8 \& 9 give the values of the constants needed to achieve simultaneous broadband fitting. Column 10 gives the values of the $\chi^2/DOF$ for each fitting. For this model fitting, 
 we have fixed the $E_{cut}$ of this model to the $E_{cut}$ of the \textit{pexrav} model for fixed $N_H$. Also, the disk temperature was set to $10^6$~Kelvin. Since, the ionization parameter $\xi$ 
 has so small value, we did not estimate the error for this parameter. The errors are estimated with 90\% confidence interval, that corresponds to 1.645$\sigma$ in XSPEC.}
 \label{tab:table6}
% \resizebox{1 \textwidth}{!}{
 \begin{tabular}{|c|c|c|c|c|c|c|c|c|c|}
 \hline
 Time       &         \multicolumn{2}{|c|}{diskbb}         &                               \multicolumn{4}{|c|}{pexriv}                                     &    \multicolumn{2}{|c|}{Fitting constants}    &    Fitting Stat  \\
\hline                                                                                                                                                                                                      
 MJD        &    $T_{in}$ (keV)     &        Norm          &        $\Gamma$       &      $rel_{frac}$     &         $\xi$           &         Norm         &       Constant1       &      Constant2        &    $\chi^2/DOF$  \\
\hline                                                                                                                                                                                                                                                  
 (1)        &         (2)        &           (3)         &         (4)          &           (5)         &          (6)          &           (7)           &          (8)         &          (9)          &         (10)        \\ 
\hline
60373.90    &    $1.6 \pm 0.2$      &    $ 106 \pm  2 $    &     $2.6 \pm 0.2 $    &   $ 0.29  \pm  0.02$  &      1.3E-08            &    $21.6 \pm 2.2$    &    $1.08 \pm 0.02$    &    $1.08 \pm0.02 $    &   1234.40/1410   \\
60374.98    &    $1.4 \pm 0.2$      &    $ 207 \pm  4 $    &     $2.6 \pm 0.4 $    &   $ 0.05  \pm  0.03$  &      9.8E-17            &    $27.9 \pm 2.4$    &    $1.01 \pm 0.01$    &    $1.01 \pm0.03 $    &   1218.89/1410   \\
60375.94    &    $1.5 \pm 0.2$      &    $ 133 \pm  3 $    &     $2.6 \pm 0.4 $    &   $ 0.06  \pm  0.01$  &      2.4E-13            &    $28.7 \pm 2.4$    &    $1.05 \pm 0.03$    &    $1.09 \pm0.01 $    &   1232.77/1410   \\
60376.96    &    $1.3 \pm 0.2$      &    $ 443 \pm  9 $    &     $2.7 \pm 0.3 $    &   $ 0.10  \pm  0.03$  &      2.2E-12            &    $30.7 \pm 2.1$    &    $1.03 \pm 0.02$    &    $1.20 \pm0.05 $    &   1286.38/1328   \\
60377.98    &    $1.3 \pm 0.2$      &    $ 419 \pm  7 $    &     $2.8 \pm 0.7 $    &   $ 0.06  \pm  0.01$  &      3.4E-15            &    $30.0 \pm 2.1$    &    $1.08 \pm 0.02$    &    $1.07 \pm0.05 $    &   1257.22/1410   \\
60378.83    &    $1.4 \pm 0.2$      &    $ 220 \pm  5 $    &     $2.9 \pm 0.4 $    &   $ 0.05  \pm  0.01$  &      1.3E-17            &    $35.1 \pm 2.3$    &    $1.08 \pm 0.01$    &    $1.12 \pm0.04 $    &   1438.47/1411   \\
60379.84    &    $1.5 \pm 0.1$      &    $  89 \pm  3 $    &     $2.8 \pm 0.3 $    &   $ 0.11  \pm  0.04$  &      1.5E-14            &    $32.3 \pm 2.9$    &    $1.04 \pm 0.04$    &    $1.15 \pm0.03 $    &   1268.14/1410   \\
60380.93    &    $1.4 \pm 0.2$      &    $  96 \pm  4 $    &     $2.8 \pm 0.3 $    &   $ 0.07  \pm  0.01$  &      1.2E-07            &    $30.6 \pm 2.7$    &    $1.04 \pm 0.01$    &    $1.01 \pm0.04 $    &   1195.73/1410   \\
60381.75    &    $1.4 \pm 0.2$      &    $ 158 \pm  3 $    &     $2.6 \pm 0.1 $    &   $ 0.05  \pm  0.01$  &      4.3E-11            &    $29.9 \pm 2.4$    &    $1.05 \pm 0.01$    &    $1.30 \pm0.06 $    &   1423.16/1411   \\
60382.93    &    $1.8 \pm 0.1$      &    $  45 \pm  2 $    &     $2.8 \pm 0.2 $    &   $ 0.18  \pm  0.02$  &      9.5E-15            &    $31.9 \pm 2.3$    &    $1.04 \pm 0.02$    &    $1.09 \pm0.03 $    &   1353.27/1410   \\
60383.85    &    $1.9 \pm 0.2$      &    $  21 \pm  1 $    &     $2.8 \pm 0.3 $    &   $ 0.24  \pm  0.02$  &      2.0E-13            &    $28.7 \pm 2.3$    &    $0.99 \pm 0.02$    &    $1.52 \pm0.12 $    &   1335.43/1410   \\
60383.98    &    $1.6 \pm 0.1$      &    $  40 \pm  2 $    &     $2.9 \pm 0.3 $    &   $ 0.13  \pm  0.01$  &      1.5E-12            &    $28.8 \pm 2.1$    &    $1.05 \pm 0.02$    &    $1.06 \pm0.04 $    &   1327.92/1410   \\
60384.12    &    $1.6 \pm 0.2$      &    $  40 \pm  3 $    &     $2.9 \pm 0.3 $    &   $ 0.15  \pm  0.02$  &      2.2E-11            &    $27.8 \pm 2.8$    &    $1.04 \pm 0.03$    &    $1.01 \pm0.03 $    &   1188.27/1409   \\
60386.59    &    $0.9 \pm 0.3$      &    $1389 \pm 23 $    &     $3.4 \pm 0.8 $    &   $ 2.43  \pm  0.23$  &      3.7E-10            &    $16.9 \pm 3.2$    &    $1.00 \pm 0.07$    &    $1.59 \pm0.16 $    &   1281.48/1410   \\
\hline 
\end{tabular}
\end{table*}
%\clearpage

\begin{table*}
\scriptsize
 \addtolength{\tabcolsep}{2.5pt}
 \centering
 \caption{Properties from spectral analysis using Model-4 by freezing the column density ($N_H$) to $5.6 \times 10^{22}~cm^{-2}$. Column 1 represents the MJD of those respective Exposure IDs for 
 which we have performed spectral analysis. Columns 2 \& 3 give the values of the parameters from the \textit{diskbb} model. Columns 4-11 give the values of the parameters from the \textit{relxill} 
 model. Columns 12 \& 13 give the values of the constants needed to achieve simultaneous broadband fitting. Column 14 gives the values of the $\chi^2/DOF$ for each fitting. The errors are estimated 
 with 90\% confidence interval, that corresponds to 1.645$\sigma$ in XSPEC.}
 \label{tab:table6}
% \resizebox{1 \textwidth}{!}{
 \begin{tabular}{|c|c|c|c|c|c|c|c|c|c|c|c|c|c|}
 \hline
 Time       &         \multicolumn{2}{|c|}{diskbb}         &                                          \multicolumn{8}{|c|}{relxill}                                                                &    \multicolumn{2}{|c|}{Fitting constants}    &    Fitting Stat  \\
\hline                                                                                                                                                                                                      
 MJD        &    $T_{in}$ (keV)     &        Norm          &     Spin    &    Incl. (i)  &   $\Gamma$     &    logxi   &     $AF_e$   &     $E_{cut}$       &  $rel_{frac}$     &     Norm         &       Constant1       &      Constant2        &    $\chi^2/DOF$  \\
\hline                                                                                                                                                                                                                                                  
 (1)        &         (2)           &         (3)          &     (4)     &       (5)     &      (6)       &     (7)    &       (8)    &         (9)         &      (10)         &     (11)         &         (12)          &         (13)          &        (14)      \\ 
\hline
60373.90    &    $ 1.54\pm 0.1$      &    $  107\pm  2 $    &     $ 0.81\pm 0.16 $    &   $ 17  \pm 4 $  &    $  2.4 \pm 0.1 $    &    $4.3 \pm 0.1$    &    $ 1.5\pm 0.1$    &    $ 76 \pm  3 $    &     $0.45 \pm 0.12$      &     $0.15 \pm 0.01$      &    $ 1.07\pm 0.01$      &   $1.07 \pm 0.02$      &  1223.49/1406    \\
60374.98    &    $ 1.47\pm 0.1$      &    $  146\pm  5 $    &     $ 0.74\pm 0.11 $    &   $ 19  \pm 4 $  &    $  2.6 \pm 0.1 $    &    $4.5 \pm 0.3$    &    $ 1.2\pm 0.1$    &    $145 \pm 12 $    &     $0.85 \pm 0.21$      &     $0.20 \pm 0.01$      &    $ 1.01\pm 0.02$      &   $1.02 \pm 0.02$      &  1246.18/1406    \\
60375.94    &    $ 1.56\pm 0.1$      &    $   87\pm  3 $    &     $ 0.81\pm 0.13 $    &   $ 20  \pm 3 $  &    $  2.6 \pm 0.1 $    &    $4.5 \pm 0.2$    &    $ 1.4\pm 0.1$    &    $133 \pm 11 $    &     $0.91 \pm 0.21$      &     $0.19 \pm 0.01$      &    $ 1.05\pm 0.02$      &   $1.09 \pm 0.03$      &  1204.20/1406    \\
60376.96    &    $ 1.36\pm 0.1$      &    $  411\pm  8 $    &     $ 0.81\pm 0.11 $    &   $ 24  \pm 4 $  &    $  2.6 \pm 0.1 $    &    $4.3 \pm 0.4$    &    $ 1.4\pm 0.1$    &    $ 97 \pm  5 $    &     $2.29 \pm 0.19$      &     $0.07 \pm 0.01$      &    $ 1.14\pm 0.03$      &   $1.18 \pm 0.05$      &  1311.53/1323    \\
60377.98    &    $ 1.24\pm 0.1$      &    $  499\pm  8 $    &     $ 0.81\pm 0.18 $    &   $ 10  \pm 3 $  &    $  2.4 \pm 0.1 $    &    $4.2 \pm 0.3$    &    $ 1.1\pm 0.1$    &    $ 30 \pm  1 $    &     $1.75 \pm 0.19$      &     $0.07 \pm 0.01$      &    $ 1.07\pm 0.02$      &   $1.23 \pm 0.08$      &  1321.49/1406    \\
60378.83    &    $ 1.31\pm 0.1$      &    $  314\pm  5 $    &     $ 0.81\pm 0.13 $    &   $ 12  \pm 2 $  &    $  2.4 \pm 0.1 $    &    $4.3 \pm 0.2$    &    $ 1.1\pm 0.1$    &    $ 33 \pm  1 $    &     $1.53 \pm 0.10$      &     $0.08 \pm 0.01$      &    $ 1.01\pm 0.01$      &   $1.36 \pm 0.04$      &  1592.23/1406    \\
60379.84    &    $ 1.34\pm 0.1$      &    $  188\pm  4 $    &     $ 0.81\pm 0.11 $    &   $ 28  \pm 2 $  &    $  2.4 \pm 0.1 $    &    $4.4 \pm 0.2$    &    $ 1.3\pm 0.1$    &    $ 64 \pm  1 $    &     $1.54 \pm 0.10$      &     $0.07 \pm 0.01$      &    $ 1.02\pm 0.02$      &   $1.26 \pm 0.04$      &  1411.99/1406    \\
60380.93    &    $ 1.28\pm 0.1$      &    $  208\pm  3 $    &     $ 0.81\pm 0.32 $    &   $ 36  \pm 4 $  &    $  2.4 \pm 0.1 $    &    $4.7 \pm 0.4$    &    $ 1.9\pm 0.3$    &    $ 56 \pm 12 $    &     $1.57 \pm 0.17$      &     $0.07 \pm 0.01$      &    $ 1.02\pm 0.02$      &   $1.11 \pm 0.04$      &  1325.13/1406    \\
60381.75    &    $ 1.46\pm 0.1$      &    $  105\pm  3 $    &     $ 0.65\pm 0.22 $    &   $ 41  \pm 5 $  &    $  2.8 \pm 0.1 $    &    $3.7 \pm 0.6$    &    $ 1.6\pm 0.1$    &    $ 19 \pm  1 $    &     $1.26 \pm 0.13$      &     $0.58 \pm 0.01$      &    $ 1.05\pm 0.02$      &   $1.22 \pm 0.06$      &  1439.98/1407    \\
60382.93    &    $ 1.79\pm 0.1$      &    $   30\pm  1 $    &     $ 0.68\pm 0.22 $    &   $ 37  \pm 5 $  &    $  2.8 \pm 0.1 $    &    $3.9 \pm 0.1$    &    $ 3.9\pm 0.1$    &    $ 32 \pm  1 $    &     $1.02 \pm 0.17$      &     $0.52 \pm 0.01$      &    $ 1.01\pm 0.01$      &   $1.22 \pm 0.06$      &  1355.12/1407    \\
60383.85    &    $ 1.90\pm 0.1$      &    $   22\pm  1 $    &     $ 0.50\pm 0.11 $    &   $ 31  \pm 3 $  &    $  2.8 \pm 0.1 $    &    $3.8 \pm 0.1$    &    $ 1.7\pm 0.2$    &    $ 62 \pm  1 $    &     $0.12 \pm 0.11$      &     $0.43 \pm 0.01$      &    $ 1.01\pm 0.02$      &   $1.44 \pm 0.01$      &  1322.27/1406    \\
60383.98    &    $ 1.59\pm 0.1$      &    $   37\pm  2 $    &     $ 0.78\pm 0.22 $    &   $ 29  \pm 9 $  &    $  2.8 \pm 0.1 $    &    $4.1 \pm 0.1$    &    $ 4.3\pm 0.7$    &    $143 \pm 21 $    &     $1.07 \pm 0.16$      &     $0.45 \pm 0.01$      &    $ 1.06\pm 0.02$      &   $1.06 \pm 0.04$      &  1323.12/1406    \\
60384.12    &    $ 1.50\pm 0.1$      &    $   37\pm  2 $    &     $ 0.50\pm 0.22 $    &   $ 24  \pm 1 $  &    $  2.8 \pm 0.1 $    &    $4.3 \pm 0.1$    &    $ 4.9\pm 0.8$    &    $248 \pm 53 $    &     $1.40 \pm 0.10$      &     $0.43 \pm 0.01$      &    $ 1.06\pm 0.01$      &   $1.02 \pm 0.04$      &  1168.19/1406    \\
60386.59    &    $ 0.89\pm 0.1$      &    $ 1532\pm 17 $    &     $ 0.50\pm 0.11 $    &   $ 30  \pm 1 $  &    $  3.0 \pm 0.1 $    &    $3.3 \pm 0.1$    &    $ 3.8\pm 0.6$    &    $954 \pm 408$    &     $0.82 \pm 0.12$      &     $0.22 \pm 0.01$      &    $ 1.11\pm 0.03$      &   $1.18 \pm 0.09$      &  1194.31/1406    \\
\hline 
\end{tabular}
\end{table*}
\clearpage

\end{appendix}

\end{document}